\documentclass[PAPER, atlasdraft=false, coverpage=false, texlive=2023, USenglish, cernpreprint=true, texmf, orcidlogo]{atlasdoc}

\usepackage{atlaspackage}
\usepackage{siunitx}
\usepackage{rotating}
\usepackage{atlasbiblatex}
 
\usepackage[hepparticle,hepprocess]{atlasphysics}

\graphicspath{{logos/}{figures/}}
 
\usepackage{ANA-HDBS-2019-29-PAPER-defs}

\AtlasTitle{Search for nonresonant pair production of Higgs bosons in the $b\bar{b}b\bar{b}$ final state in $pp$ collisions at $\sqrt{s}= 13$~TeV with the ATLAS detector}

\AtlasAbstract{
A search for nonresonant Higgs boson pair production in the $b\bar{b}b\bar{b}$ final state is presented.
The analysis uses 126~fb$^{-1}$ of $pp$ collision data at $\sqrt{s}={13}$~TeV collected with the ATLAS detector at the Large Hadron Collider, and targets both the gluon--gluon fusion and vector-boson fusion production modes.
No evidence of the signal is found and the observed (expected) upper limit on the cross-section for nonresonant Higgs boson pair production is determined to be 5.4 (8.1) times the Standard Model predicted cross-section at 95\% confidence level.
Constraints are placed on modifiers to the $HHH$ and $HHVV$ couplings.
The observed (expected) $2\sigma$ constraints on the $HHH$ coupling modifier, $\kappa_\lambda$, are determined to be  $[-3.5, 11.3]$ ($[-5.4, 11.4]$), while the corresponding constraints for the $HHVV$ coupling modifier, $\kappa_{2V}$, are $[-0.0, 2.1]$ ($[-0.1, 2.1]$).
In addition, constraints on relevant coefficients are derived in the context of the Standard Model effective field theory and Higgs effective field theory, and upper limits on the $HH$ production cross-section are placed in seven Higgs effective field theory benchmark scenarios.
}

\PreprintIdNumber{CERN-EP-2022-235}

\AtlasJournal{Phys.\ Rev.\ D.}
\AtlasJournalRef{Phys. Rev. D 108 (2023) 052003}
\AtlasDOI{10.1103/PhysRevD.108.052003}

\hypersetup{pdftitle={ATLAS document},pdfauthor={The ATLAS Collaboration}}
 
\begin{document}
 
\maketitle

\section{Introduction}
\label{sec:intro}

% The next lines are included from the .//sections/intro.tex input file
 
The discovery of the \SI{125}{\GeV} Higgs boson (\(H\))~\cite{HIGG-2012-27,CMS-HIG-12-028,HIGG-2018-04,CMS-HIG-18-016} at the Large Hadron Collider (LHC) has prompted a broad research program to investigate its properties and compare the measurements with the Standard Model (SM) predictions.
Of particular interest is the search for nonresonant Higgs boson pair production, also known as di-Higgs (\HH) production.
This process has a strong dependence on the Higgs self-coupling, which is a key ingredient of the electroweak symmetry breaking mechanism and a sensitive probe for physics beyond the SM (BSM physics) in various scenarios, such as two-Higgs-doublet models~\cite{THDM}, composite Higgs models~\cite{Grober:2010yv}, twin Higgs models~\cite{Chacko:2005vw}, and the minimal supersymmetric extension of the SM~\cite{Fayet:1976et,Fayet:1977yc}.
The Higgs self-coupling also plays a fundamental role in understanding the stability of the universe~\cite{Degrassi_2012}.
 
\begin{figure}[!tbh]
\centering
\subfloat[]{
\includegraphics[width=0.48\textwidth]{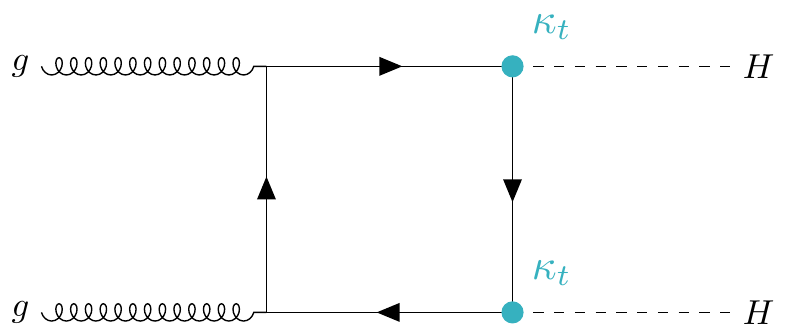}
\label{"fig:box"}
}
\subfloat[]{
\includegraphics[width=0.48\textwidth]{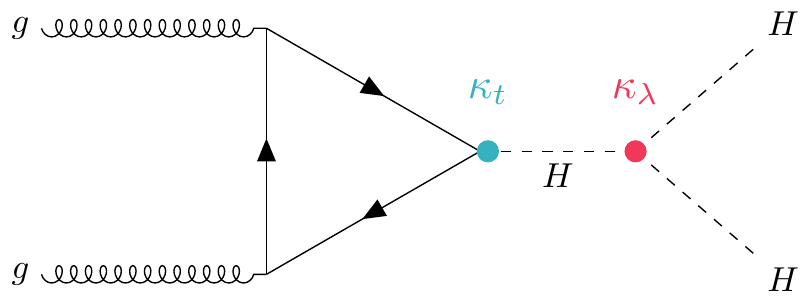}
\label{"fig:triangle"}
}
\caption{The two leading-order gluon--gluon fusion di-Higgs production Feynman diagrams: (a) the box diagram; (b) the triangle diagram.}
\label{fig:ggF_feyn_dias}
\end{figure}

The dominant SM \HH production process is gluon--gluon fusion (ggF).
Its cross-section, for a Higgs boson mass $m_H =\SI{125}{\GeV}$, calculated at next-to-next-to-leading order (NNLO) including finite top-quark-mass effects~\cite{Grazzini:2018bsd}, is \SI{31.05}{fb} at a center-of-mass energy $\sqrt{s}=\SI{13}{\TeV}$.
The two dominant leading-order Feynman diagrams contributing to this process are shown in Figure \ref{fig:ggF_feyn_dias}, where Figure~\ref{"fig:box"} is commonly referred to as the \textit{box} diagram and Figure \ref{"fig:triangle"} as the \textit{triangle} diagram.
The triangle diagram introduces the dependence on the trilinear Higgs self-coupling, $\lambda$, shown by the red vertex in Figure~\ref{"fig:triangle"}, which can be expressed in terms of its modifier, \kl.\footnote{A coupling modifier, $\kappa$, is defined as the ratio of the modified coupling to its SM value, $\kappa = c / c^\text{SM}$. By definition, $\kappa=1$ denotes the value of the coupling predicted by the SM.}
In the SM, these two diagrams interfere destructively.
As a result, the \HH production cross-section and kinematic properties depend critically on the value of \kl.

\begin{figure}[!bth]
\centering
\subfloat[]{
\includegraphics[width=0.32\textwidth]{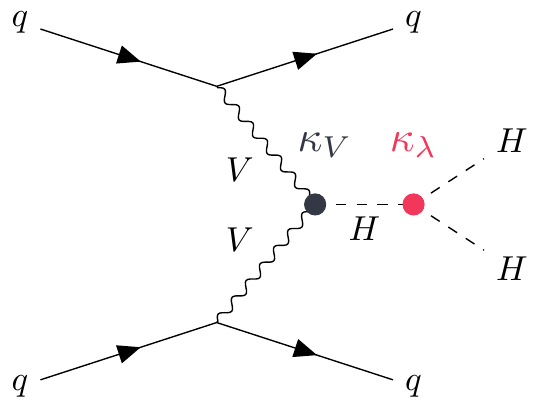}
}
\subfloat[]{
\includegraphics[width=0.32\textwidth]{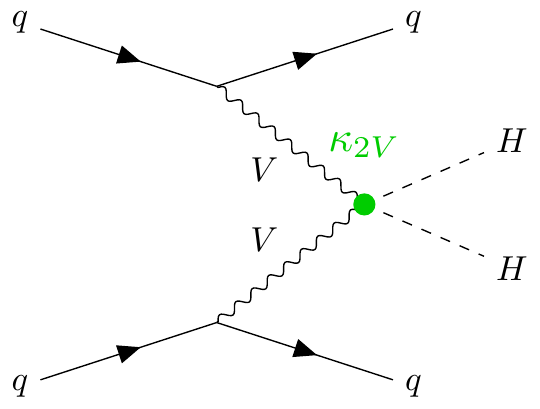}
\label{"fig:VBF_k2v"}
}
\subfloat[]{
\includegraphics[width=0.32\textwidth]{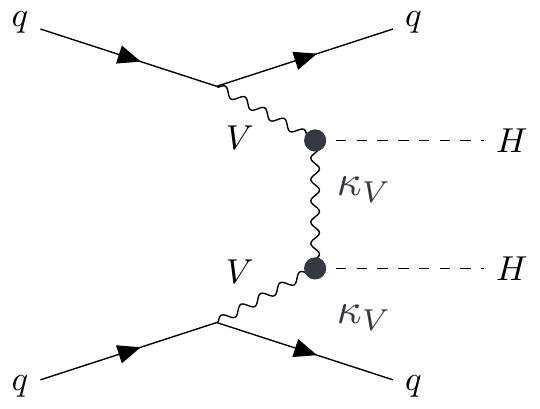}
\label{"fig:VBF_kvkv"}
}
\caption{The three tree-level vector-boson fusion di-Higgs production Feynman diagrams.}
\label{fig:VBF_feyn_dias}
\end{figure}

The \HH production process with the second-highest cross-section in the SM is vector-boson fusion (VBF), with a calculated value of \SI{1.73}{fb} at next-to-next-to-next-to-leading order (N$^3$LO)~\cite{Dreyer:2018qbw}, for  $m_H =\SI{125}{\GeV}$ at $\sqrt{s}=\SI{13}{\TeV}$.
Figure~\ref{fig:VBF_feyn_dias} illustrates the Feynman diagrams involved in di-Higgs production via vector-boson fusion at leading order (LO). The coupling modifiers \kl, \kv and \kvv are respectively shown at the \HHH, \PH{}\textit{VV} and \HHVV\ interaction vertices, where $V$ stands for the gauge vector bosons $\PW$ or $\PZ$.
In the SM, the divergences in the Figures~\ref{"fig:VBF_k2v"} and \ref{"fig:VBF_kvkv"} diagrams exactly cancel out due to perturbative unitarity. 
As \kv and \kvv depart from their SM value of one, this canceling out no longer occurs, introducing a linear dependence of the cross-section on the effective center-of-mass energy of the incoming vector bosons~\cite{Bishara_2017}.
Therefore, the Higgs bosons produced in non-SM \kv/\kvv scenarios are expected to be more energetic and more central in the detector on average.
This increase in the energy of Higgs bosons with increasing deviation from the SM continues up to the scale of some new physics, which is required to unitarize the total amplitude.

The analysis described in this paper targets the \HH process in the \bbbb final state, in both the ggF and VBF production modes, using the data collected by ATLAS between 2016 and 2018, during \RunTwo of the LHC.
Assuming the SM branching ratio of 58.2\% for  \Hbb~\cite{deFlorian:2016spz, Djouadi:1997yw}, about one third of di-Higgs events decay into \bbbb, making it the most abundant di-Higgs final state.
However, as this is a fully hadronic final state, the analysis faces the challenge of large backgrounds, which originate mostly from nonresonant QCD production of multiple heavy ($\Pqb/\Pqt$) quarks, as well as from light-quark-initiated jets misidentified as originating from heavy quarks.
 
The results are interpreted in terms of constraints on the \kl and \kvv coupling modifiers, assuming $\kv = 1$.
The analysis also provides one- and two-dimensional constraints on relevant couplings in the SM effective field theory (SMEFT)~\cite{Buchmuller:1985jz,Leung:1984ni,Brivio:2017vri} and Higgs effective field theory (HEFT)~\cite{Alonso:1499850, Buchalla:2013rka} frameworks.
In the SMEFT framework, the effects of new physics may be described with an effective Lagrangian:
\begin{equation}
\mathcal{L}_{\text{SMEFT}} = \mathcal{L}_{\text{SM}} + \frac{1}{\Lambda^2}\sum_{k}c_{k}^{(6)}O_{k}^{(6)}\,,
\end{equation}
where $\mathcal{L}_{\text{SM}}$ represents the SM Lagrangian, $O_{k}$ are higher-dimensional local operators, $c_k$ are the Wilson coefficients, and $\Lambda$ is the mass scale of the new physics phenomena (set to \SI{1}{\TeV} for this result).
The analysis considers operators $O_{k}$ in the Warsaw basis, which provides a complete set of operators allowed by SM gauge symmetries at dimension six~\cite{smeft2010} (dimension-five operators introduce lepton and baryon number violation, and are therefore ignored in this result).
The five operators relevant to the \HH process and their coefficients, $c_{\higgs}$, $c_{\higgs\square}$, $c_{t\higgs}$, $c_{tG}$, and $c_{\higgs G}$, are listed in Table~\ref{tab:smeft}~\cite{ATL-PHYS-PUB-2022-037}.
The computation of amplitudes from the above Lagrangian includes three terms: a pure SM term, a ``quadratic'' term of order $(1/{\Lambda^4})$ including purely new physics, and a ``linear'' term of order $(1/\Lambda^2)$ accounting for the interference between the SM and new physics.
The SMEFT constraints calculated in this analysis include both the linear and quadratic new physics terms.

% The next lines are included from the .//tables/eft_coefs.tex input file
\begin{table}[bth]
\centering
\caption{The five relevant SMEFT coefficients and their corresponding dimension-6 operators, as defined in the Warsaw basis~\cite{smeft2010,ATL-PHYS-PUB-2022-037}.}
\begin{tabular}{c c}
\toprule
\textbf{Wilson Coefficient} & \textbf{Operator} \\
 
\hline{$c_{\higgs}$}         &  $(\higgs^{\dag}\higgs)^3$ \\
{$c_{\higgs\square}$}  &  $(\higgs^{\dag}\higgs)\square(\higgs^{\dag}\higgs)$ \\
{$c_{\Pqt\higgs}$}     &  $(\higgs^{\dag}\higgs)(\bar{Q} \tilde{\higgs} t)$ \\
{$c_{\higgs G}$}       &  $\higgs^{\dag}\higgs G^{A}_{\mu\nu}G_{A}^{\mu\nu} $ \\
{$c_{\Pqt G}$}         &  $(\bar{Q}\sigma^{\mu\nu}T^{A}\Pqt )\tilde{\higgs}G^{A}_{\mu\nu} $ \\

\bottomrule
\end{tabular}
\label{tab:smeft}
\end{table}

% End of text imported from the .//tables/eft_coefs.tex input file

In the HEFT framework, new physics in the electroweak sector is described through anomalous couplings of the Higgs boson.
The organization of the HEFT Lagrangian is guided by chiral perturbation theory~\cite{Buchalla:2013eza}, with the low-energy dynamics of electroweak symmetry breaking described using a nonlinear realization of the gauge symmetry group $SU(2)_{L}\times U(1)_{Y}$.
One advantage of the HEFT framework is that the anomalous single-Higgs-boson and \HH couplings are defined separately, allowing simplified \HH interpretations.
In the HEFT Lagrangian, ggF \HH production is described at LO by five relevant operators and their associated Wilson coefficients: $c_{\higgs\higgs\higgs}$, $c_{t\overline{t}\higgs}$, $c_{gg\higgs}$, $c_{gg\higgs\higgs}$, and $c_{t\overline{t}\higgs\higgs}$.
In this formalism, $c_{\higgs\higgs\higgs}$ is equivalent to \kl and $c_{t\overline{t}\higgs}$ is equivalent to the modifier for the coupling between the Higgs boson and top quark, $\kappa_{t}$, shown by the light blue vertex in Figure~\ref{fig:ggF_feyn_dias}.
Fixing $c_{t\overline{t}\higgs} = c_{\higgs\higgs\higgs} = 1$ and $c_{gg\higgs} = c_{gg\higgs\higgs} = c_{t\overline{t}\higgs\higgs} = 0$ restores the SM.
At next-to-leading order (NLO), seven HEFT benchmark models (BM)~\cite{Capozi:2019xsi} have been defined using cluster analysis~\cite{Carvalho:2015ttv} to probe a wide variety of characteristic shapes of the \mhh spectrum resulting from different BSM scenarios.
The values of the coefficients used to define these scenarios are given in Table~\ref{tab:heft_bm}.
 
\begin{table}[tbh!]
\caption{The values of the HEFT Wilson coefficients in the SM and in seven BSM benchmark models, as defined in Ref.~\cite{Capozi:2019xsi}.}
\begin{center}
\begin{tabular}{c| SSSSS[table-format=1.2,table-number-alignment=center]}
\toprule
\textbf{Benchmark Model} & $c_{HHH}$ & $c_{ttH}$ & $c_{ggH}$ & $c_{ggHH}$ & $c_{ttHH}$ \\
\midrule
SM  & 1 & 1 & 0 & 0 & 0 \\
BM1 & 3.94 & 0.94 & {1/2} & {1/3} & {$-$1/3}~~~~ \\
BM2 & 6.84 & 0.61 &0 & {$-$1/3}~~~  & {1/3}~ \\
BM3 & 2.21 & 1.05 & {1/2} & {1/2} & {$-$1/3}~~~~ \\
BM4 & 2.79 & 0.61 & {$-$1/2}~~~ & {1/6}  & {1/3}~ \\
BM5 & 3.95 & 1.17 & {1/6} & {$-$1/2}~~~ & {$-$1/3}~~~~ \\
BM6 & 5.68 & 0.83 & {$-$1/2}~~~ & {1/3}  & {1/3}~ \\
BM7 & -0.10 & 0.94 & {1/6} & {$-$1/6}~~~  & 1 \\
\bottomrule
\end{tabular}
\end{center}
\label{tab:heft_bm}
\end{table}

The ATLAS Collaboration has previously published search results for nonresonant \hhb production using \SI{27}{\ifb} of early \RunTwo data~\cite{EXOT-2016-31}, and a dedicated search for VBF \HH production in \SI{126}{\ifb} of data collected between 2016 and 2018~\cite{HDBS-2018-18-witherratum}.
The present analysis benefits from the use of the 2016--2018 data for both production channels and also
takes advantage of improvements in jet reconstruction and in the identification of jets arising from the hadronization of \Pqb-quarks (``\(b\)-tagging'') achieved by the ATLAS Collaboration since the publication of Ref.~\cite{EXOT-2016-31}.
In addition, the analysis employs a fully data-driven technique for the background estimation, which uses an artificial neural network to perform a kinematic reweighting of data from an alternative control region of the data to model the background in the region of interest.
The CMS Collaboration has also published results of a search for nonresonant \hhb with its full \RunTwo dataset~\cite{CMS:2022cpr}, setting the observed (expected) upper limit on the \HH cross-section at 3.9 (7.8) times the SM predicted cross-section, and restricting the allowed interval for \kl to $[-2.3, 9.4]$ ($[-5.0, 12.0]$), both at 95\% confidence level (CL).
A more recent CMS \hhb publication~\cite{CMS:2022nmn}, in which the analysis exploits topologies arising from highly energetic Higgs boson decays into $b\bar b$,  sets the observed (expected) upper limit at 9.9 (5.1) times the SM cross-section expectation, and restricts the allowed interval for \kvv to [0.62, 1.41] ([0.66, 1.37]), at  95\% CL.
Other searches for nonresonant $HH$ production were performed by ATLAS and CMS in the $b\bar b\tau^+\tau^-$~\cite{ATLAS:2022xzm,CMS-HIG-17-002}, $b\bar b\gamma\gamma$~\cite{HDBS-2018-34,CMS-HIG-19-018}, $b\bar b\ell^+\nu\ell^-\nu$~\cite{HDBS-2018-33,CMS-HIG-17-006} decay channels, as well as by ATLAS in the $b\bar bqq\ell\nu$~\cite{HIGG-2016-27}, $WW^*\gamma\gamma$~\cite{HIGG-2016-20} and $WW^*WW^*$~\cite{HIGG-2016-24} decay channels. 
Amongst them, the most sensitive results to date from  ATLAS come from the $b\bar b\gamma\gamma$ analysis, which sets the observed (expected) 95\% CL upper limit on the SM nonresonant $HH$ cross-section at 4.2 (5.7) times the SM expectation and restricts the corresponding \kl interval to $[-1.5, 6.7]$ ($[-2.4, 7.7]$).
The most sensitive results to date from CMS come from the combination of the $b\bar{b}ZZ$, multilepton, $b\bar b\gamma\gamma$, $b\bar b\tau\tau$, and \bbbb analyses, which set the observed (expected) 95\% CL upper limit on the SM nonresonant $HH$ cross-section at 3.4 (2.5) times the SM expectation and restricts the corresponding observed \kl interval to $[-1.24, 6.49]$~\cite{CMS:2022natureHiggs}.
 
This document is structured as follows. The ATLAS detector and the data and simulated events used in the analysis are described in Sections~\ref{sec:detector} and~\ref{sec:data-mc-samples}, respectively. Section~\ref{sec:reco} presents the reconstruction and identification of physics objects in this analysis and Section~\ref{sec:analysis} details the event selection and categorization.
The background modeling method is described in Section~\ref{sec:background}, the systematic uncertainties are detailed in Section~\ref{sec:systematics} and, finally, the results are reported in Section~\ref{sec:result} and the conclusion is given in Section~\ref{sec:conclusion}.

% End of text imported from the .//sections/intro.tex input file

\section{ATLAS detector}
\label{sec:detector}

% The next lines are included from the .//sections/detector.tex input file
 
The ATLAS detector~\cite{PERF-2007-01} at the LHC covers nearly the entire solid angle around the collision point.\footnote{ATLAS uses a right-handed coordinate system with its origin at the nominal interaction point (IP)
in the center of the detector and the \(z\)-axis along the beam pipe.
The \(x\)-axis points from the IP to the center of the LHC ring,
and the \(y\)-axis points upwards.
Cylindrical coordinates \((r,\phi)\) are used in the transverse plane,
\(\phi\) being the azimuthal angle around the \(z\)-axis.
The pseudorapidity is defined in terms of the polar angle \(\theta\) as \(\eta = -\ln \tan(\theta/2)\).
Angular distance is measured in units of \(\Delta R \equiv \sqrt{(\Delta\eta)^{2} + (\Delta\phi)^{2}}\).}
It consists of an inner tracking detector surrounded by a thin superconducting solenoid, electromagnetic and hadron calorimeters,
and a muon spectrometer incorporating three large superconducting air-core toroidal magnets.
 
The inner-detector (ID) system is immersed in a \SI{2}{\tesla} axial magnetic field and provides charged-particle tracking in the range \(|\eta| < 2.5\).
The high-granularity silicon pixel detector covers the vertex region and typically provides four space-point measurements per track, the first hit normally being in the insertable B-layer installed before Run~2~\cite{ATLAS-TDR-19,PIX-2018-001}.
Following the pixel detector is the silicon microstrip tracker, which usually provides eight measurements per track.
These silicon detectors are surrounded by the transition radiation tracker, which enables radially extended track reconstruction up to \(|\eta| = 2.0\).
 
The calorimeter system covers the pseudorapidity range \(|\eta| < 4.9\).
Within \(|\eta|< 3.2\), electromagnetic calorimetry is provided by barrel and endcap high-granularity lead/liquid-argon (LAr) calorimeters, with an additional thin LAr presampler covering \(|\eta| < 1.8\) to correct for energy loss in material upstream of the calorimeters.
Hadron calorimetry is provided by the steel/scintillator-tile calorimeter, segmented into three barrel structures within \(|\eta| < 1.7\), and two copper/LAr hadron endcap calorimeters.
The solid angle coverage is completed with forward copper/LAr and tungsten/LAr calorimeter modules optimized for electromagnetic and hadronic energy measurements respectively.
 
The muon spectrometer (MS) comprises separate trigger and high-precision tracking chambers measuring the deflection of muons in a magnetic field generated by the superconducting air-core toroidal magnets.
The field integral of the toroids ranges between \num{2.0} and \SI{6.0}{\tesla}\(\cdot\)\SI{}{\metre} across most of the detector.
A set of precision chambers covers the region \(|\eta| < 2.7\) with three layers of monitored drift tubes, complemented by cathode-strip chambers in the forward region, where the background is highest.
The muon trigger system covers the range \(|\eta| < 2.4\) with resistive-plate chambers in the barrel, and thin-gap chambers in the endcap regions.
 
Interesting events are selected by the first-level trigger system implemented in custom hardware, followed by selections made by algorithms implemented in software in the high-level trigger~\cite{TRIG-2016-01}.
The first-level trigger accepts events from the \SI{40}{\MHz} bunch crossings at a rate below \SI{100}{\kHz},
which the high-level trigger reduces in order to record events to disk at about \SI{1}{\kHz}.
 
An extensive software suite~\cite{ATL-SOFT-PUB-2021-001} is used in data simulation, in the reconstruction and analysis of real and simulated data, in detector operations, and in the trigger and data acquisition systems of the experiment.

% End of text imported from the .//sections/detector.tex input file

\section{Data and simulated samples}
\label{sec:data-mc-samples}

% The next lines are included from the .//sections/data-mc-samples.tex input file
 
\subsection{Data sample}
 
This analysis is performed in LHC proton--proton (\pp) collision data at \(\sqrt{s} = \SI{13}{\TeV}\) collected between 2016 and 2018.
Only data collected during stable beam conditions are used, with all relevant detector systems functional~\cite{DAPR-2018-01}, corresponding to an integrated luminosity of \SI{126}{\ifb}. During 2016 data taking, a fraction of the data (\SI{8.3}{\ifb}) was affected by an inefficiency in the online primary vertex reconstruction, which reduced the efficiency of the $b$-tagging algorithms in the trigger; those events were not retained for further analysis, resulting in an integrated luminosity of \SI{24.6}{\ifb} for the 2016 dataset. The integrated luminosities of the 2017 and 2018 datasets are \SI{43.7}{\ifb} and \SI{57.7}{\ifb}, respectively.
 
The analysis uses events that satisfy either of two types of trigger signatures, each with different requirements on the number of jets and their \(b\)-tagging status~\cite{TRIG-2018-08}.
The jets used are reconstructed with the \antikt algorithm~\cite{Cacciari:2008gp,Fastjet}, with a radius parameter of \(R=0.4\).
The \(b\)-tagging is performed at the trigger level with the MV2c20 algorithm in 2016 and the MV2c10 algorithm in 2017 and 2018~\cite{TRIG-2018-08}, with a range of \Pqb-jet identification efficiency operating points from 40\% to 70\% (as calculated from simulated \ttbar samples.)
The first of the two trigger signatures used for selecting \bbbb events requires two \(b\)-jets plus one additional jet (``2\(b\)1j''), while the second requires two \(b\)-jets plus two additional jets (``2\(b\)2j'').
The minimum transverse energy (\ET) requirement on the jets is \SI{35}{\GeV} for all jets used in the 2\(b\)2j trigger.
In the 2\(b\)1j trigger, the \Pqb-tagged jets must have $\ET > \SI{55}{\GeV}$, while the requirement on the minimum \ET of the additional jet is between 100 and \SI{150}{\GeV}, depending on the year of data taking.

\subsection{Simulated samples}
 
Monte Carlo (MC) simulation is used for the modeling of signal events, as well as to produce event samples of background processes for cross-checks and validation studies.
The Higgs boson mass is set to \SI{125}{\GeV} in the simulation.
All samples were processed by the ATLAS simulation framework~\cite{SOFT-2010-01} and the detector response was simulated with \GEANT~\cite{Agostinelli:2002hh}.
 
The ggF signal process was simulated using the \POWHEGBOX[v2] generator~\cite{Powheg1, Powheg2, Powheg3} at NLO, including finite top-quark-mass effects, using the \PDFforLHC[15]~\cite{Butterworth:2015oua} parton distribution function (PDF) set.
Parton showers and hadronization were simulated with \PYTHIA[8.244]~\cite{Sjostrand:2014zea} with the A14 set of tuned parameters~\cite{ATL-PHYS-PUB-2014-021} and the \NNPDF[2.3lo] PDF set~\cite{Ball:2012cx}.
The SM ggF \HH cross-section was taken as $\sigma_{\text{ggF}}=\SI{31.05}{\fb}$, calculated at NNLO including finite top-quark-mass effects~\cite{Grazzini:2018bsd}.
Signal samples for the ggF process were generated explicitly for coupling modifier values of $\kl=1$ and $10$. 
A reweighting method is used to obtain a ggF signal sample at each $\kl$ value, as described in Ref.~\cite{ATL-PHYS-PUB-2019-007}:
scale factors are derived as a function of \kl in bins of the generator-level invariant mass of the \HH system by performing a linear combination of generator-level samples at three different \kl values ($\kl=0,1,$ and $20$).
The $\kl=10$ ggF signal sample is used to validate the derived scale factors; this generated sample and the signal sample obtained from the reweighting method are found to agree within the statistical precision of the simulated sample.
Additional generator-level ggF \HH signal samples without parton showering were produced with \POWHEGBOX[v2] for the $\kl=0$ and $20$ coupling modifier configurations to provide a basis for the \kl reweighting, along with the SM ggF sample.
For the reweighted ggF signal, the NNLO cross-section as a function of \kl is taken from Ref.~\cite{Grazzini:2018bsd}.
In order to assess parton showering uncertainties, alternative ggF samples were generated using the \POWHEGBOX[v2] generator at NLO with the \PDFforLHC[15] PDF set, interfaced to \HERWIG[7.1.6]~\cite{Herwigpp} for parton showering and hadronization using the \HERWIG[7.1]-default set of tuned parameters~\cite{Gieseke:2012ft} and \MMHT[lo] PDF set~\cite{Harland-Lang:2014zoa}. 
 
To extract SMEFT coefficient constraints, parton-level ggF \HH samples were generated with \MGNLO~\cite{ATL-PHYS-PUB-2015-048,Alwall:2014hca,Alwall:2011uj} with the SMEFT@NLO model~\cite{PhysRevD.103.096024} for a variety of SMEFT coefficients.
A finely spaced multidimensional grid of signal samples was obtained using a LO-derived reweighting procedure in the generator-level invariant mass of the \HH system; this procedure is similar to that used to obtain \kl variations for the ggF signal, as described above.
To extract HEFT coefficient constraints, a similar NLO-derived reweighting procedure was applied to the simulated reconstruction-level ggF signal sample to produce a variety of HEFT signal scenarios, including the seven benchmark scenarios defined in Section~\ref{sec:intro}, following the prescription outlined in Refs.~\cite{Buchalla:2018yce,ATL-PHYS-PUB-2022-019}.
Additional $K$-factors were applied to the SMEFT samples; these $K$-factors were derived using the ratio of the NLO cross-section to the LO cross-section at the equivalent HEFT point, as obtained using the HEFT to SMEFT translation from Ref.~\cite{Capozi:2019xsi}.\footnote{Variations in the $c_{tG}$ Wilson coefficient were neglected when calculating $K$-factors because the corresponding chromomagnetic operator does not appear at LO within HEFT.}

The VBF signal process was simulated using \MADGRAPH[2.7.3]~\cite{Alwall:2014hca} at LO with the \NNPDF[3.0nlo] PDF set~\cite{Ball:2014uwa}, interfaced with \PYTHIA[8.244] for parton showering and hadronization using the A14 set of tuned parameters and \NNPDF[2.3lo] PDF set.
Signal samples for the VBF process were generated explicitly for coupling modifier values of $(\kl,\kvv,\kv)=(1,1,1)$, $(1,1.5,1)$, $(2,1,1)$, $(10,1,1)$, $(1,1,0.5)$, $(-5,1,0.5)$, $(0,1,1)$, $(1,0,1)$, and $(1,3,1)$.
A linear combination of the first six of the listed samples is used to derive distributions for a finer granularity of \kvv values, following a technique used previously to generate \kl distributions~\cite{HDBS-2018-58}.
The specific basis of six samples utilized is chosen to avoid large statistical uncertainties in the reweighted signal samples resulting from sparsely populated areas of kinematic phase space.
The generated VBF signal samples not included in the linear combination basis –  $(\kl,\kvv,\kv)=(0,1,1)$, $(1,0,1)$, and $(1,3,1)$ – were used to validate the performance of the combination method.
These generated samples and the corresponding signal samples obtained from the combination method were found to agree within the statistical precision of the simulated samples.
The cross-section for the VBF $HH$ process, evaluated at N$^3$LO in QCD, is \SI{1.73}{\fb} in the SM~\cite{Baglio:2012np,Liu-Sheng:2014gxa,Dreyer:2018qbw,Dreyer:2018rfu}. 
For the reweighted VBF signal points, the N$^3$LO to LO cross-section ratio at the SM value is calculated, and this factor is applied to the cross-sections at each \kl, \kvv, and \kv point.
In order to assess parton showering uncertainties, alternative LO samples were generated using \MADGRAPH[2.7.3] with the \NNPDF[3.0nlo] PDF set, interfaced to \HERWIG[7.0.4] with the \HERWIG[7.1]-default set of tuned parameters and \MMHT[lo] PDF set for parton showering and hadronization.

Top-quark pair production (\ttbar) and multijet background processes were simulated in order to validate the background modeling procedure.
The \ttbar sample was simulated at NLO in \(\alphas\) using \POWHEGBOX[v2]~\cite{Campbell:2014kua}.
Parton showering, hadronization, and the underlying event were modeled using \PYTHIA[8.230]. 
The matrix element calculation uses \NNPDF[3.0nlo] as the PDF set,
while the parton shower and underlying-event modeling uses \NNPDF[2.3lo] and the A14 set of tuned parameters.
The damping parameter \hdamp, which effectively regulates radiation at high \pt, was set to 1.5 times the top quark's mass.
The \ttbar simulation is normalized using the value of the inclusive cross-section calculated with \TOPpp[2.0]~\cite{Czakon:2011xx,ATL-PHYS-PUB-2018-009}.
This accounts for NNLO corrections in \(\alphas\), including next-to-next-to-leading logarithmic (NNLL) resummation of soft gluon terms.
The multijet background samples were modeled using \PYTHIA[8.235].
This simulates pure QCD 2-to-2 interactions at LO in \(\alphas\).
Events were showered using the parton shower native to \PYTHIA, which includes radiation and splitting that can result in additional jets. 
The A14 set of tuned parameters and the \NNPDF[2.3lo] PDF set were used.
 
Other background processes, such as SM Higgs boson, \HH (in other final states) and electroweak diboson production, have been estimated to give negligible contributions to the selected event yields and are therefore not included.
 
The effect of multiple interactions in the same and neighboring bunch crossings (\pileup) was modeled by overlaying each simulated hard-scattering event with inelastic \pp events generated with \PYTHIA[8.186] using the \NNPDF[2.3lo] PDF set and the A3 set of tuned parameters~\cite{ATL-PHYS-PUB-2016-017}. Additionally, for all \HH signal samples, heavy-flavor decays were modeled using \EVTGEN[1.7.0]~\cite{Lange:2001uf}.

% End of text imported from the .//sections/data-mc-samples.tex input file

\section{Object reconstruction}
\label{sec:reco}

% The next lines are included from the .//sections/reco.tex input file
Primary vertices from \pp interactions are reconstructed~\cite{ATL-PHYS-PUB-2015-026} using at least two charged-particle tracks with transverse momentum (\pt) above \SI{500}{\MeV} measured with the ID.
The vertex with the largest sum of squared track momenta (\(\sum \pt^2)\) is taken as the hard-scatter primary vertex.
 
Hadronic jets are reconstructed using the \antikt algorithm 
with radius parameter \(R=0.4\).
The jet clustering uses particle-flow objects as inputs~\cite{PERF-2015-09}. Particle-flow objects are charged-particle tracks matched to the hard-scatter vertex and calorimeter energy clusters after applying an energy subtraction algorithm that removes the calorimeter deposits associated with good-quality tracks from any vertex.
The tracking information helps to improve the energy resolution of the calorimeter clusters and reduce the impact from \pileup.
The momenta of reconstructed jets are calibrated in a multistep procedure~\cite{JETM-2018-05}.
Jets with \(\pt < \SI{60}{\GeV}\) and \(|\eta| < 2.4\) must also satisfy a requirement based on the output of the multivariate ``jet vertex tagger'' (JVT) algorithm~\cite{PERF-2014-03}, which is used to identify and reject jets in which much of the energy originates from \pileup interactions.
Correction factors are applied to the simulated events to compensate for differences between the JVT efficiencies in data and simulation.
In the \hhb analysis, jets are discarded if they fail the ``Tight'' JVT working point, corresponding to an average efficiency of 96\% for jets from the hard-scatter vertex.
 
Jets with radius parameter \(R=0.4\) are also reconstructed from topological clusters of energy deposits in the calorimeter~\cite{PERF-2014-07} and calibrated in the same way as the jets reconstructed from particle-flow objects.
These jets are used exclusively for the purpose of applying quality criteria to identify events which are consistent with noise in the calorimeter or noncollision background~\cite{ATLAS-CONF-2015-029}.
Events containing at least one such jet with \(\pt > \SI{20}{\GeV}\), satisfying the JVT requirement, but not these quality criteria, are rejected.
 
The identification of jets originating from \(b\)-quarks is performed by the DL1r algorithm~\cite{FTAG-2019-07}, which is applied to all jets with \(|\eta| < 2.5\).
DL1r is based on a multivariate classification technique  combining information from the impact parameters of ID tracks, the presence of displaced secondary vertices, and the reconstructed flight paths of \(b\)- and \(c\)-hadrons inside the jet.
The DL1r working point used in the \hhb analysis is the one that gives 77\% efficiency for jets associated with true \(b\)-hadrons in simulated \ttbar events.
At this working point, the light-jet (charm-jet) rejection measured in \ttbar simulation is about a factor of 130 (4.9).
The calibration of the DL1r algorithm is performed separately for each jet type~\cite{ATLAS-CONF-2018-006,ATLAS:2021cxe} and correction factors are derived and applied to the simulated samples to compensate for differences between the \(b\)-tagging efficiencies in data and simulation.
 
Muons are reconstructed by matching ID tracks with either MS tracks or aligned individual hits in the MS and performing a combined track fit~\cite{MUON-2018-03}.
They are required to have \(\pt > \SI{4}{\GeV}\) and \(|\eta| < 2.5\), and to satisfy ``Medium'' identification criteria based on track-quality variables.
Muons are used only to apply energy corrections to jets.
 
A momentum correction is applied to \(b\)-tagged jets to account for energy lost to soft out-of-cone radiation and to muons and neutrinos in semileptonic \(b\)-hadron decays.
This correction follows the procedure used in Ref.~\cite{HIGG-2016-29} and consists of two steps.
First, a search is performed for muons located near the jet which fall within a cone of variable size $\Delta R(\mu, \text{jet}) < \min \left( 0.4, 0.04 + 10 / p_\text{T}^{\mu} \ \GeV \right)$ around the jet axis.
If a muon is found, its four-momentum is added to that of the jet, and the energy deposited in the calorimeter by the muon is subtracted from the jet to avoid double counting; this is computed according to the description in Ref.~\cite{PERF-2015-10}.
In the second step, a global scale factor is applied to each \(b\)-tagged jet according to its \pt and whether or not it has a muon associated with it.
These scale factors are derived from simulation.

% End of text imported from the .//sections/reco.tex input file

\section{Analysis selection and categorization}
\label{sec:analysis}

% The next lines are included from the .//sections/analysis.tex input file
 
The analysis utilizes a set of criteria to select \hhb candidate events, including dedicated requirements to separate events into orthogonal ggF and VBF signal regions.
``Forward'' and ``central'' jets are used with the following selection criteria:
\begin{itemize}
\item central jets: $|\eta|<2.5$ and \(\pt > \SI{40}{\GeV}\);
\item forward jets: $2.5<|\eta|< 4.5$ and \(\pt > \SI{30}{\GeV}\).
\end{itemize}

An initial ``preselection'' is applied to all events, 
which requires at least four central jets with $\pt > \SI{40}{\GeV}$, at least two of which are \(b\)-tagged.
As described in Section~\ref{sec:data-mc-samples}, the events considered in this analysis are selected online through the 2$b$2j or 2$b$1j trigger signatures.
In order to simplify the modeling of trigger efficiencies, a further selection is applied using offline kinematic quantities.
Events are selected if they have a leading\footnote{In this document, terms  like ``leading'', ``subleading'' etc for physics objects refer to the ordering of these objects in decreasing \pt.} jet with \(\pt > \SI{170}{\GeV}\), a third leading jet with \(\pt > \SI{70}{\GeV}\), and pass the 2$b$1j trigger, or if they fail either of the two jet-\pt requirements and pass the 2$b$2j trigger.
This selection step retains about 90\% of signal efficiency, and it enables the reliable calculation of simulation-to-data correction factors for estimating the trigger efficiency in the remaining \hhb signal events, depending on which of the above two \textit{trigger classes} they belong to.
 
Events passing the above preselection are required to contain at least four central jets passing the \Pqb-tagging requirement outlined in Section~\ref{sec:reco}.
The four highest-\pT \Pqb-tagged jets are chosen to reconstruct the decays of the two Higgs bosons.
In about $\SI{75}\%$ of simulated signal events reaching this selection stage, these four jets can be matched one-to-one (within $\dr<0.3$) to the four \Pqb-quarks from the decays of the Higgs bosons.
In signal events where this matching fails, one of the \Pqb-quarks from the Higgs boson decays typically produces a jet that is outside the analysis acceptance.
 
From the four selected \Pqb-tagged jets, there are three possible combinatorial pairings to form the two Higgs boson candidates.
Of those three configurations, the analysis selects the one in which the higher-\pT jet pair has the smallest \dr separation.
In the simulated samples with SM coupling values, for which the analysis was mainly optimized, this method gives the correct pairing in around 90\% of those signal events in which the four \Pqb-tagged jets are correctly matched to the \Pqb-quarks from the decays of the Higgs bosons.
While the pairing accuracy drops for values of the coupling modifiers \kl and \kvv that result in softer \pT spectra for the produced Higgs bosons, this pairing method leads to a smoothly varying distribution of the expected background in the plane of the invariant masses of the two Higgs boson candidates, which facilitates the 
data-driven background estimation described in Section~\ref{sec:background}.

Events are then subjected to additional selections designed to separate out those consistent with the VBF production mode.
For this, events must contain at least two additional jets, central or forward; \Pqb-tagged jets are excluded.
The two jets forming the pair with the largest invariant mass (\mjj) are chosen as the ``VBF jets''. 
The VBF jet pair is required to satisfy $\mjj>\SI{1}{\TeV}$, and the pseudorapidity separation between the two jets, \detajj, must satisfy $\detajj>3$.
Lastly, the transverse component of the momentum vector sum of the two VBF jets and the four jets forming the Higgs boson candidates is required to be less than \SI{65}{\GeV}.
Events satisfying the above criteria enter the VBF signal region, while those failing to satisfy any of these criteria are considered further in the ggF signal region.
 
Events satisfying either the ggF or VBF selections are required to satisfy additional selection criteria designed to reduce the background and improve the analysis sensitivity.
In order to suppress the \ttbar background, a top-veto discriminant \Xwt is defined as:
\begin{equation}
\Xwt = \text{min} \left[  \sqrt{\left( \frac{m_{\text{jj}} - m_{\PW}}{0.1 m_{\text{jj}}}\right)^{2} + \left(\frac{m_{\text{jj}b} - m_{\Pqt}}{0.1 m_{\text{jj}b}}\right)^{2}}\,  \right ]\,,
\end{equation}
where $m_W = \SI{80.4}{\GeV}$ and $m_t = \SI{172.5}{\GeV}$ are the nominal \PW boson and top quark masses, and $m_{\text{jj}}$ and $m_{\text{jj}b}$ are the invariant masses of \PW boson and top quark candidates formed from jet combinations in each event.
The ``minimum'' refers to the minimum value from all possible jet combinations (of one \Pqb-tagged jet and two additional untagged jets) that would give a \PW boson candidate and a corresponding top candidate.
The factor of $0.1$ in the denominators is chosen to approximate the experimental dijet mass resolution.
The \PW boson candidates are formed from any pair of central jets in the event 
and the top quark candidates are then reconstructed by pairing the \PW boson candidates with any remaining \Pqb-tagged Higgs boson candidate jets.
The \Xwt discriminant is designed to quantify the likelihood that an event contains a hadronic top quark decay.
Events with $\Xwt<1.5$ are rejected.
This reduces the \ttbar background by a factor of about 2 in simulated events, for a small loss of signal efficiency, of around 15\%, and a similar reduction in the non-\ttbar, multijet background.
 
In order to further reduce the overall background contamination, events in the ggF signal region are also required to have reconstructed Higgs bosons that satisfy a pseudorapidity separation $\deta<1.5$\,.
No such requirement is imposed in the VBF signal region, since SM VBF \HH signal events tend to have a larger $\deta$.
 
A final analysis selection criterion to test the compatibility of events with the \HH decay is applied in both the ggF and VBF selections.
A discriminant \Xhh is defined as:
\begin{equation}
\Xhh =  \sqrt{\left(\frac{m_{\PH1} - \SI{124}{\GeV}}{0.1 \ m_{\PH1}}\right)^{2} + \left(\frac{m_{\PH2} - \SI{117}{\GeV}}{0.1 \ m_{\PH2}}\right)^{2}} \,,
\end{equation}
where $m_{H1}$ and $m_{H2}$ are the masses of the leading and subleading reconstructed Higgs boson candidates respectively.
The values of \SI{124}{\GeV} and \SI{117}{\GeV} in the \Xhh definition are chosen in accord with the centers of the $m_{H1}$ and $m_{H2}$ distributions for correctly paired signal events from simulation.
Events are required to have $\Xhh < 1.6$ to be included in the \textit{Signal Region} (SR) of the analysis.

Both the ggF and VBF signal regions are subdivided into a number of orthogonal categories in order to better isolate the \HH signal and improve the analysis sensitivity.
The \Xhh and \deta  quantities are used to define six orthogonal ggF categories.
The categories are defined by two intervals in \Xhh, with boundaries at 0, 0.95, and 1.6, and three in \deta, with boundaries at 0, 0.5, 1, and 1.5\,.
In the VBF signal region, two categories are defined using the \deta quantity, with the dividing boundary at $\deta=1.5$\,.
The $\deta<1.5$ category is more sensitive to VBF signals with non-SM couplings, while the $\deta>1.5$ category is more sensitive to SM VBF production.

The reconstructed invariant mass of the Higgs boson candidate pair, \mhh, is used as the discriminating variable for all analysis regions and categories when extracting results, as detailed in Section~\ref{sec:result}.
The \mhh distribution is found to have significant separation power between background and signal, for all the different values of coupling modifiers.
The binning of the \mhh distributions may vary between categories and is chosen in order to both maintain discrimination power and limit the expected statistical uncertainty in each bin to less than approximately $30\%$.
This 30\% limit ensures that the assumptions used in the statistical procedure, outlined in Section~\ref{sec:result}, are satisfied.
In the VBF signal region, only events with $\mhh>\SI{400}{\GeV}$ are considered, as the background in the lower \mhh region was found to be inadequately modeled by the data-driven method described in Section~\ref{sec:background} in validation studies with control data samples.
For the ggF signal region, no requirements on \mhh are applied.
 
\begin{figure}[!tbh]
\centering
\includegraphics[width=0.9\textwidth]{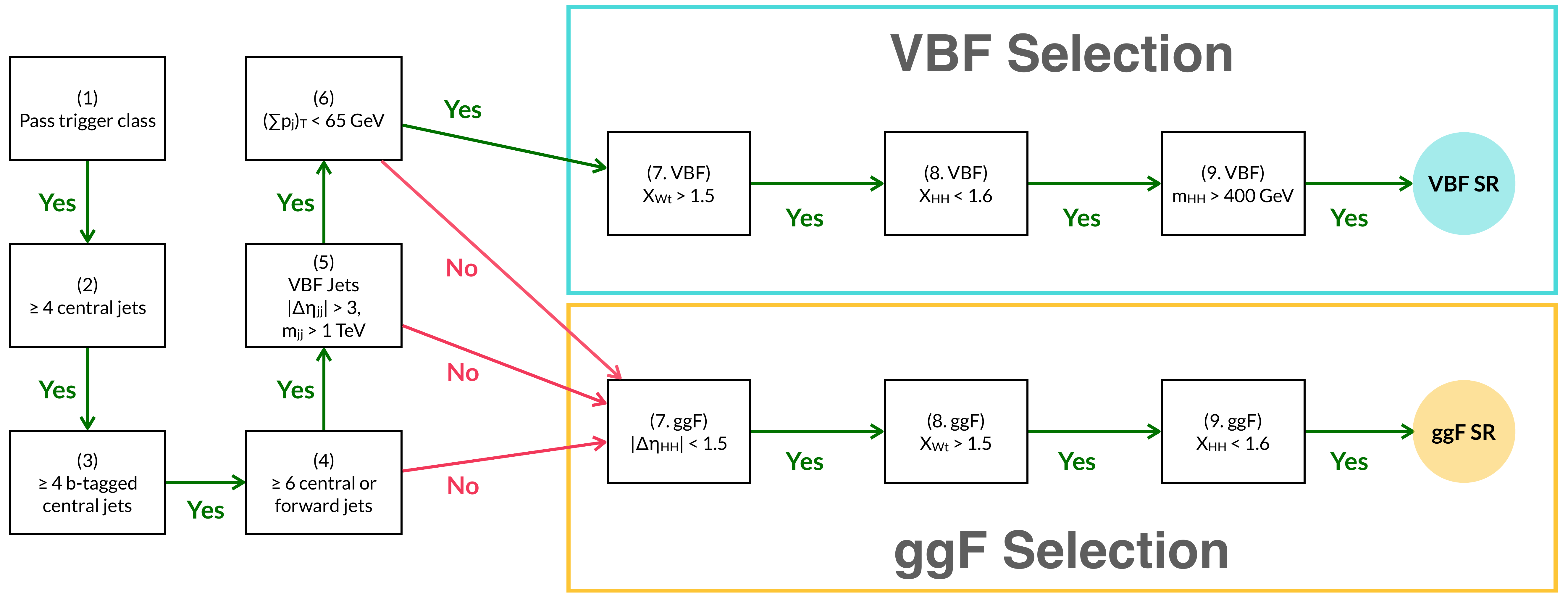}
\caption{A flowchart summarizing the nine selection criteria used for the VBF and ggF analysis selections.
Events must satisfy selection criteria 1--3 in order to be considered for either analysis signal region.
Events failing to satisfy any of the selection criteria 4--6 are considered for inclusion in the ggF signal region, while those satisfying selection criteria 4--6 are considered for the VBF signal region.
}
\label{fig:selection}
\end{figure}

All the selection steps of the analysis are summarized in Figure~\ref{fig:selection}.
The yields in the data and the simulated signal samples for some typical coupling values are shown in Table~\ref{tab:cutflow_comb}. This sample of data events is referred to as \(4b\) events hereafter.
 
% The next lines are included from the .//tables/cutflow_comb.tex input file

\begin{table}[!tbh]
\centering
\caption{The yields of data and various example ggF and VBF \HH signal models at each step of the analysis selection.
The ``Preselection'' entry denotes an initial selection requiring at least four jets with $\pt > \SI{40}{\GeV}$, at least two of which are \(b\)-tagged.
Events which satisfy the ``VBF selection'' requirements are considered as part of the VBF signal region of the analysis, while the rest are considered for the ggF signal region.
The signal yields are taken from simulation and are normalized by their theoretical cross-sections and the integrated luminosity of \SI{126}{\ifb}.
Corrections for differences in the \(b\)-tagging efficiency and trigger acceptance between data and simulation are applied starting from the ``Trigger class'' requirement.
}
\begin{tabular}{lSSSSS[table-format=4.1,table-number-alignment=center]}
\toprule
\multicolumn{1}{c}{ } & \multicolumn{1}{c}{\hspace*{1.2cm}\textbf{Data}} & \multicolumn{2}{c}{\textbf{ggF Signal}} & \multicolumn{2}{c}{\textbf{VBF Signal}} \\
\multicolumn{1}{c}{ } & \multicolumn{1}{c}{ } & \multicolumn{1}{c}{\textbf{SM}} & \multicolumn{1}{c}{\textbf{$\kl=10$}} & \multicolumn{1}{c}{\textbf{SM}} & \multicolumn{1}{c}{\textbf{$\kvv=0$}} \\

\hline
\multicolumn{1}{c}{ {Common preselection} }&\multicolumn{5}{c}{} \\
\hline
 
Preselection                                          & 5.70e8                & 530          & 7300       & 22             & 630 \\
Trigger class                                         & 2.49e8                & 380          & 5300       & 16             & 410 \\

\hline
\multicolumn{1}{c}{ {ggF selection} }&\multicolumn{5}{c}{} \\
\hline

Fail VBF selection                                     & 2.46e8                 & 380            & 5200       & 14             & 330 \\
At least 4 \Pqb-tagged central jets            & 1.89e6                   & 86            & 1000       & 1.9             & 65 \\
$\deta < 1.5$                                             & 1.03e6                   & 72              & 850        & 0.94            & 46 \\
$\Xwt > 1.5$                                              & 7.51e5                   & 60              & 570        & 0.74            & 43  \\
$\Xhh < 1.6$ (ggF signal region)               & 1.62e4                   & 29            & 180        & 0.24              & 23 \\
 
\hline
\multicolumn{1}{c}{ {VBF selection} }&\multicolumn{5}{c}{} \\
\hline
 
Pass VBF selection                                         & 3.30e6                & 5.2            & 81         & 2.2              & 71 \\
At least 4 \Pqb-tagged central jets                   & 2.71e4               & 1.1             & 15         & 0.74              & 28 \\
$\Xwt > 1.5$                                                     & 2.18e4               & 1.0             & 11         & 0.67              & 26  \\
$\Xhh < 1.6$                                                     & 5.02e2               & 0.48             & 3.1        & 0.33              & 17 \\
$\mhh>$ \SI{400}{GeV} (VBF signal region)    & 3.57e2               & 0.43             & 1.8          & 0.30              & 16 \\
 
\bottomrule
\end{tabular}
\label{tab:cutflow_comb}
\end{table}

% End of text imported from the .//tables/cutflow_comb.tex input file

% End of text imported from the .//sections/analysis.tex input file

\section{Background modeling}
\label{sec:background}

% The next lines are included from the .//sections/background.tex input file
After the selection described above, about 90\% of the background events come from multijet processes (excluding top quark production), with the approximately 10\% remainder almost entirely composed of \ttbar events.
This background composition was determined by applying the full event selection to simulated samples of the various processes and comparing the yields with the total background estimate in the SR; it is purely meant to be indicative and is not used for deriving any results.
The background is modeled using the fully data-driven technique described below.
 
The background estimation makes use of an alternative set of events, which pass the same \Pqb-jet triggers and satisfy all the same selection criteria as the $4\Pqb$ events, with one difference: they are required to contain exactly two \Pqb-tagged jets.
This sample, referred to hereafter as ``$2\Pqb$'', has about two orders of magnitude more events than the $4\Pqb$ sample, hence the presence of any \hhb signal in it is negligible, making it suitable for the background estimation.
The jets selected to form the two Higgs boson candidates in the $2\Pqb$ events are the two \Pqb-tagged jets and the two untagged central jets with the highest \pT (excluding the VBF jets in the VBF categories).
 
The kinematic properties of the \(2b\) and \(4b\) events are not expected to be identical, partly due to different processes contributing to the two samples, but also due to differences in the trigger acceptance and because the probability of tagging a \Pqb-jet varies as a function of jet \pT and $\eta$.
Therefore, a reweighting function is required, which, when applied to the \(2b\) events, maps their kinematic distributions onto the corresponding \(4b\) distributions.
This function is derived using the \(2b\) and \(4b\) events in a \textit{Control Region} (CR) surrounding the SR in the reconstructed ($m_{H1}$, $m_{H2}$) plane and then applied to the \(2b\) events in the SR to produce the background estimate.
The ``inner edge'' of the CR is defined by $\Xhh = 1.6$ and the ``outer edge'' by the circle:
\begin{equation}
R_{\text{CR}}=\sqrt{ \left(m_{\PH1} - 1.05 \cdot \SI{124}{\GeV}\right)^2 +  \left(m_{\PH2} - 1.05 \cdot \SI{117}{\GeV}\right)^2 } = \SI{45}{\GeV}\,.
\end{equation}
 
\begin{figure}[tbh]
\centering
\subfloat[]{ 
\includegraphics[width=0.49\textwidth]{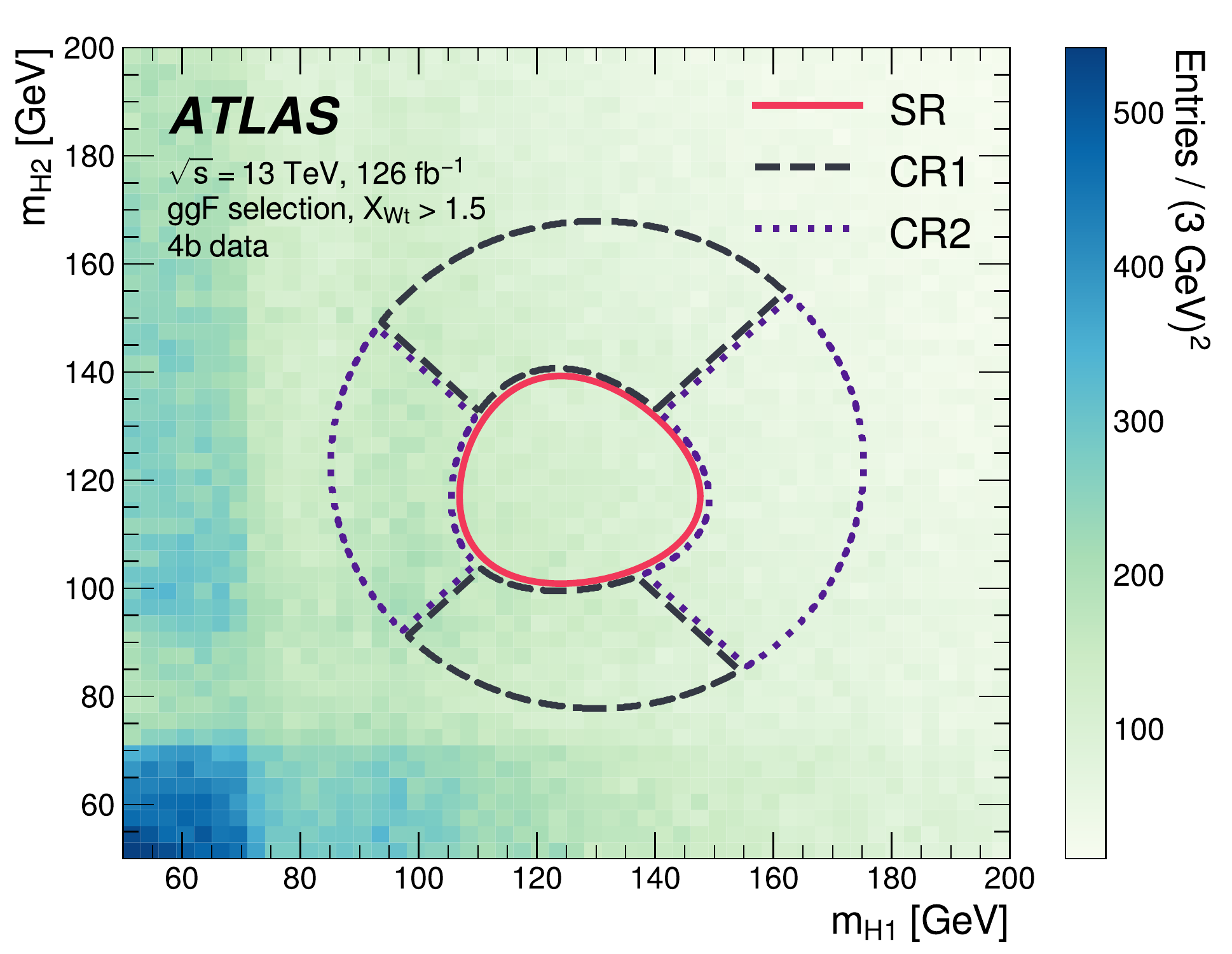}
}
\subfloat[]{ 
\includegraphics[width=0.49\textwidth]{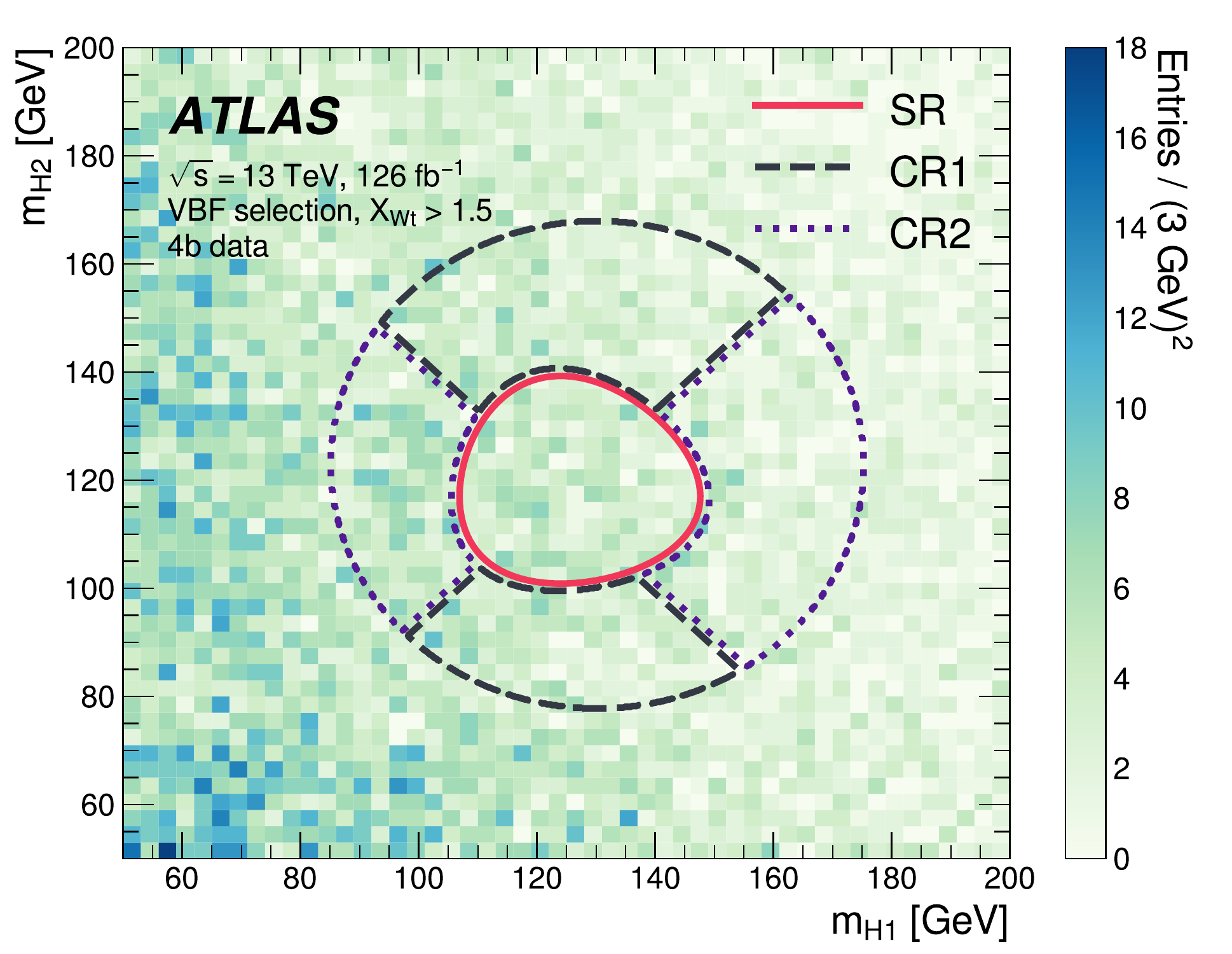}
}\\
\caption{The mass planes of the reconstructed Higgs boson candidates for the (a) ggF and (b) VBF signal regions of the analysis, shown for the $4\Pqb$ data events.
In (a), the analysis selection up to step~8 (as outlined in Figure~\ref{fig:selection}) of the ggF selection has been applied, while in (b), the analysis selection up to step~7 of the VBF selection has been applied.
The continuous red line describes the Signal Region (SR), the dashed line describes Control Region~1 (CR1) and the dotted line describes Control Region~2 (CR2).
} 
\label{fig:massplanes-allYrs-data}
\end{figure}

\begin{figure}[tbh]
\centering
\subfloat[]{ 
\includegraphics[width=0.49\textwidth]{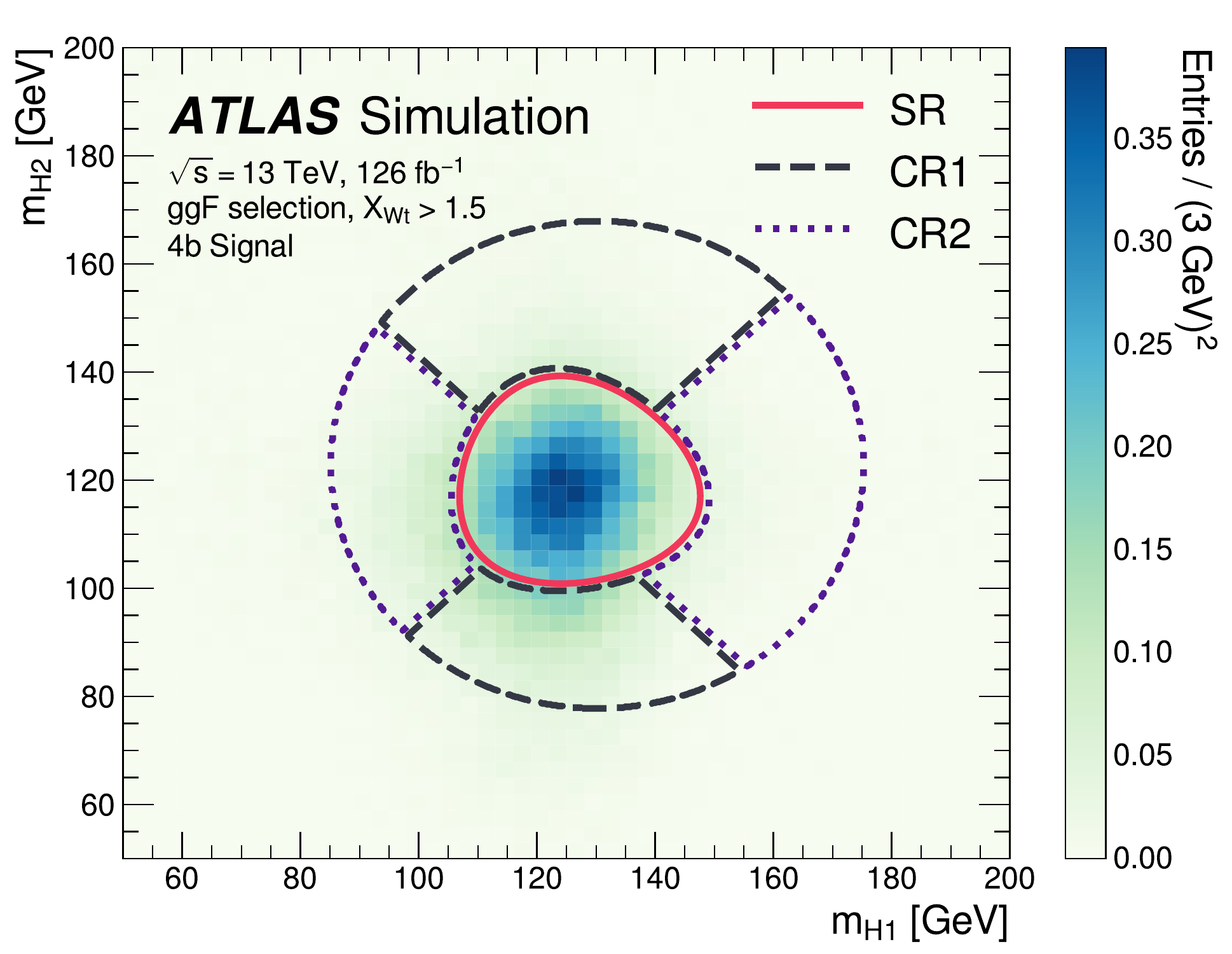}
}
\subfloat[]{ 
\includegraphics[width=0.49\textwidth]{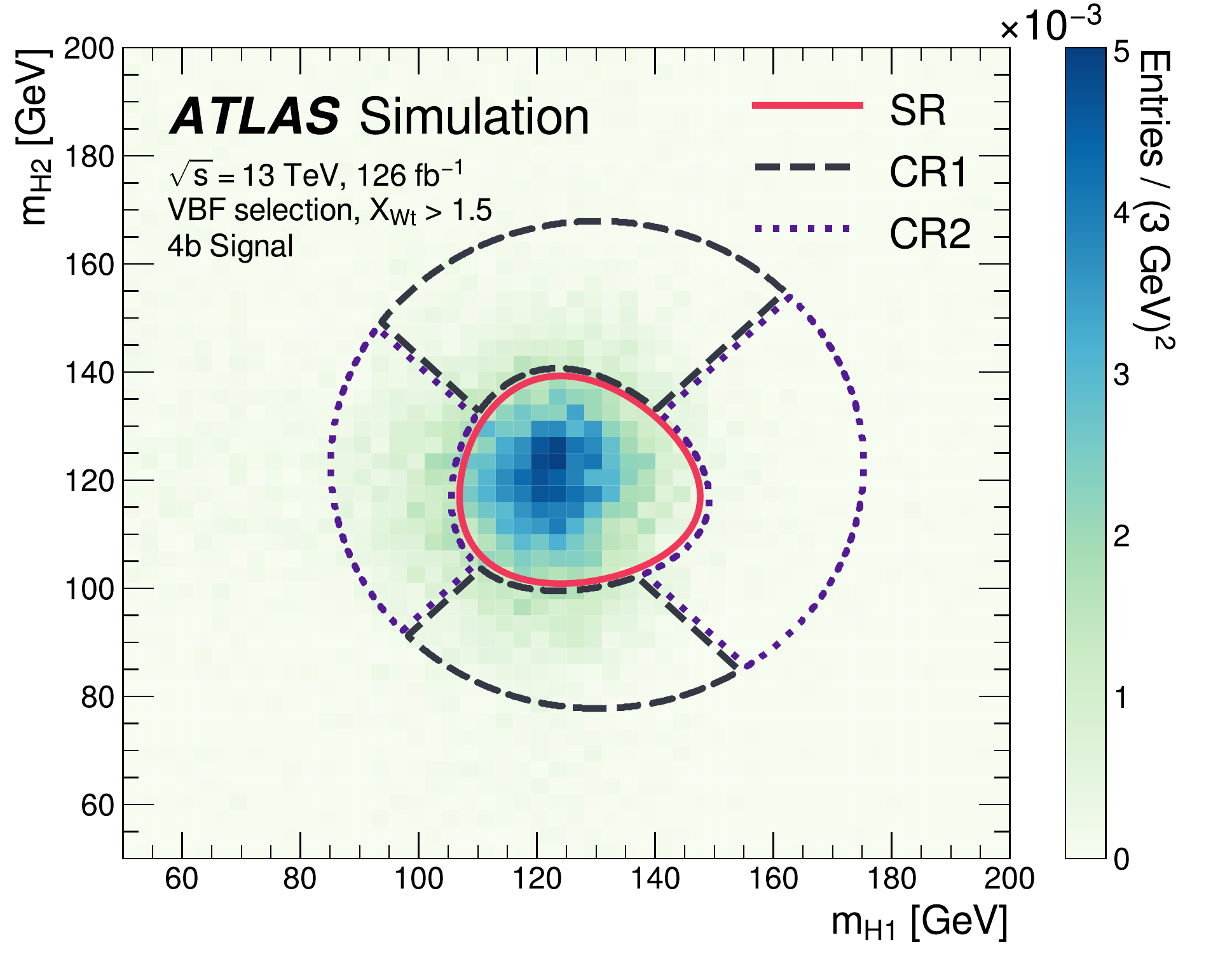}
}\\
\caption{The mass planes of the reconstructed Higgs boson candidates for the (a) simulated ggF \HH signal in the ggF signal region and (b) simulated VBF \HH signal in the VBF signal region of the analysis.
In (a), the analysis selection up to step~8 (as outlined in Figure~\ref{fig:selection}) of the ggF selection has been applied, while in (b), the analysis selection up to step~7 of the VBF selection has been applied.
The continuous red line describes the Signal Region (SR), the dashed line describes Control Region~1 (CR1) and the dotted line describes Control Region~2 (CR2).
} 
\label{fig:massplanes-allYrs-mcsig}
\end{figure}

\noindent
The shift of the center of the above circle by a factor of 1.05, relative to $\Xhh = 0$, is found to be the optimal trade-off between having a good number of events outside of the SR and avoiding the low $m_{H1}$/$m_{H2}$ regions, where the differences between \(2b\) and \(4b\) kinematic distributions are larger.
The CR is split into four roughly equal directional quadrants, defined by $45^{\circ}$ and $135^{\circ}$ lines passing through the SR center, (124, 117)~\GeV.
The four quadrants are given labels based on compass directions: the upper quadrant \QN, the lower \QS, the left \QW, and the right \QE.
The above lines also define four quadrants, with the same names as above, in the SR.
Events in  CR \QN and \QS, hereafter referred to as CR1, are used to derive the reweighting function for the nominal background estimate, while an alternative reweighting function, derived from the CR events in \QE and \QW (referred to hereafter as CR2) is used to define a systematic uncertainty related to the reweighting function interpolation into the SR, as detailed in Section~\ref{sec:systematics}.
The boundaries of the SR, CR1, and CR2 in the reconstructed ($m_{H1}$, $m_{H2}$) plane are shown in Figure~\ref{fig:massplanes-allYrs-data}. 
The horizontal and vertical bands of lower event density around \SI{80}{\GeV} visible in these plots are caused by the \Xwt selection criterion.
For comparison, the distributions of the simulated ggF and VBF \HH signals in the reconstructed ($m_{H1}$, $m_{H2}$) plane are presented in Figure~\ref{fig:massplanes-allYrs-mcsig}.

The reweighting function has the form:
 
\begin{equation} \label{eq:reweight}
w(\vec{x}) = \frac{p_{4b}(\vec{x})}{p_{2b}(\vec{x})},
\end{equation}
 
\noindent where \(p_{4b}(\vec{x})\) and \(p_{2b}(\vec{x})\) are the probability density functions for \(4b\) and \(2b\) data, respectively, over a set of kinematic variables \(\vec{x}\).
The computation of \(w(\vec{x})\) is a density ratio estimation problem, for which a variety of approaches exist.
The method employed in this analysis is modified from Refs.~\cite{NNloss2,NNLoss} and makes use of an artificial neural network (NN).
This NN is trained on \(2b\) and \(4b\) CR1 data (or CR2 data, for determining systematic uncertainties, as described Section~\ref{sec:systematics}).
The training minimizes the following loss function:
 
\begin{equation}\label{eq:loss}
\mathcal{L}(w(\vec{x})) = \int d\vec{x} \left[\sqrt{w(\vec{x})}p_{2b}(\vec{x})
+ \frac{1}{\sqrt{w(\vec{x})}}p_{4b}(\vec{x})\right] \,,
\end{equation}
\noindent The function in Eq.~(\ref{eq:reweight}) minimizes the loss function in Eq.~(\ref{eq:loss}) by equalizing the contributions from the two terms.
The kinematic variables used to make up \(\vec{x}\) are listed in Table~\ref{tbl:rw-vars} for the ggF and VBF signal regions; they are among those kinematic variables that exhibit larger differences between the \(2b\) and \(4b\) events.
The NN used in the ggF signal region has three densely connected hidden layers of 50 nodes, each with a rectified linear unit activation function~\cite{Nair:2010}, and a single-node linear output.
A similar architecture is chosen for the NN used in the VBF signal region, except that only 20 nodes are used in each of the three hidden layers.
This reflects the fact that the \(2b\) and \(4b\) sample sizes in the VBF signal region are nearly two orders of magnitude smaller than the corresponding ones in the ggF signal region.
This is also the reason behind the choice to perform the NN training in the VBF signal region for all data-taking years together, with the year index as a one-hot encoded input feature.\footnote{One-hot encoding is a standard technique in machine learning. For example, for the data-taking years in the VBF reweighting, instead of presenting the year numbers as input features to the NN, one-hot encoding uses three input features: $(1,0,0)$ for 2016, $(0,1,0)$ for 2017, and $(0,0,1)$ for 2018.}
For the ggF signal region, a dedicated reweighting is derived for each year separately, which, thanks to the adequate sample sizes, deals better with the different levels of disparity between 2\Pqb and 4\Pqb distributions, due to the differences in the trigger conditions from year to year.
Finally, in order to ensure that there are adequate numbers of 4\Pqb events for both the ggF and VBF NN trainings, these trainings are performed inclusively, before separating the events into the \deta categories (the CR events have $\Xhh>1.6$, hence it would not be possible to separate them into the \Xhh categories defined for the SR events).
Both \deta and \Xhh are found to be insensitive to the kinematic reweighting, and so the inclusive training is not expected to introduce any additional bias when separating the events into the various categories.

\begin{table}[hbt]
\centering
\caption{\label{tbl:rw-vars} The set of input variables used for the 2\Pqb to 4\Pqb reweighting in the
ggF and VBF channels respectively.
}
\begin{tabular}{p{7cm}|p{7cm}}
\toprule
\multicolumn{1}{c|}{\textbf{ggF}} & \multicolumn{1}{c}{\textbf{VBF}} \\
\hline
\begin{enumerate}
\vspace*{-5mm}
\setlength\itemsep{-0.5em}
\item $\log(p_\mathrm{T} )$ of the $2^\mathrm{nd}$ leading Higgs boson candidate jet
\item $\log(p_\mathrm{T} )$ of the $4^\mathrm{th}$ leading  Higgs boson candidate jet
\item $\log(\Delta R)$ between the closest two Higgs boson candidate jets
\item $\log(\Delta R)$ between the other two Higgs boson candidate jets
\item Average absolute $\eta$ value of the Higgs boson candidate jets
\item $\log(p_\mathrm{T}$) of the di-Higgs system
\item $\Delta R$ between the two Higgs boson candidates
\item $\Delta \phi$ between jets in the leading Higgs boson candidate
\item $\Delta \phi$ between jets in the subleading Higgs boson candidate
\item $\log(\Xwt)$
\item Number of jets in the event
\item Trigger class index as one-hot encoder
\vspace*{-5mm}
\end{enumerate}
&
\begin{enumerate}
\vspace*{-5mm}
\setlength\itemsep{-0.5em}
\item Maximum dijet mass from the possible pairings of the four Higgs boson candidate jets
\item Minimum dijet mass from the possible pairings of the four Higgs boson candidate jets
\item Energy of the leading Higgs boson candidate
\item Energy of the subleading Higgs boson candidate
\item Second-smallest $\Delta R$ between the jets in the leading Higgs boson candidate (from the three possible pairings for the leading Higgs candidate)
\item Average absolute $\eta$ value of the four Higgs boson candidate jets
\item $\log(\Xwt)$
\item Trigger class index as one-hot encoder
\item Year index as one-hot encoder (for years inclusive training)
\vspace*{-5mm}
\end{enumerate}\\
\bottomrule
\end{tabular}
\end{table}

\begin{figure}[bth]
\centering
\subfloat[]{ 
\includegraphics[width=0.49\textwidth]{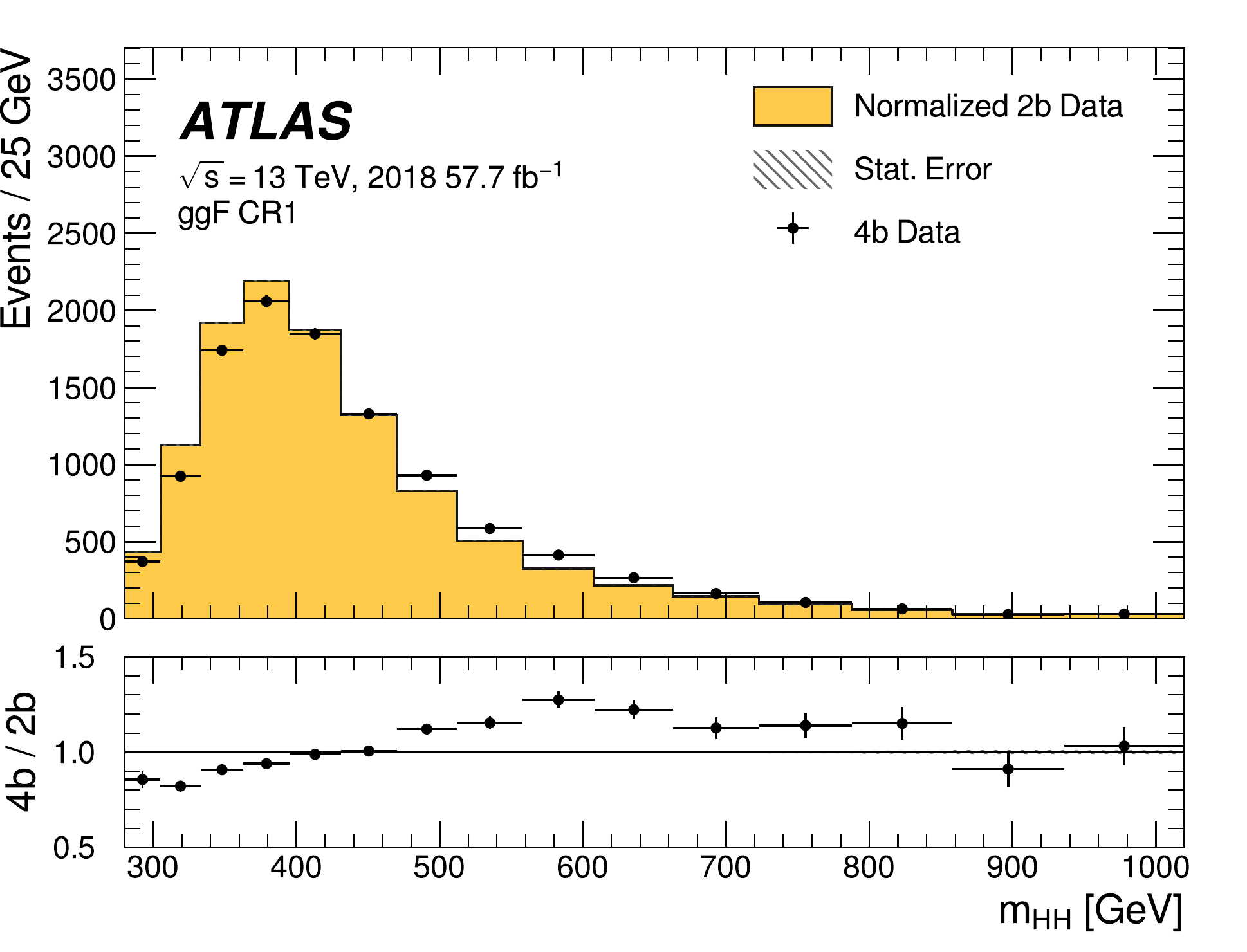}
}
\subfloat[]{ 
\includegraphics[width=0.49\textwidth]{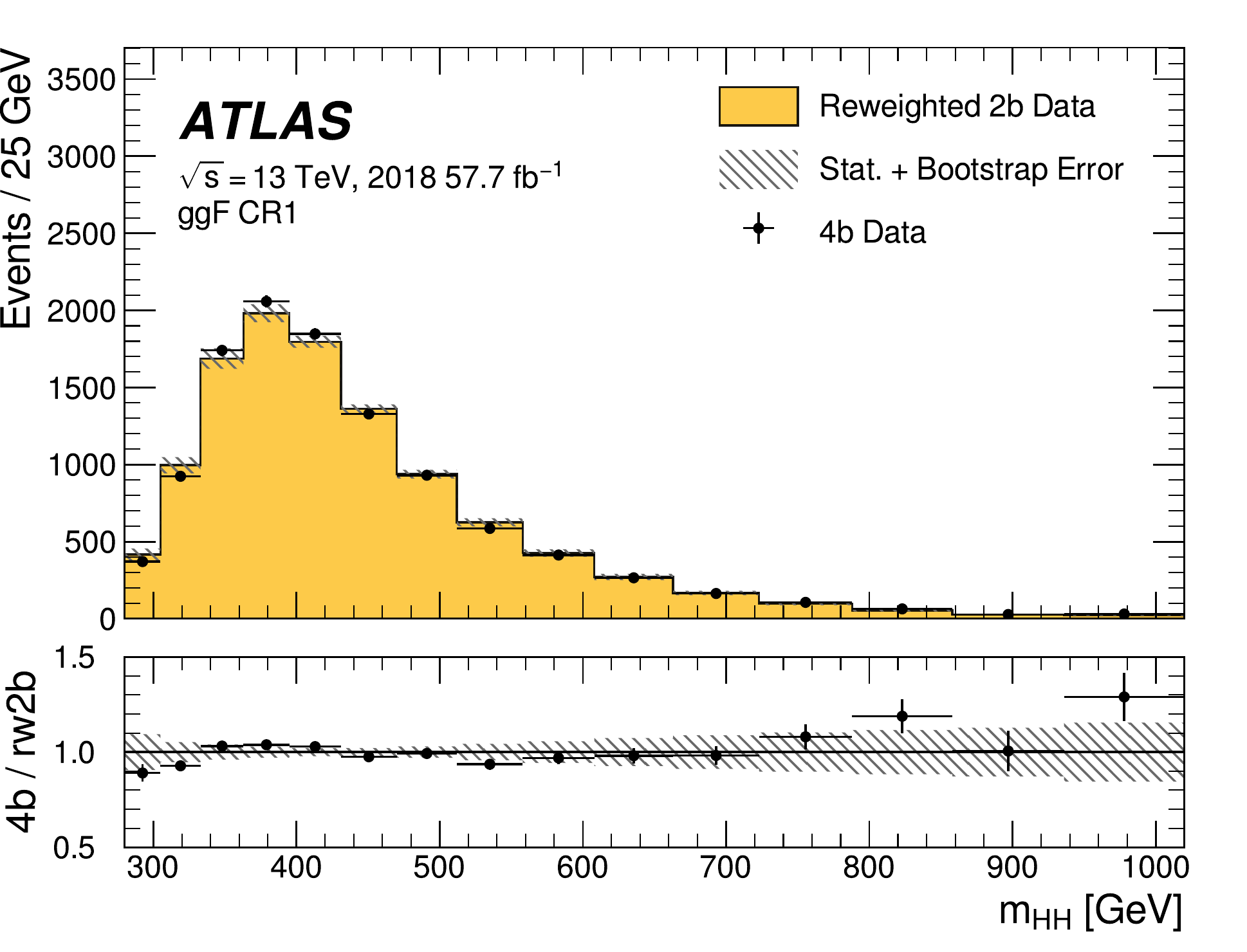}
}
\caption{
Comparison of the $2\Pqb$ (yellow histogram with hatching) and $4\Pqb$ (black points with error bars) \mhh distributions, for events in Control Region 1 (CR1) of the ggF signal region from the 2018 data: (a) before the kinematic reweighting of the $2\Pqb$ events, with only a normalization factor applied; and (b) after the kinematic reweighting of the $2\Pqb$ events.
The error bars indicate the statistical uncertainty of the $4\Pqb$ data, while the hatching indicates the statistical uncertainty of the $2\Pqb$ data.
The latter is only the Poisson uncertainty of the $2\Pqb$ data, in (a), while in (b), it also includes the uncertainty from the bootstrap procedure described in Section~\ref{sec:systematics}.
The hatching in (a) is narrower than the line width of the plotted histogram. }
\label{fig:ggf_bkgNN_mhh}
\end{figure}

\begin{figure}[tbh]
\centering
\subfloat[]{ 
\includegraphics[width=0.49\textwidth]{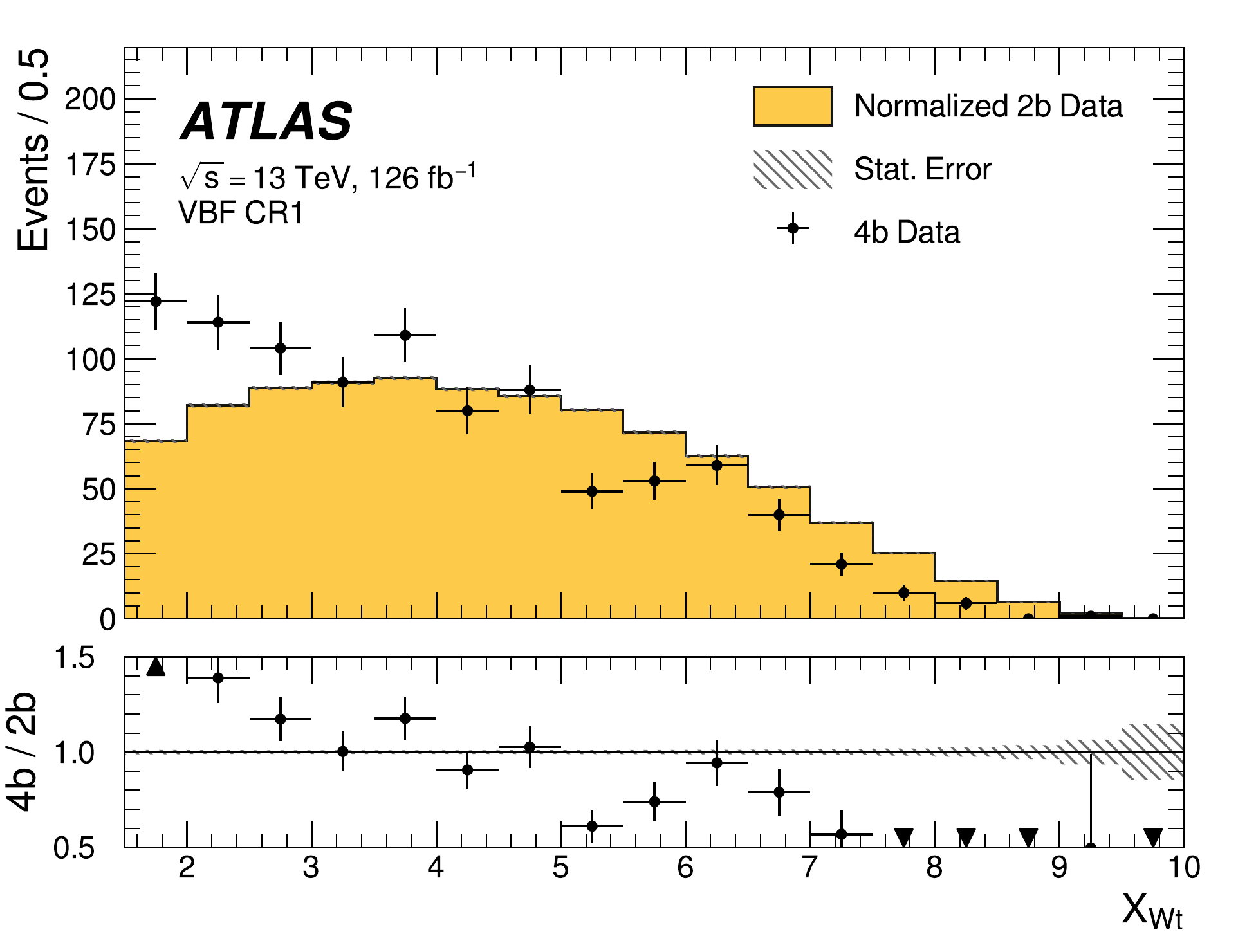}
}
\subfloat[]{ 
\includegraphics[width=0.49\textwidth]{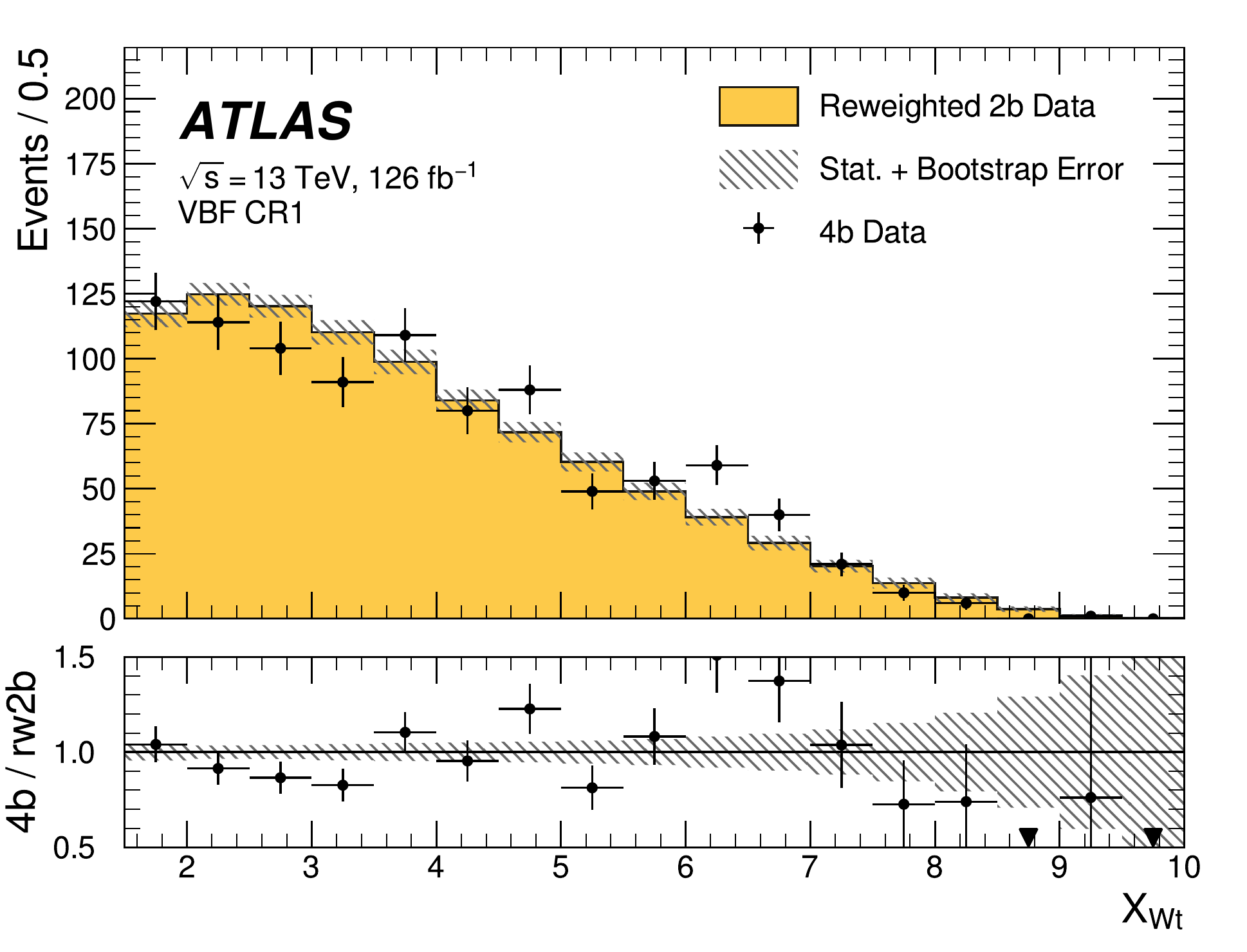}
}
\caption{
Comparison of the $2\Pqb$ (yellow histogram with hatching) and $4\Pqb$ (black points with error bars) \Xwt distributions, for events in Control Region 1 (CR1) of the VBF signal region: (a) before the kinematic reweighting of the $2\Pqb$ events, with only a normalization factor applied; and (b) after the kinematic reweighting of the $2\Pqb$ events.
The error bars indicate the statistical uncertainty on the $4\Pqb$ data, while the hatching indicates the statistical uncertainty on the $2\Pqb$ data.
The latter is only the Poisson uncertainty on the $2\Pqb$ data, in (a), while in (b), it also includes the uncertainty from the bootstrap procedure described in Section~\ref{sec:systematics}.
}
\label{fig:vbf_bkgNN_Xwt}
\end{figure}
 
In order to estimate and mitigate the impact of the varying initial conditions and limited size of the training samples on the NN training, the deep ensembles technique~\cite{lakshminarayanan2017simple} is used together with a bootstrap resampling~\cite{Bootstrap} of the training data. This entails constructing a set of training datasets by sampling with replacement from the original dataset. In this analysis, this is approximated by the usage of different random training weights, following a Poisson distribution with $\mu = 1$, for each event in each training.
The NN is trained independently on each element of this set, using different initial conditions each time.
This results in an ensemble of reweighting functions.
Each reweighting function is further multiplied by a normalization factor, such that the number of  reweighted \(2b\) events is equal to the number of \(4b\) events in the region where the NN is trained.
In this analysis, the ensembles comprise 100 reweighting functions each, hence 100 weights are calculated for each \(2b\) event in the SR.
The background estimate uses the mean of these weights for each event, and the variation of the background predictions from the ensemble of reweighting functions is used to estimate a systematic uncertainty for the stability of the NN training procedure, as described in Section~\ref{sec:systematics}.
 
The effect of the above reweighting procedure in CR1, where the reweighting function is derived, is illustrated in Figure~\ref{fig:ggf_bkgNN_mhh} for the \mhh distribution of the ggF-selected events and in Figure~\ref{fig:vbf_bkgNN_Xwt} for the \Xwt distribution of the VBF-selected events.
The reweighted ``$2\Pqb$'' distributions agree with the corresponding ``$4\Pqb$'' distributions to within about 10\% for most of the phase space, with some larger deviations observed in bins near the tails of the distributions where fewer data events are available.
A large number of additional kinematic variables were also studied before and after applying the reweighting in order to validate the performance of the NN.
For all variables, the level of agreement, as quantified by the $\chi^2$ metric, either improves after the reweighting or, for variables where the ``$2\Pqb$'' and ``$4\Pqb$'' distributions are already similar, changes only slightly.
 
The background modeling procedure was tested and found to produce good results in a large simulated \ttbar sample and a much smaller sample of simulated (non-\ttbar) multijet events in the SR.
The procedure was also tested in several control data samples orthogonal to the nominal event selection, where the presence of any \HH signal is negligible and the \(4b\) events in the corresponding SR can be compared with the reweighted SR \(2b\) events without any bias.
These samples, summarized in Table~\ref{tab:cr_summary}, include: (a) events satisfying all the \(2b\)/\(4b\) ggF selection criteria, with the difference that the $\deta<1.5$ cut is inverted; (b) events satisfying all the \(2b\)/\(4b\) selection criteria, except that the center of the SR (and hence also of CR1 and CR2) is shifted, to avoid any overlap with the nominal SR; and (c) events that satisfy all the same \(4b\) selection criteria, except that, in terms of \Pqb-tagging, they contain exactly three \Pqb-tagged jets, and all other jets fail a looser working point of the \Pqb-tagging algorithm (one that gives 85\% efficiency for \Pqb-jets in simulated \ttbar events); from those jets, the one with the highest \pt is chosen as the fourth jet.
The latter sample, hereafter referred to as \bbbf, has about one order of magnitude more events than the \(4b\) sample and a negligible amount of \HH signal; hence it is used to derive a nonclosure systematic uncertainty for the reweighting procedure, as discussed in Section~\ref{sec:systematics}.
No significant background modeling nonclosure was observed in the other control data samples.
 
\begin{table}
\centering
\caption{ A summary of all the data samples used in the analysis. For each control sample, the variations from the nominal analysis are noted explicitly. }
\begin{tabular}{ p{2.4cm} p{6.5cm} p{5.8cm} }
\toprule
 
\multicolumn{1}{l}{\textbf{Data Sample}} & \multicolumn{1}{l}{\textbf{Definition}}& \multicolumn{1}{l}{\textbf{Usage}} \\
 
\hline{}&{}&{}\\
 
Signal Region (SR) & Events with $\Xhh < 1.6$    & Defines signal region in the $m_{\PH1}$--$m_{\PH2}$ plane \\
{}&{}&{}\\
 
Control Region (CR) & Events with $\Xhh > 1.6$ and $R_{\text{CR}}<45$~\GeV   & Defines control region in the $m_{\PH1}$--$m_{\PH2}$ plane for background estimation (ggF and VBF) \\
{}&{}&{}\\
 
Shifted validation regions & Shift the center of the SR in the $m_{\PH1}$--$m_{\PH2}$ plane to avoid overlap with the nominal SR & Background estimation validation (ggF only) \\
 
{}&{}&{}\\
\hline{}&{}&{}\\
 
$4\Pqb$ &  Events with at least 4 \Pqb-tagged central jets & Final analysis sample (ggF and VBF) \\
{}&{}&{}\\
 
$2\Pqb$ &  Events with exactly 2 \Pqb-tagged central jets plus at least two additional untagged central jets & Background estimation (ggF and VBF) \\
{}&{}&{}\\
 
\bbbf &  Events with exactly 3 \Pqb-tagged central jets plus at least one central jet failing a looser \Pqb-tagging requirement & Background estimation validation (ggF and VBF), additional background modeling uncertainty (ggF only) \\
{}&{}&{}\\
 
\hline{}&{}&{}\\
 
Reverse \deta & $2\Pqb$ and $4\Pqb$ events with $\deta>1.5$ & Background estimation validation (ggF only) \\
{}&{}&{}\\
 
\bottomrule
\end{tabular}
\label{tab:cr_summary}
\end{table}

% End of text imported from the .//sections/background.tex input file

\section{Systematic uncertainties}
\label{sec:systematics}

% The next lines are included from the .//sections/systematics.tex input file
 
The uncertainties with the greatest impact on the analysis sensitivity are those arising from the data-driven background estimate described in Section~\ref{sec:background}.
These uncertainties have two main sources: the limited sample sizes in the CR and SR, and physical differences between the CR, where the 2\Pqb reweighting function is derived, and the SR, where it is applied.
 
As described in Section~\ref{sec:background}, the ensemble of 100 reweighting functions results in 100 separate background predictions.
An \mhh histogram can be constructed from each of these predictions, and the standard deviation of the predictions in each bin is taken as the bootstrap uncertainty.
The uncertainty is treated as uncorrelated across \mhh bins.
 
An additional statistical uncertainty results from the limited sample size of the $2\Pqb$ SR dataset in which the trained background reweighting network is applied to obtain the final background estimate.
A Poisson uncertainty is taken for each \mhh bin, which is combined in quadrature with the bootstrap uncertainty described above.

For the background estimate, the uncertainty component related to the kinematic differences between the SR and CR1 is evaluated by using alternative predictions from the CR2 region.
Four alternative background estimates are produced by applying the CR1-derived weights to three of the SR quadrants, and CR2-derived weights to the one remaining SR quadrant.
For example, one alternative background estimate is obtained by applying CR1-derived weights to Q$_\text{S}$, Q$_\text{E}$ and Q$_\text{W}$, and CR2-derived weights to Q$_\text{N}$.
Each of these four background predictions is symmetrized around the nominal \mhh distribution to construct a two-sided uncertainty.
Since the \mhh distribution differs across the four SR quadrants, substituting the CR2-based prediction for the CR1-based prediction in each of the four SR quadrants separately and utilizing a four-component uncertainty gives the fit model greater flexibility to describe these \mhh variations with finer granularity.
In the ggF signal region, these uncertainties are taken to be uncorrelated across the datasets from the three different years.
In both the ggF and VBF signal regions, the uncertainty is treated as correlated across the analysis categories.
 
An additional closure uncertainty is estimated by applying the full background modeling procedure to the \bbbf sample instead of the 4\Pqb sample.
The predicted \bbbf \mhh distribution in the various analysis categories is then compared with the observed \bbbf data in the SR.
For the VBF signal region, no statistically significant difference between the prediction and observation is found, and hence no additional uncertainty is applied.
For the ggF signal region, an additional uncertainty is evaluated in each category from the observed differences between the predicted and observed \bbbf \mhh distributions.
For \mhh bins in which the predicted and observed values differ by less than $1\sigma$, where $\sigma$ is obtained from all other background modeling uncertainties combined, no additional uncertainty is applied.
For \mhh bins where the predicted and observed values differ by more than $1\sigma$, the amount beyond $1\sigma$ is averaged with the corresponding amounts in the two adjacent bins, to limit the impact of statistical fluctuations, and is symmetrized around the nominal prediction to construct a two-sided uncertainty.
This nonclosure uncertainty has a much smaller impact on the analysis sensitivity than the other sources of background modeling uncertainty.
 
Several detector modeling uncertainties are evaluated and included.
These affect only the signal description, as the background is estimated entirely from data.
Uncertainties in the jet energy scale and resolution, as well as the JVT, are treated according to the prescription in Refs.~\cite{JETM-2018-05, PERF-2014-03}. 
Additional uncertainties arising from the correction of the simulated \pileup distribution are treated according to the prescription in Ref.~\cite{STDM-2015-05}.
Uncertainties in the \(b\)-tagging efficiency are treated according to the prescription in Ref.~\cite{FTAG-2018-01}.
Uncertainties in the trigger efficiencies are evaluated from measurements of per-jet online efficiencies for both jet reconstruction and \(b\)-tagging, which are used to compute event-level uncertainties.
These are then applied to the simulated events as overall weight variations.
The uncertainty in the integrated luminosity used in this analysis is in the range 2.0\%--2.4\% for the three years of data taking and 1.7\% for the entire dataset~\cite{ATLAS-CONF-2019-021}, obtained using the LUCID-2 detector for the primary luminosity measurements~\cite{Avoni:2018iuv}.
 
Several sources of theoretical uncertainty affecting the signal models are considered as described below.
Uncertainties due to modeling of the parton shower and underlying event are evaluated by comparing results between two generators for these parts of the calculation:
the nominal \PYTHIA[8] and the alternative \HERWIG[7].
This is found to have an effect of roughly $10\%$ on the ggF and VBF signal acceptances, and a negligible impact on the shape of the \mhh distributions.
The parton showering uncertainty is derived within each analysis SR category; the uncertainty is observed to reach approximately $40\%$ for a given production mode in some categories in which the acceptance is small for that mode.
Uncertainties in the matrix element calculation are evaluated by raising and lowering the factorization and renormalization scales used in the generator by a factor of two, both independently and simultaneously.
This results in an effect of typically $2\%$ for both ggF and VBF, with a maximum effect of about $6\%$ in certain analysis categories.
PDF uncertainties are evaluated using the PDF4LHC\_NLO\_MC set~\cite{Butterworth:2015oua} by calculating the signal acceptance for each replica and taking the standard deviation.
The magnitude of this uncertainty is typically found to be less than $1\%$ in both the ggF and VBF signal acceptances, with a maximum magnitude of approximately $2\%$.
Theoretical uncertainties in the \hbb branching ratio~\cite{deFlorian:2016spz} are included, amounting to an approximately $3.5\%$ overall uncertainty in the signal normalization.
The dependence of the branching ratio uncertainty on \kl is neglected.
Theoretical uncertainties in the ggF and VBF \HH cross-sections arising from uncertainties in the PDF and $\alphas$, as well as the choice of renormalization scheme and the scale of the top quark's mass, are taken from Refs.~\cite{deFlorian:2016spz,Baglio:2020wgt,Grazzini:2018bsd}. 
The cross-section uncertainties are included in the derivation of the upper limits on the ggF, VBF, and combined \HH signal strengths, as well as the likelihood-based constraints on the values of the \kl and \kvv modifiers, as presented in Section~\ref{sec:result}.
 
An additional signal modeling systematic uncertainty is evaluated for the SMEFT and HEFT measurements.
The \mhh spectra of reweighted SMEFT/HEFT signal samples are compared against explicitly generated samples for a select number of coefficient variations.
A two-component normalization uncertainty is derived by taking the average of the relative deviations across the \mhh bins in the ranges of $280<\mhh< \SI{936}{\GeV}$ and $\mhh>\SI{936}{\GeV}$.
The use of separate components in the low- and high-\mhh regions prevents the level of agreement in the more populated low-\mhh region from overconstraining the uncertainty in the more sparsely populated high-\mhh region.

% End of text imported from the .//sections/systematics.tex input file

\section{Results}
\label{sec:result}

% The next lines are included from the .//sections/result.tex input file
 
The analysis results are obtained using a maximum-likelihood fit performed in bins of reconstructed \mhh.
For the ggF signal region, the fit is performed simultaneously across the different data-taking years (2016, 2017, and 2018), while for the VBF signal region, the fit is performed inclusively on the data from all years.
 
The likelihood function used to construct the test statistic has a standard form, consisting of a product of Poisson distributions for the yields in each bin and constraint functions for nuisance parameters describing systematic uncertainties.
For uncertainties due to the limited sample size in data or simulation, the constraint is a Poisson distribution.
For all other systematic uncertainties, the constraint is a Gaussian distribution.
Where systematic uncertainties are deemed to be uncorrelated, independent nuisance parameters are introduced.
Uncertainties in the luminosity and signal modeling are treated as fully correlated between the analysis categories and, for ggF, the data-taking years.
Each component of the quadrant-derived uncertainty covering the kinematic differences between the SR and CR1 regions is correlated across the data-taking years for the ggF region.
The components are correlated across analysis categories within the ggF and VBF signal regions, but not between the ggF and VBF signal regions.
All other uncertainties in the background model are treated as uncorrelated across the different categories and data-taking years.
The statistical model is implemented using RooFit~\cite{RooFit}.

The hypothesis of the presence of a signal is tested using the profile likelihood ratio~\cite{Cowan:2010js}.
The signal strength of the combined ggF and VBF signal process, $\mu_{\text{ggF}+\text{VBF}}$ ($=\sigma_{\text{ggF}+\text{VBF}}/\sigma_{\text{ggF}+\text{VBF}}^{\text{SM}}$), is chosen as the parameter of interest (POI) and is a free parameter in the fit.
The relative contributions of the ggF and VBF signals to the total signal model are fixed to their predicted values.
The profile likelihood ratio takes the following form:
\begin{equation}
-2 \Delta \ln\lambda(\mu)  = -2 \ln\Bigg(\frac{ L( \mu,\hat{\hat{ \vect{\theta} }} ) }{L(\hat{\mu},\hat{\vect{\theta}})}\Bigg)\,,
\end{equation}
where $\mu$ is the POI and $\vect{\theta}$ represents the nuisance parameters.
The numerator represents the conditional maximum-likelihood fit, in which the nuisance parameters are set to their profiled values $\hat{\hat{\vect{\theta}}}$ for which the likelihood is maximized for a fixed value of $\mu$.
The denominator represents the unconditional likelihood fit, where both $\mu$ and $\vect{\theta}$ are set to the values which jointly maximize the likelihood, $\hat{\mu}$ and $\hat{\vect{\theta}}$, respectively.

\begin{figure}[!tbh]
\centering
\subfloat[]{ 
\includegraphics[width=0.32\textwidth]{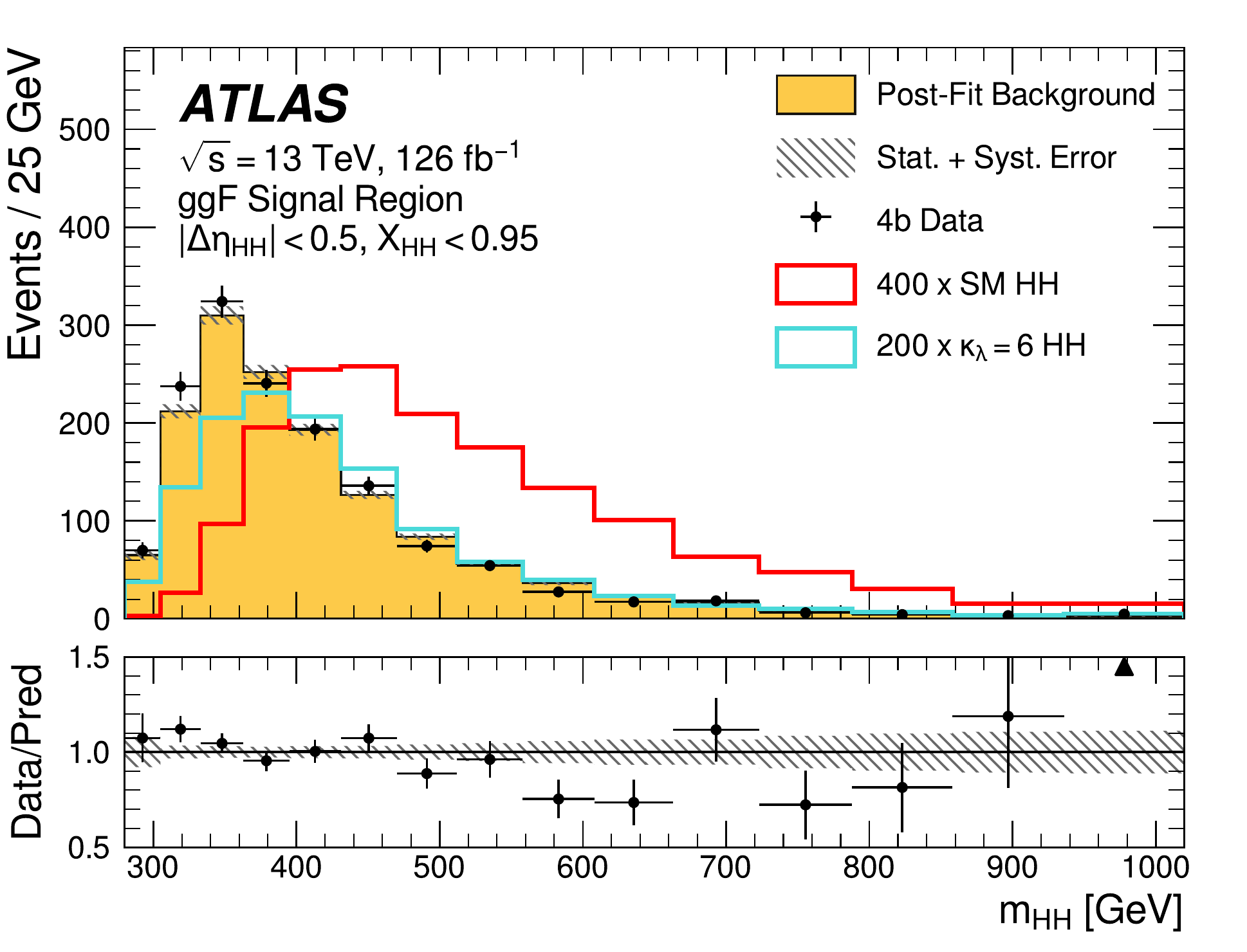}
}
\subfloat[]{ 
\includegraphics[width=0.32\textwidth]{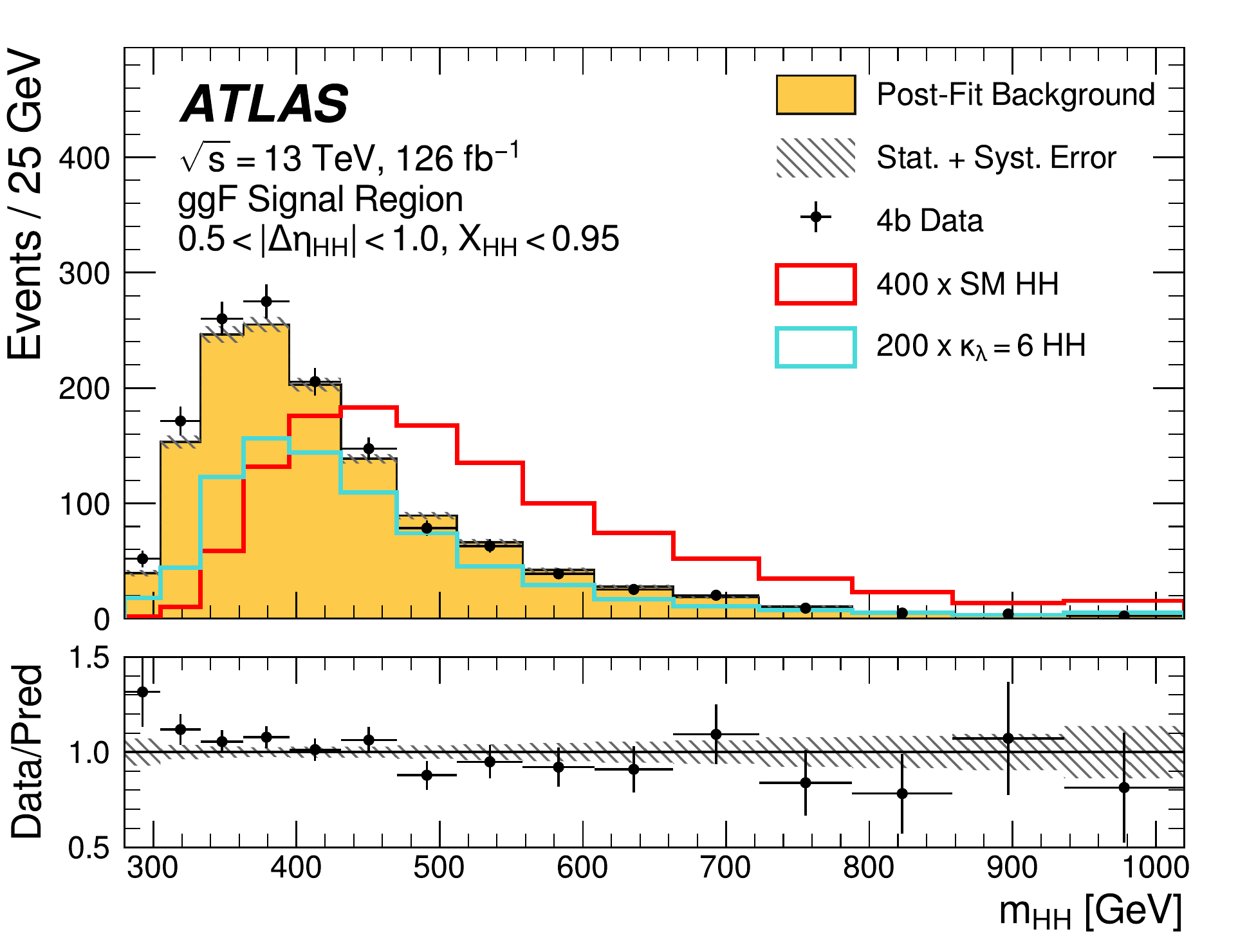}
}
\subfloat[]{ 
\includegraphics[width=0.32\textwidth]{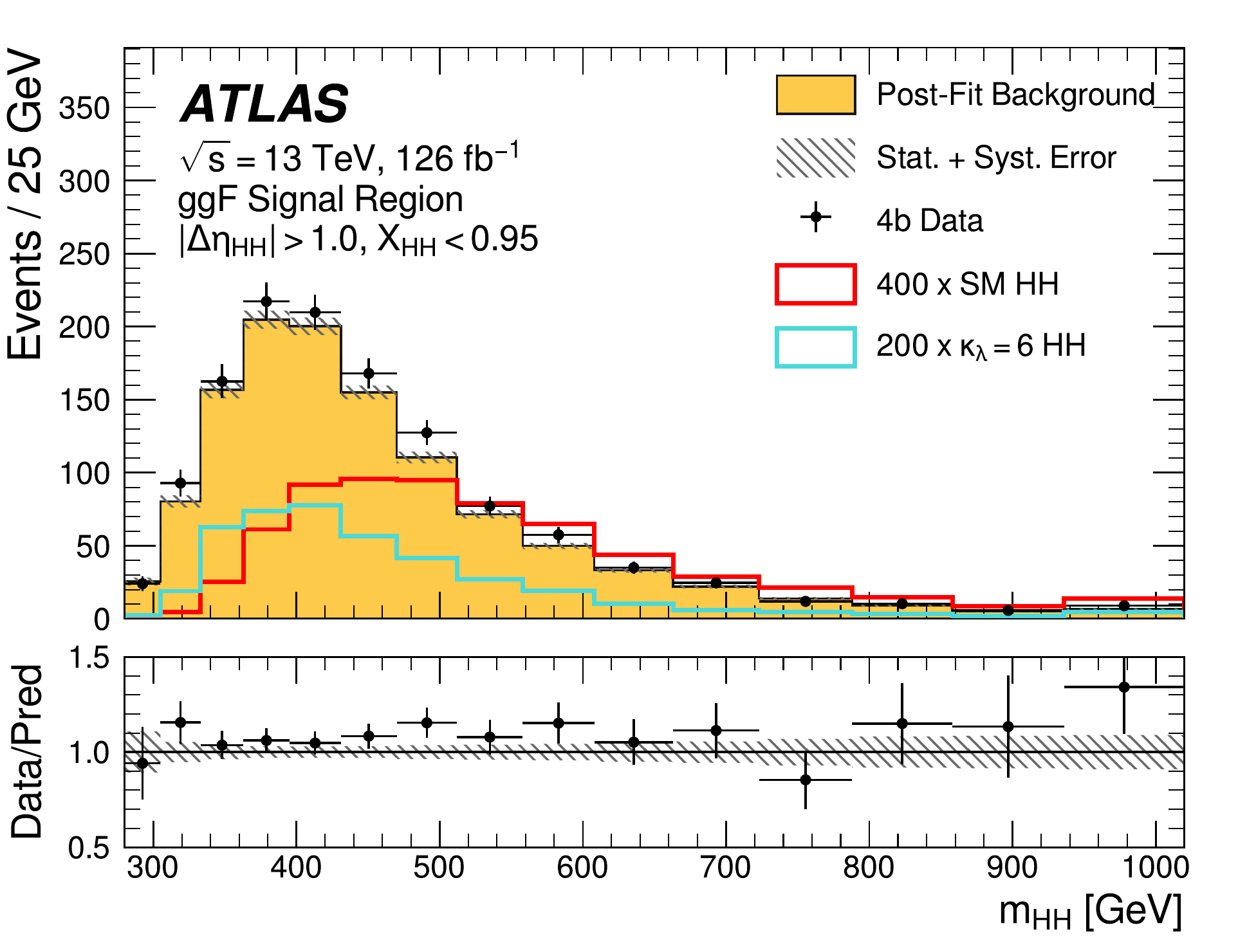}
}
\\
\subfloat[]{ 
\includegraphics[width=0.32\textwidth]{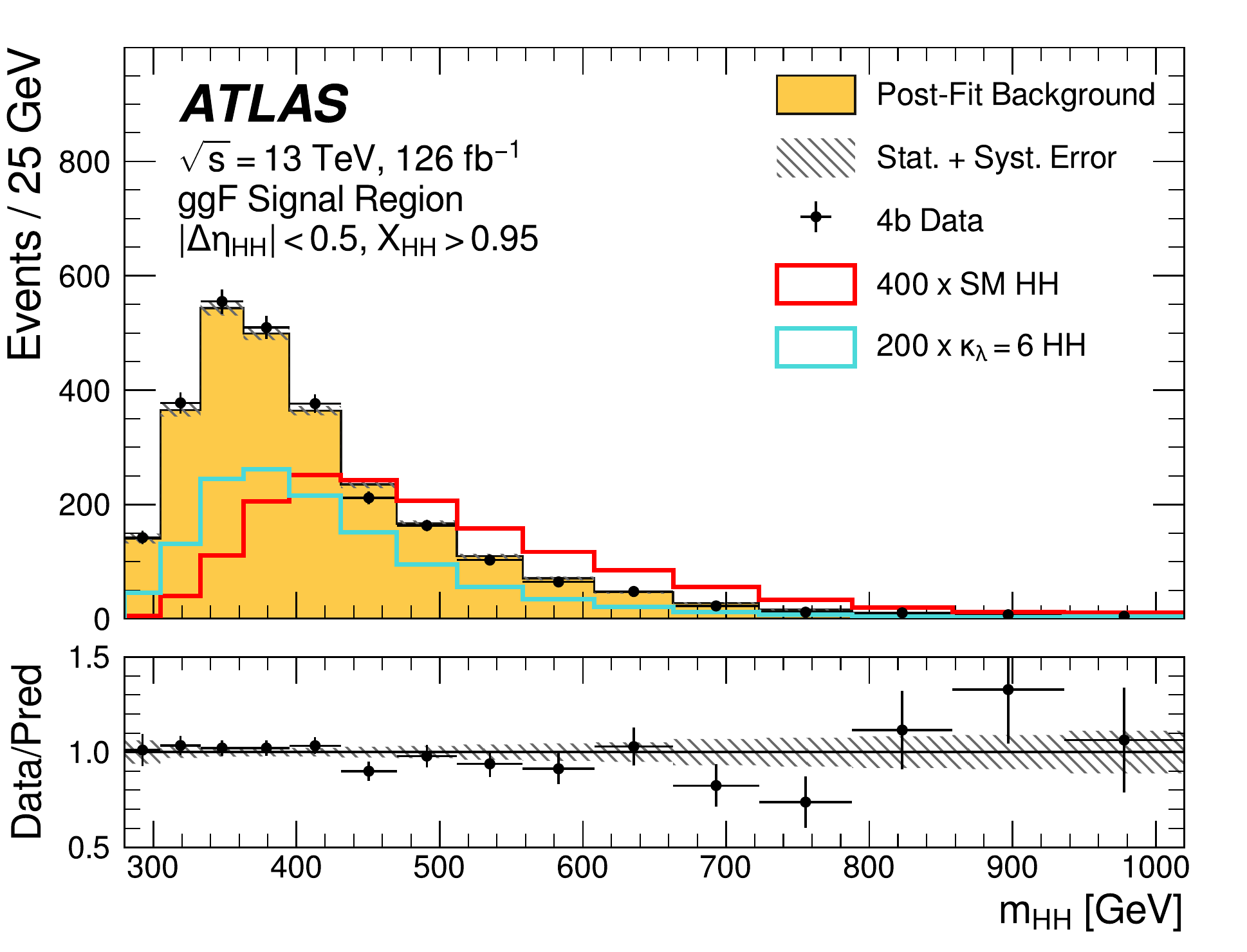}
}
\subfloat[]{ 
\includegraphics[width=0.32\textwidth]{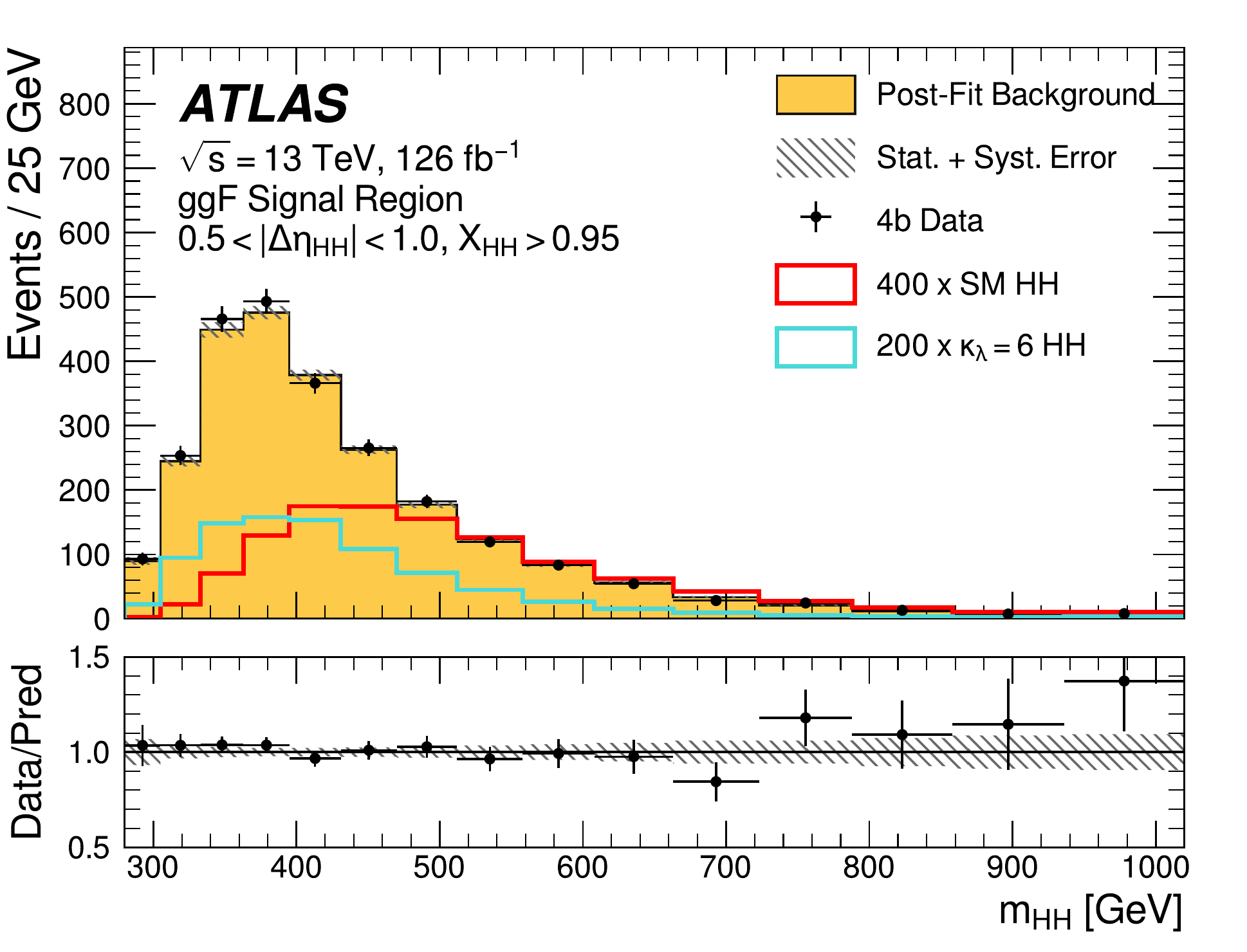}
}
\subfloat[]{ 
\includegraphics[width=0.32\textwidth]{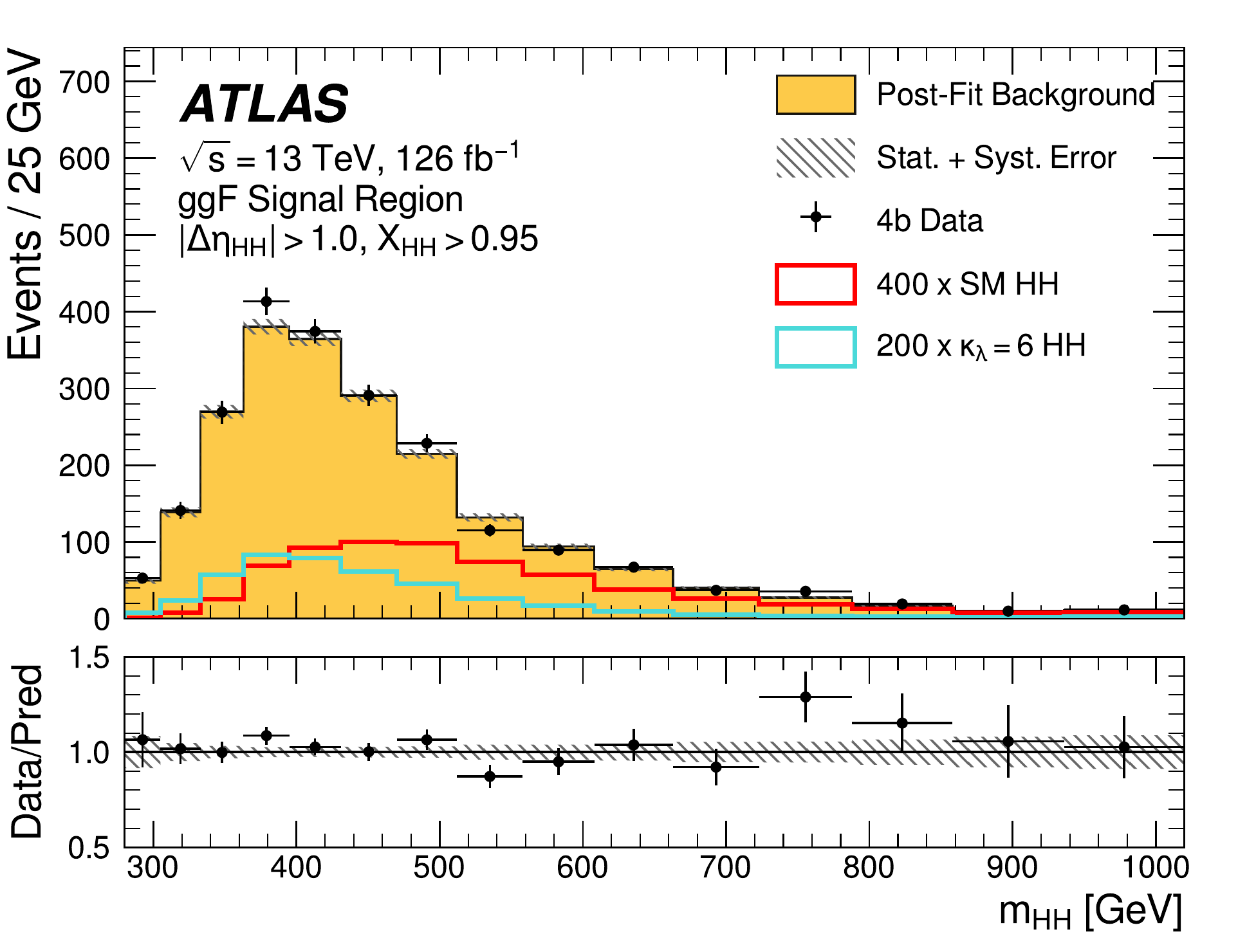}
}
\caption{
Distributions of the reconstructed \mhh in data (shown by the black points) and the estimated background (shown by the yellow histograms), in each of the six $\deta$,\,\Xhh categories in the ggF signal region: (a) $\deta<0.5$, $\Xhh<0.95$; (b) $0.5<\deta<1.0$, $\Xhh<0.95$; (c) $\deta>1.0$, $\Xhh<0.95$; (d) $\deta<0.5$, $\Xhh>0.95$; (e) $0.5<\deta<1.0$, $\Xhh>0.95$; and (f) $\deta>1.0$, $\Xhh>0.95$.
The contributions from the different data-taking years are combined in each category for presentation purposes.
The hatching shows the total uncertainty of the background estimate. 
The distribution of the expected background is obtained using the best-fit values of the nuisance parameters in the fit to the data with the background-only hypothesis.
Distributions of the SM and $\kl= 6$ signal models are overlaid, scaled so as to be visible on the plot, and the scaling for each signal model is the same across the six categories.
The lower panels show the ratio of the observed data yield to the predicted background in each bin.
Events in the underflow and overflow bins are counted in the yields of the initial and final bins respectively.
}
\label{fig:ggF-4b-disc}
\end{figure}

\begin{figure}[!tbh]
\centering
\subfloat[]{ 
\includegraphics[width=0.45\textwidth]{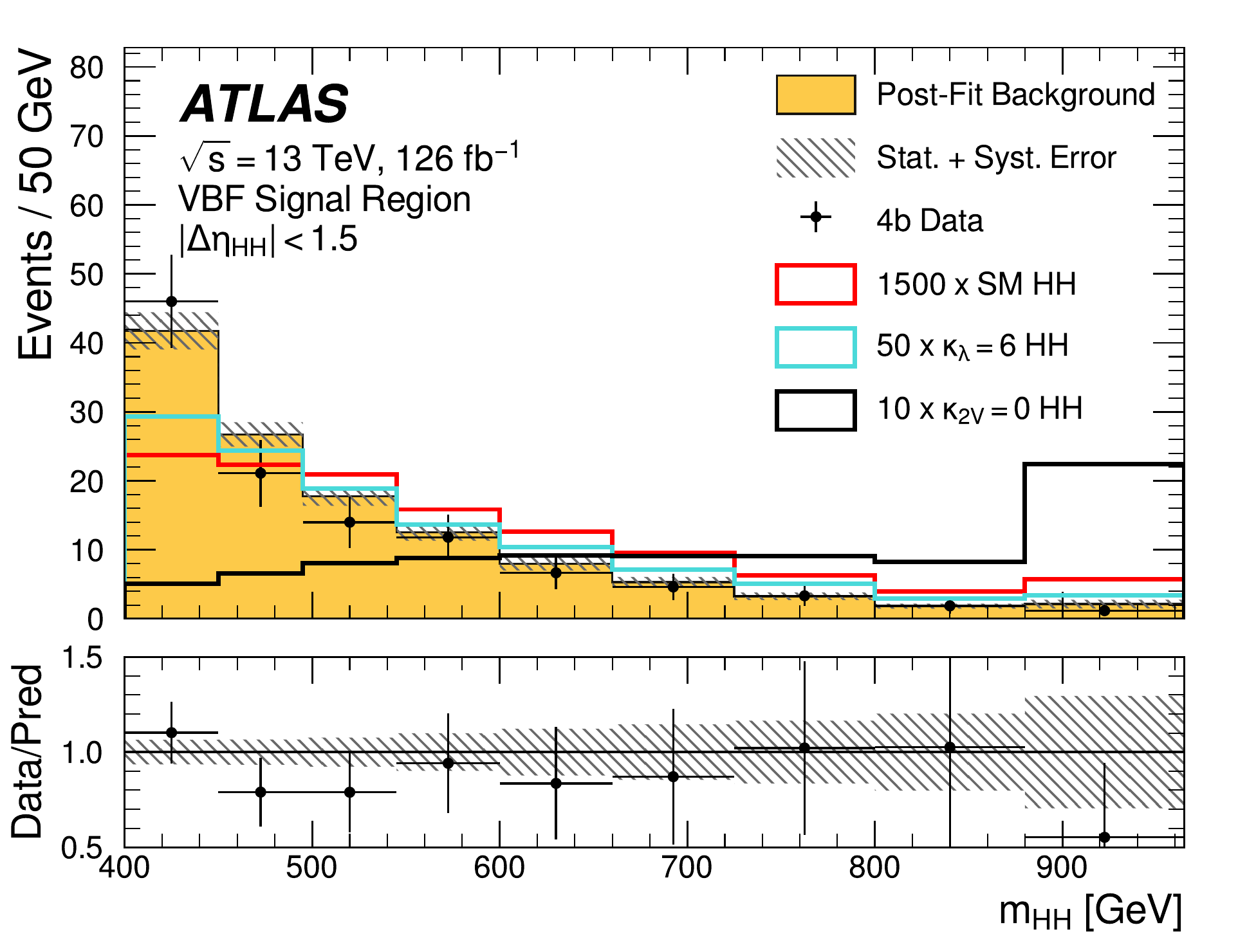}
}
\subfloat[]{ 
\includegraphics[width=0.45\textwidth]{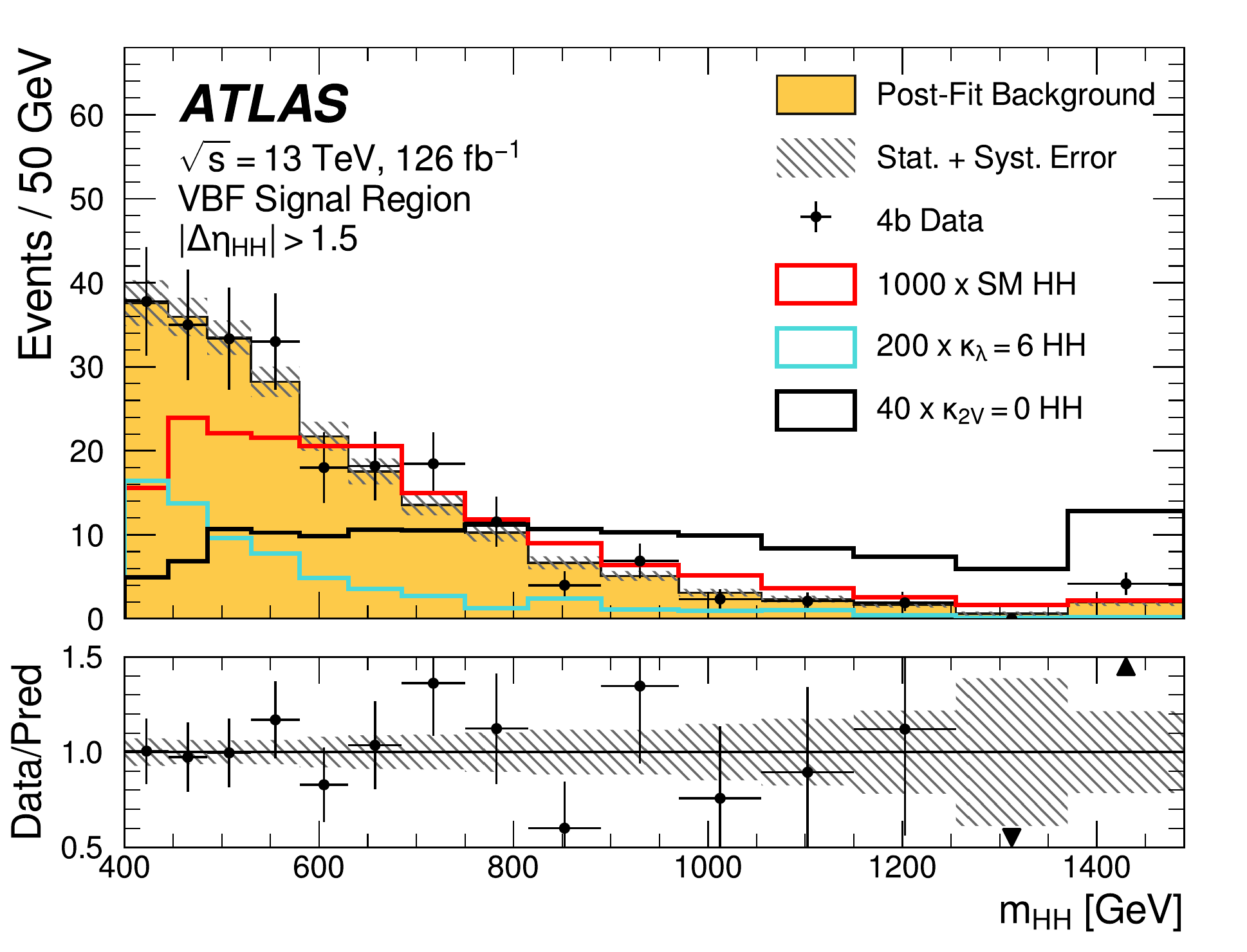}
}
\caption{Distributions of the reconstructed \mhh in data (shown by the black points), the estimated background (shown by the yellow histograms), in each of the two $\deta$ categories in the VBF signal region: (a) $\deta < 1.5$ and (b) $\deta > 1.5$.
The hatching shows the total uncertainty of the background estimate. 
The distribution of the expected background is obtained using the best-fit values of the nuisance parameters in the fit to the data with the background-only hypothesis.
Distributions for three choices of couplings are shown: the SM, $\kl= 6$,  and $\kvv = 0$ (with all other couplings set to their SM values in the last two models), scaled so as to be visible on the plot. 
The lower panels show the ratio of the observed data yield to the predicted background in each bin.
Events in the overflow bins are counted in the yields of the final bins.}
\label{fig:vbf-mhh-same}
\end{figure}

The observed distributions in \mhh, as well as the predicted background and example signal shapes, are presented in Figure~\ref{fig:ggF-4b-disc} for each of the six ggF categories (with all data-taking years combined, for presentation purposes). 
The distributions of the expected background are obtained using the best-fit values of the nuisance parameters in the fit to the data with the background-only hypothesis.
The corresponding \mhh distributions in the two VBF categories are shown in Figure~\ref{fig:vbf-mhh-same}.
The signal shape for $\kvv = 0$ in Figure~\ref{fig:vbf-mhh-same}(a) clearly shows the impact of the divergences in Figures \ref{"fig:VBF_k2v"} and \ref{"fig:VBF_kvkv"} not canceling out as discussed in Section~\ref{sec:intro}.
While the deviations from the SM studied in this analysis are below the level that violates unitarity, this behavior makes the VBF topology in this analysis particularly sensitive to \kvv.
The observed number of data events, predicted number of background events, and expected number of signal events for the SM ggF and VBF signals are summarized for each of the analysis categories in Table~\ref{tab:yields}.

% The next lines are included from the .//tables/yields.tex input file
 
\begin{table}[!tbh]
\centering
\caption{The yields in each analysis category of the data, expected background, and expected SM ggF and VBF signals.
The expected background yields are obtained using a fit to the data with the background-only hypothesis; the quoted uncertainties are the sum in quadrature of all the per-bin systematic uncertainties.
The expected signal yields are obtained from simulation.}
\begin{tabular}{lSSSS[table-format=4.4,table-number-alignment=center]}
\toprule
\multicolumn{1}{c}{\textbf{Category}}     & \multicolumn{1}{c}{\textbf{Data}}  & \multicolumn{1}{c}{\textbf{Expected}}    & \multicolumn{1}{c}{\textbf{ggF Signal}} & \multicolumn{1}{c}{\textbf{VBF Signal}} \\
&                                    & \multicolumn{1}{c}{\textbf{Background}}  & \multicolumn{1}{c}{\textbf{SM}}         & \multicolumn{1}{c}{\textbf{SM}}   \\
 
\hline
\multicolumn{1}{c}{ggF signal region} & \multicolumn{4}{c}{} \\
\hline

$\deta<0.5$, $\Xhh<0.95$                  & \num{1940}                     & \num{1935+-25}                        & \num{7.0}                             & 0.038   \\                               
$\deta<0.5$, $\Xhh>0.95$                  & \num{3602}                     & \num{3618+-37}                        & \num{6.5}                             & 0.036   \\                               
 
$0.5<\deta<1.0$, $\Xhh<0.95$              & \num{1924}                     & \num{1874+-21}                        & \num{5.1}                             & 0.037   \\                               
$0.5<\deta<1.0$, $\Xhh>0.95$              & \num{3540}                     & \num{3492+-35}                        & \num{4.7}                             & 0.040   \\                               
 
$\deta>1.0$, $\Xhh<0.95$                  & \num{1880}                     & \num{1739+-22}                        & \num{2.9}                             & 0.043   \\                               
$\deta>1.0$, $\Xhh>0.95$                  & \num{3285}                     & \num{3212+-37}                        & \num{2.8}                             & 0.041   \\

\midrule
\multicolumn{1}{c}{VBF signal region} & \multicolumn{4}{c}{} \\
\hline
 
$\deta<1.5$                               & \num{116}                      & \num{125.3+-4.4}                          & \num{0.37}                             & 0.090   \\                               
$\deta>1.5$                               & \num{241}                      & \num{230.6+-5.3}                          & \num{0.06}                             & 0.21   \\

\bottomrule
\end{tabular}
\label{tab:yields}
\end{table}

% End of text imported from the .//tables/yields.tex input file

An upper limit on the combined ggF and VBF \HH signal strength $\mu_{\text{ggF+VBF}}$ is computed using the asymptotic formula~\cite{Cowan:2010js} and based on the \CLs method~\cite{Read:2002hq}.
The observed (expected) 95\% CL upper limit on $\mu_{\text{ggF+VBF}}$ is found to be 5.4 (8.1).
The expected upper limits are obtained using a background-only hypothesis, excluding a $HH$ signal.
The upper limit on the combined $\mu_{\text{ggF+VBF}}$, as well as upper limits on the individual $\mu_{\text{ggF}}$ ($=\sigma_\text{ggF}/\sigma_\text{ggF}^\text{SM}$) and $\mu_{\text{VBF}}$ ($=\sigma_\text{VBF}/\sigma_\text{VBF}^\text{SM}$), are summarized in Table~\ref{tab:all-sm-xs-tab}.
For the individual $\mu_{\text{ggF}}$ and $\mu_{\text{VBF}}$ limits, the results are derived by treating the other production mode (VBF when placing limits on $\mu_{\text{ggF}}$, and vice-versa) as a background process, with its normalization only loosely constrained in the fit.
Compared to the previous ATLAS measurement of ggF \HH production in the \bbbb decay channel (using \SI{27}{\ifb} of early \RunTwo data)~\cite{EXOT-2016-31}, the upper limit on the ggF cross-section is over 50\% lower, with approximately 20\% of this improvement arising from advances in analysis techniques and object reconstruction.
Similarly, compared to the previous ATLAS measurement of VBF \HH production in the \bbbb decay channel, which used \SI{126}{\ifb} of data collected between 2016 and 2018~\cite{HDBS-2018-18-witherratum}, the upper limit on the VBF \HH cross-section is over 75\% lower, with this improvement arising entirely from advances in analysis technique and object reconstruction.

% The next lines are included from the .//tables/xsec_results.tex input file
\begin{table}[!tbh]
\renewcommand{\arraystretch}{1.5} 
\centering
\caption{The observed and expected upper limits on the SM ggF \HH production cross-section $\sigma_{\text{ggF}}$, SM VBF \HH production cross-section $\sigma_{\text{VBF}}$, and combined SM ggF and VBF \HH production cross-section $\sigma_{\text{ggF+VBF}}$ at the 95\% CL, expressed as multiples of the corresponding SM cross-sections.
The expected values are shown with corresponding one- and two-standard-deviation error bounds, and they are obtained using a background-only fit to the data.
When extracting the limits on $\sigma_{\text{ggF+VBF}}$, the relative contributions of ggF and VBF production to the total cross-section are fixed to the SM prediction.}
\begin{tabular}{c S S S S S S[table-format=3.1,table-number-alignment=center]}
\toprule
& \textbf{Observed Limit} & \textbf{$-2\sigma$} & \textbf{$-1\sigma$} & \textbf{Expected Limit} & \textbf{$+1\sigma$} & \textbf{$+2\sigma$}\\
\hline

$\mu_{\text{ggF}}$ & 5.5 & 4.4 & 5.9 & 8.2 & 12.4 & 19.6  \\
$\mu_{\text{VBF}}$ & 130  & 70 & 100 & 130 & 190 & 280  \\
$\mu_{\text{ggF}+\text{VBF}}$ & 5.4 & 4.3 & 5.8 & 8.1 & 12.2 & 19.1  \\
 
\bottomrule
 
\end{tabular}
\label{tab:all-sm-xs-tab}
\renewcommand{\arraystretch}{1.0} 
\end{table}

% End of text imported from the .//tables/xsec_results.tex input file

The total uncertainty in the upper limit of the cross-section is dominated by the uncertainty sources related to the background modeling procedure and theoretical predictions.
With only the statistical uncertainties of the reweighted $2\Pqb$ data, observed $4\Pqb$ data, and simulated signal samples included in the fit, the expected upper limit on $\mu_{\text{ggF+VBF}}$ is found to be $6.0$ times the SM prediction.
Including the uncertainty sources resulting from the background estimation (the bootstrap uncertainty, the uncertainty from the kinematic differences between the SR and CR1, and, in the ggF signal region, the \bbbf nonclosure uncertainty), the expected upper limit on $\mu_{\text{ggF+VBF}}$ is relaxed to $7.1$ times the SM prediction.
The further reduction of sensitivity to the value of 8.1, as quoted in Table~\ref{tab:all-sm-xs-tab}, is driven primarily by the uncertainties arising from theoretical predictions.
The relative impact of the various sources of systematic uncertainty on the expected upper limit on $\mu_\text{ggF+VBF}$ is summarized in Table~\ref{tab:systematics}.

% The next lines are included from the .//tables/systematics_table.tex input file
\begin{table}[!tbh]
\centering
\caption{Breakdown of the dominant systematic uncertainties.
The impact of the uncertainties on the expected upper limit on $\mu_\text{ggF+VBF}$ when re-evaluating the profile likelihood ratio after fixing the nuisance parameter(s) in question to its (their) best-fit value(s), while all remaining nuisance parameters remain free to float.
The impact is shown in \%.
Only (groups of) systematic uncertainties that have an impact of at least 1\% are shown. The impact of each experimental source of systematic uncertainty described in the text, as well as of all of them together, is less than 1\%.
}
\begin{tabular}{l r}
\toprule
\textbf{Source of Uncertainty} & $ \Delta \mu / \mu$ \\
\hline
 
\hline
\textbf{Theory  uncertainties}                                        &   \\
Theory uncertainty in signal cross-section                                                     & $-9.0$\%  \\
All other theory uncertainties                                                    & $-1.4$\%  \\
\hline
\textbf{Background modeling uncertainties}                                &   \\
Bootstrap uncertainty                                                               &  $-7.1$\%  \\
CR to SR extrapolation uncertainty                                          &  $-7.5$\%  \\
\bbbf nonclosure uncertainty                                                   &  $-2.0$\%  \\
\bottomrule
 
\end{tabular}
\label{tab:systematics}
\end{table}

% End of text imported from the .//tables/systematics_table.tex input file

Constraints are placed on the \kl and \kvv modifiers using two different interpretations, the first named the ``$95\%$ CL'' method and the second named the ``profile likelihood ratio'' method.
The former uses the signal strength $\mu$ as the POI, while the latter uses the vector of coupling modifiers $\vect{\kappa}=(\kl,\kvv)$.
The $95\%$ CL method allows for interpretation as a traditional search for an arbitrarily normalized set of signals with different shapes against an estimated background, while the profile likelihood ratio method allows for interpretation as to whether the data are compatible with the specific cross-section and shape predictions of the $\kappa$ framework.
The $95\%$ CL results presented here offer a consistent comparison with previous ATLAS \HH measurements.
The constraints obtained from the two interpretations are not expected to be identical, as the two strategies employ slightly different physical assumptions.
In the profile likelihood ratio interpretation, the signal strength is fixed to the prediction obtained for a specific coupling modifier configuration, while for the $95\%$ CL interpretation, the signal strength is allowed to float.
The profile likelihood ratio method utilizes a hypothesis consisting of the predicted background plus the SM \HH signal, while the $95\%$ CL results utilize a hypothesis containing only the predicted background and no \HH signal.
Given the relatively small size of the SM \HH signal compared to the predicted background, the use of different hypotheses is not expected to have a significant effect.
Additionally, $2\sigma$-level constraints are quoted from the profile likelihood ratio interpretation, as opposed to $95\%$ CL constraints.

\begin{figure}[!tbh]
\centering
\subfloat[]{ 
\includegraphics[width=0.45\textwidth]{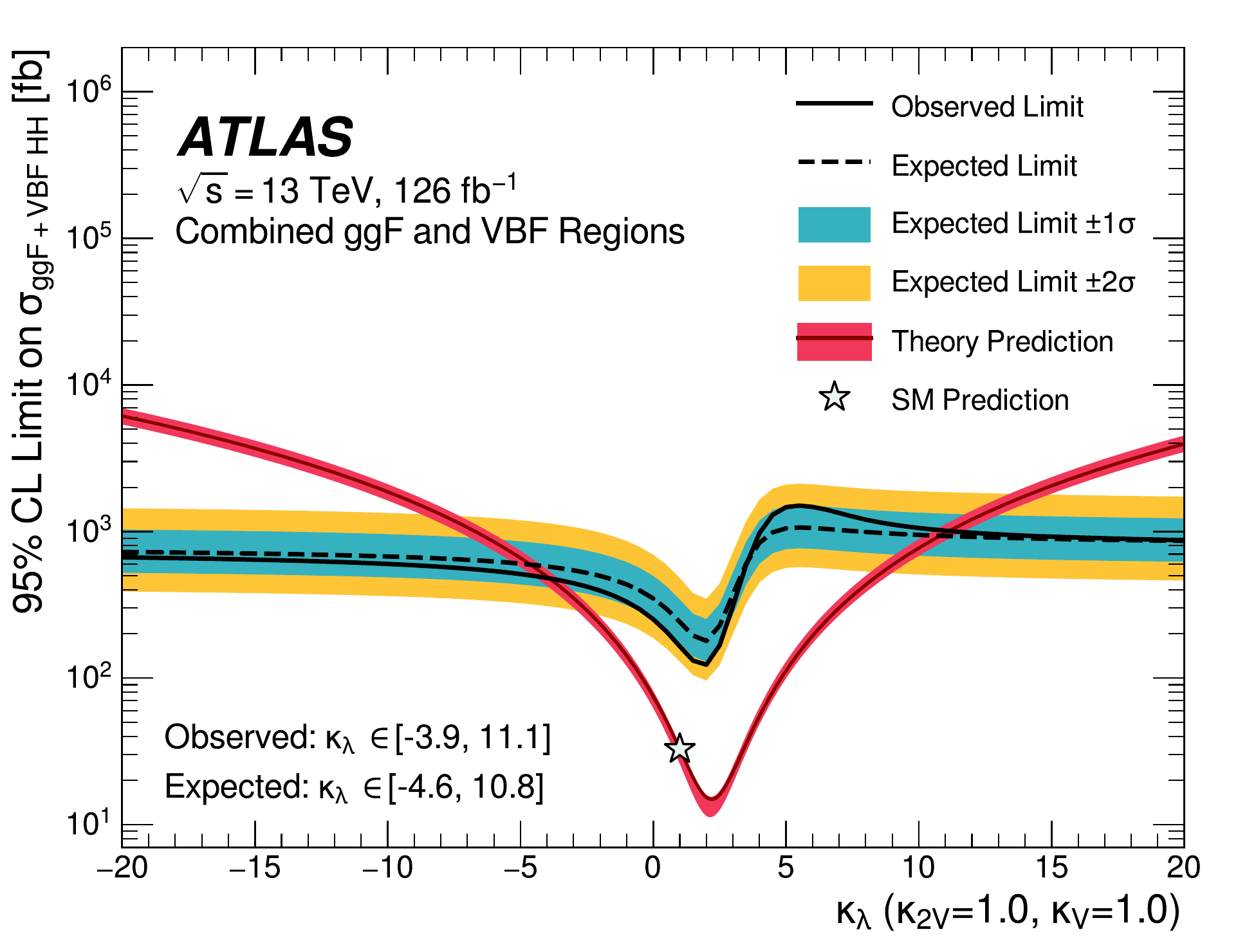}
}
\subfloat[]{ 
\includegraphics[width=0.45\textwidth]{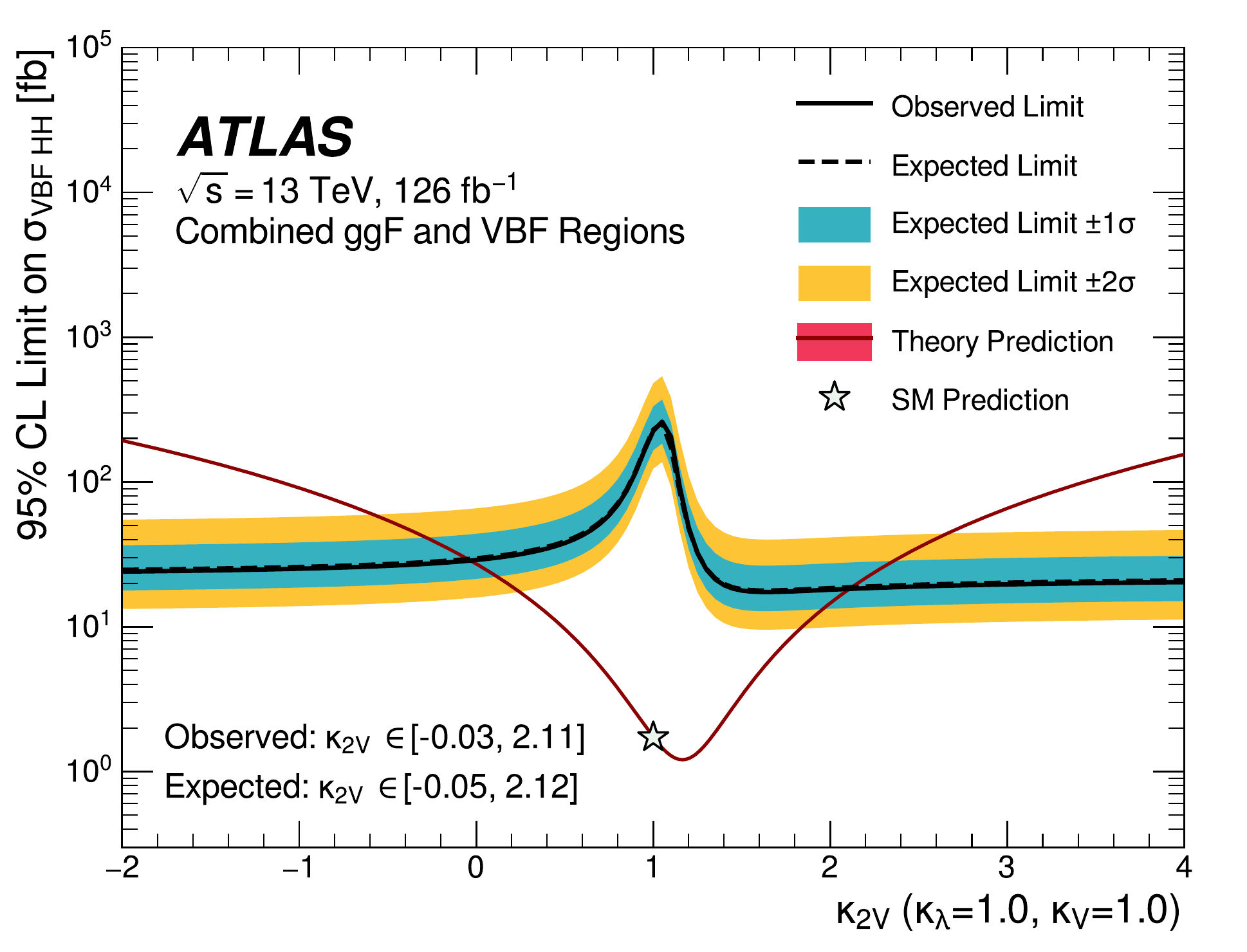}
}
\caption{The observed 95\% CL exclusion limits as a function of (a) \kl (obtained using the signal strength $\mu_{\text{ggF}+\text{VBF}}$ as the POI) and (b) \kvv (obtained using the signal strength $\mu_{\text{VBF}}$ as the POI) from the combined ggF and VBF signal regions, as shown by the solid black line.
In each case, the value of the other modifier is fixed to 1.
The blue and yellow bands show respectively the $1\sigma$ and $2\sigma$ bands around the expected exclusion limits, which are shown by the dashed black line.
The expected exclusion limits are obtained using a fit to the data with the background-only hypothesis.
The dark red line shows in (a) the predicted combined ggF and VBF \HH cross-section as a function of \kl and in (b) the predicted VBF \HH cross-section as a function of \kvv.
The dark pink bands surrounding the predicted cross-section lines indicate the theoretical uncertainty of the cross-section, as taken from Ref.~\cite{Baglio:2020wgt}.
The band in (b) is smaller than the width of the plotted line. }
\label{fig:res_1d_kl_kvv_scans}
\end{figure}

The $95\%$ CL constraints on \kl and \kvv are obtained by determining the 95\% CL upper limits on the cross-section as a function of these coupling modifiers, $\mu_{\text{ggF}+\text{VBF}}(\kl,\kvv)$.
Values of the coupling modifiers $(\kl, \kvv)$ are excluded if the predicted cross-section of the signal model obtained with that configuration is excluded at the 95\% CL.
The \hbb branching ratio is fixed to the SM prediction in the likelihood fit and any dependence on \kl is ignored.
Upper limits on the \HH signal strength as a function of \kl and \kvv are shown in Figure~\ref{fig:res_1d_kl_kvv_scans}, and the exclusion boundaries are summarized in Table~\ref{tab:1dkappa-xs-tab}.
With the values of the other modifiers (\kv and either \kvv or \kl, respectively) fixed to their SM value of 1, values of \kl beyond $[-3.9,11.1]$ and values of \kvv beyond $[-0.03,2.11]$ are excluded.

% The next lines are included from the .//tables/kappa_limits.tex input file
\begin{table}[bth]
\centering
\caption{The observed and expected constraints on the \kl and \kvv coupling modifiers at 95\% CL. For each modifier, the constraints were extracted with all other modifiers fixed to the SM prediction.}
\begin{tabular}{l SSSS[table-format=2.2,table-number-alignment=center]}
\toprule
\textbf{Parameter} & \multicolumn{2}{c}{\textbf{Expected Constraint}} & \multicolumn{2}{c}{\textbf{Observed Constraint}} \\
& {Lower} & {Upper}  & {Lower}  & {Upper}  \\
 
\hline{\kl}     & -4.6 & 10.8 & -3.9 & 11.1  \\
{\kvv}     & -0.05 & 2.12 & -0.03 & 2.11  \\
 
\bottomrule
\end{tabular}
\label{tab:1dkappa-xs-tab}
\end{table}

% End of text imported from the .//tables/kappa_limits.tex input file

\begin{figure}[!tbh]
\centering
\subfloat[]{ 
\includegraphics[width=0.45\textwidth]{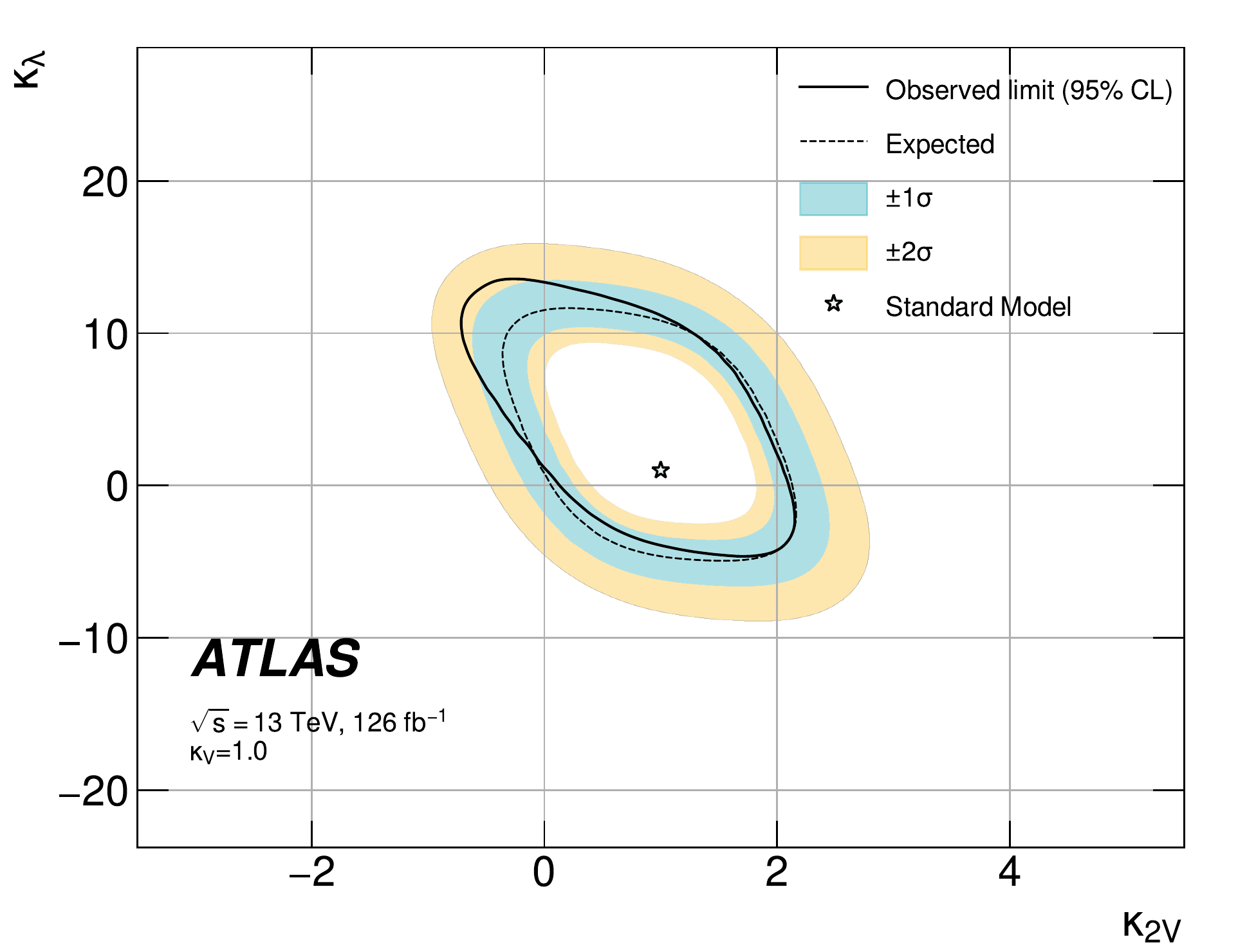}
}
\caption{The observed $95\%$ CL exclusion limit obtained using the signal strength $\mu_{\text{ggF}+\text{VBF}}$ as the POI in the two-dimensional \kl vs.\ \kvv space, obtained from the combined ggF and VBF signal model, as shown by the solid black line.
The blue and yellow bands show respectively the $1\sigma$ and $2\sigma$ bands around the expected exclusion limits, which are shown by the dashed black line.
The star denotes the SM prediction ($\kl=\kvv=1$). }
\label{fig:res_2dlims}
\end{figure}

Figure~\ref{fig:res_2dlims} shows the $95\%$ CL exclusion limits in the two-dimensional plane of the \kl--\kvv modifier space.

\begin{figure}[!tbh]
\centering
\subfloat[]{ 
\includegraphics[width=0.45\textwidth]{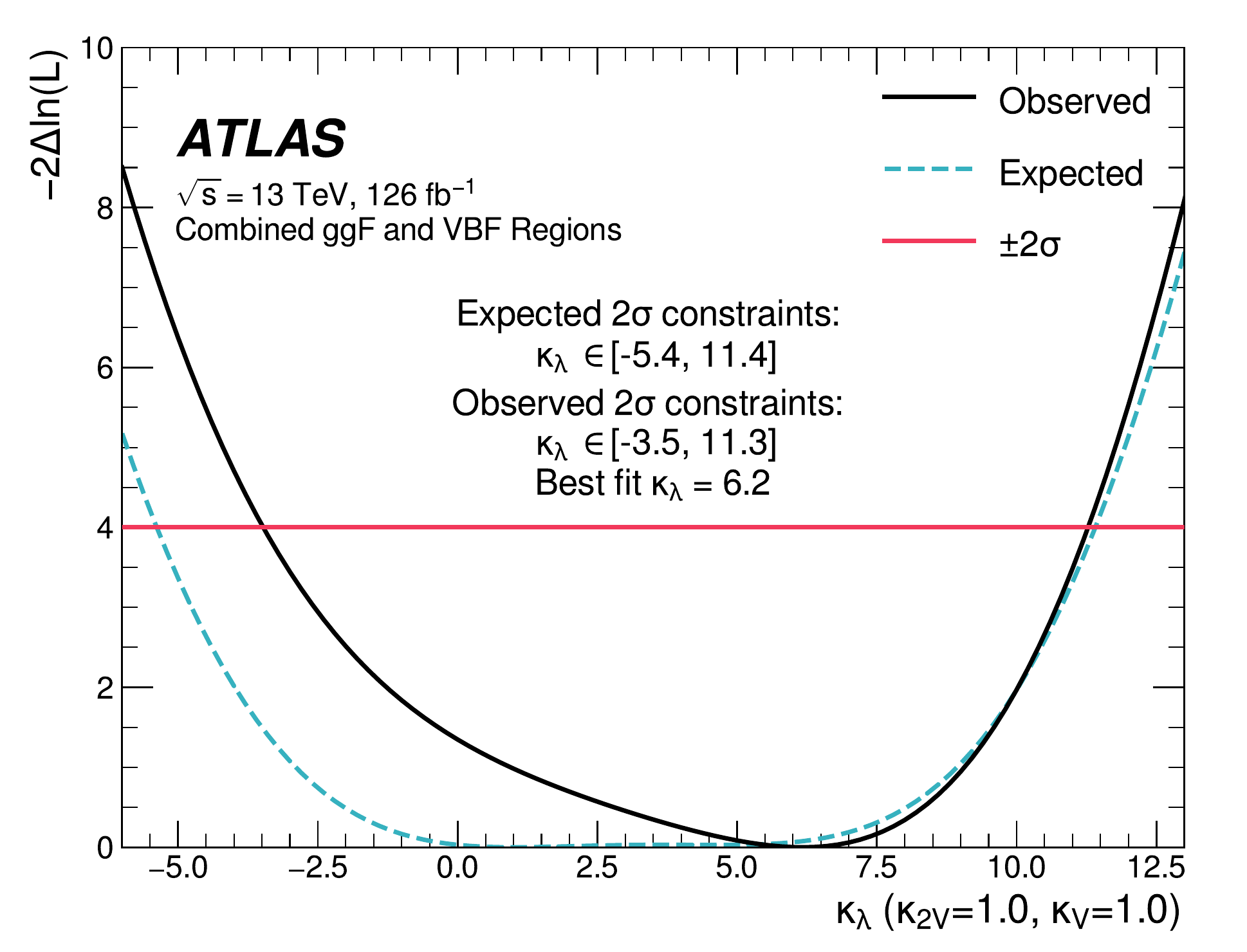}
}
\subfloat[]{ 
\includegraphics[width=0.45\textwidth]{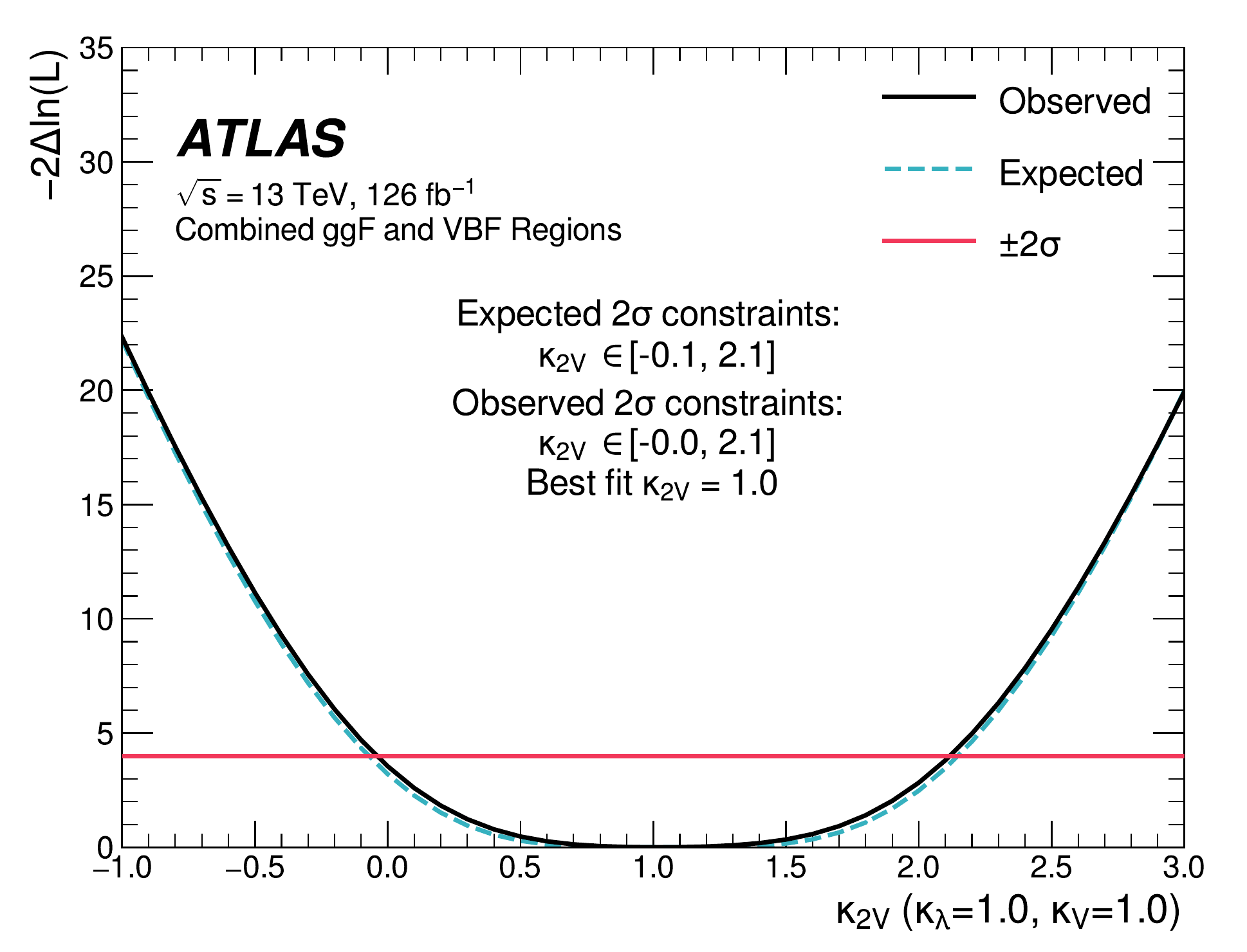}
}
\caption{The observed profile likelihood ratio scans for the (a) \kl and (b) \kvv coupling modifiers, shown by the solid black line, using the coupling modifiers $\vect{\kappa}$ as the POIs.
In each case, the value of the other parameter is fixed to 1.
The dashed blue line shows the expected profile likelihood ratio, as obtained using a fit to the data with the background-only hypothesis.
The pink line indicates the $2\sigma$ exclusion boundary.}
\label{fig:res_llh_scan_kl_kvv}
\end{figure}

\begin{figure}[!tbh]
\centering
\subfloat[]{
\includegraphics[width=0.45\textwidth]{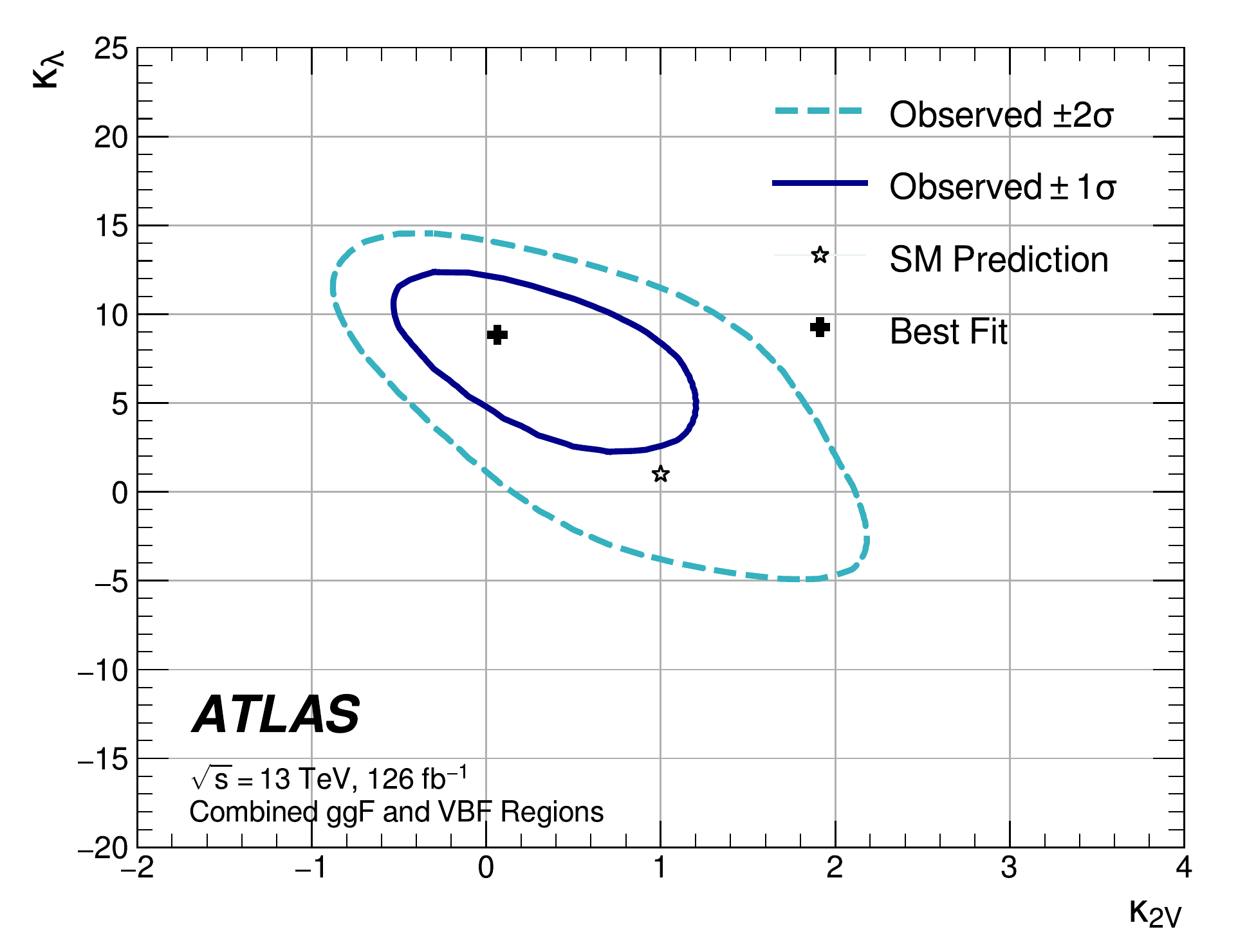}
}
\subfloat[]{
\includegraphics[width=0.45\textwidth]{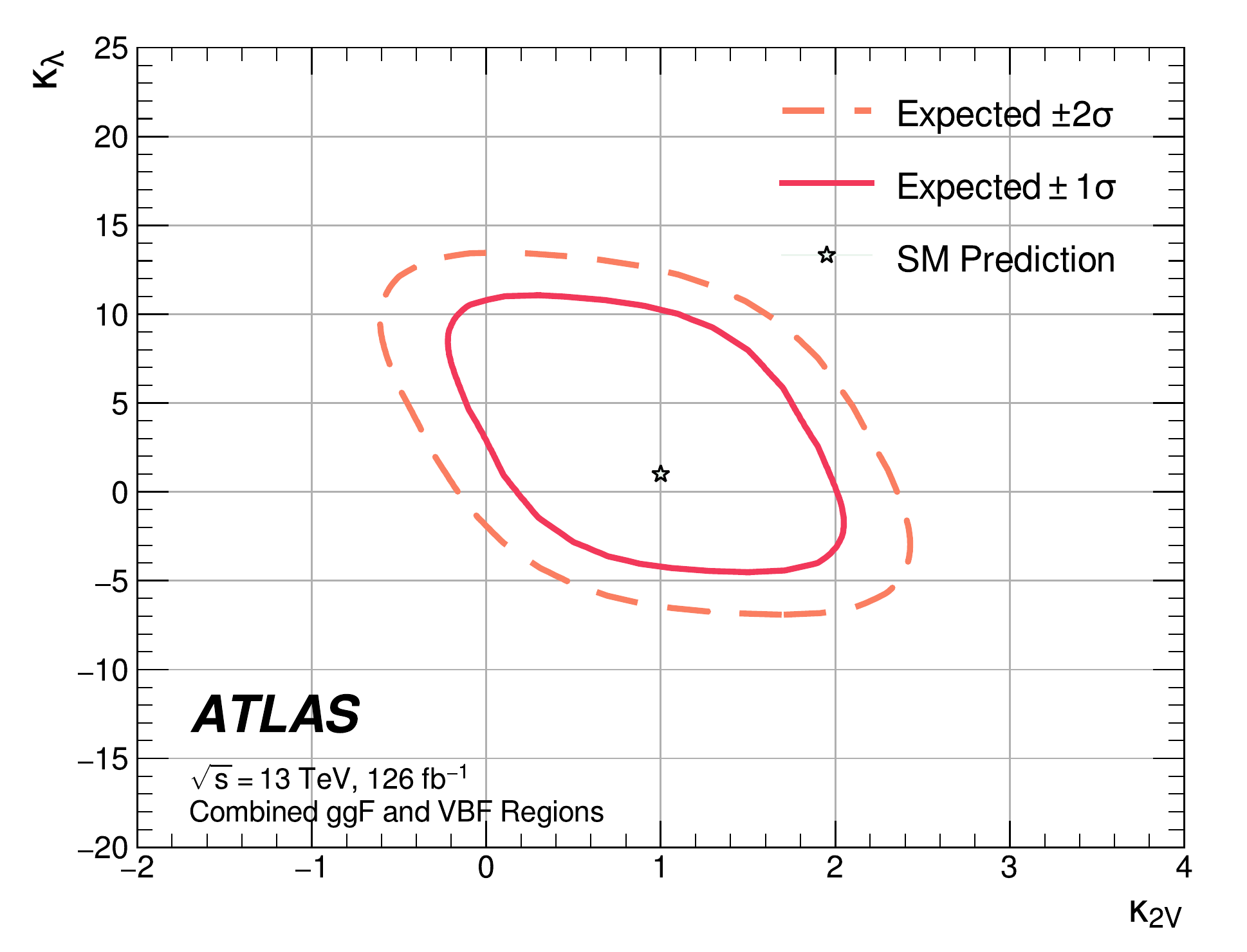}
}
\caption{
(a) The observed profile likelihood ratio exclusion limits for the two-dimensional \kl vs.\ \kvv modifier space, 
shown by the solid dark purple line at the $1\sigma$ level and the dashed turquoise line at the $2\sigma$ level.
The black cross denotes the best-fit values of $(\kl,\kvv)$. 
The expected exclusion limits are presented in (b), where the solid pink line denotes the $1\sigma$-level exclusion and the dashed orange line denotes the $2\sigma$-level exclusion.
For both the expected and observed limit plots, the black star indicates the SM prediction ($\kl=\kvv=1$).
}
\label{fig:res_2dlims_llh}
\end{figure}

The alternative coupling modifier constraints are obtained using the profile likelihood ratio interpretation, with the coupling modifiers $\vect{\kappa}=(\kl,\kvv)$ as the POIs, rather than the signal strength $\mu$:
\begin{equation}
-2 \Delta \ln L(\vect{\kappa})  = -2 \ln\Bigg(\frac{L(\vect{\kappa},\hat{\hat{\vect{\theta}}})}{L(\hat{\vect{\kappa}},\hat{\vect{\theta}})}\Bigg)\,.
\end{equation}
A scan of the profile likelihood ratio is performed as a function of the coupling modifiers at discrete points to produce the curves shown in Figure~\ref{fig:res_llh_scan_kl_kvv}.
The best-fit value of \kl is found to be 6.2 from the profile likelihood scan. 
The observed pull of the best-fit \kl value away from the SM value is due to a slight excess in the observed data in the ggF signal region, specifically in the low-\mhh range.
The particular signal model in which \kl is close to $6$ is favored due to a balance between two competing effects: the \mhh spectrum becomes softer as \kl increases away from the SM, but the cross-section also grows beyond the magnitude of the excess as \kl increases much further.
This slight excess also results in the deviation of the observed limits in Figure~\ref{fig:res_llh_scan_kl_kvv} from the expected limits by about $1\sigma$.
No such excess is observed in the VBF signal region, and the best-fit value of \kvv from the likelihood scan is found to be~$1.0$.
With the values of the other modifiers (\kv and either \kvv or \kl, respectively) fixed to their SM value of~1, the observed (expected) $2\sigma$ allowed range for \kl is found to be $[-3.5, 11.3]$ ($[-5.4, 11.4]$) and the corresponding range for \kvv is $[-0.0, 2.1]$ ($[-0.1, 2.1]$).

The exclusion constraints obtained using the profile likelihood ratio method are also presented in the two-dimensional \kl--\kvv coupling modifier space, similarly to the $95\%$ CL constraints described above.
The excluded regions are presented in Figure~\ref{fig:res_2dlims_llh}. With both modifiers able to float in the two-dimensional fit that combines both the ggF and VBF signal regions, the fit converges to \kl and \kvv values slightly different from the ones where the minimum is found in the fits with a single parameter free.

\begin{figure}[!tbh]
\centering
\subfloat[]{
\includegraphics[width=0.45\textwidth]{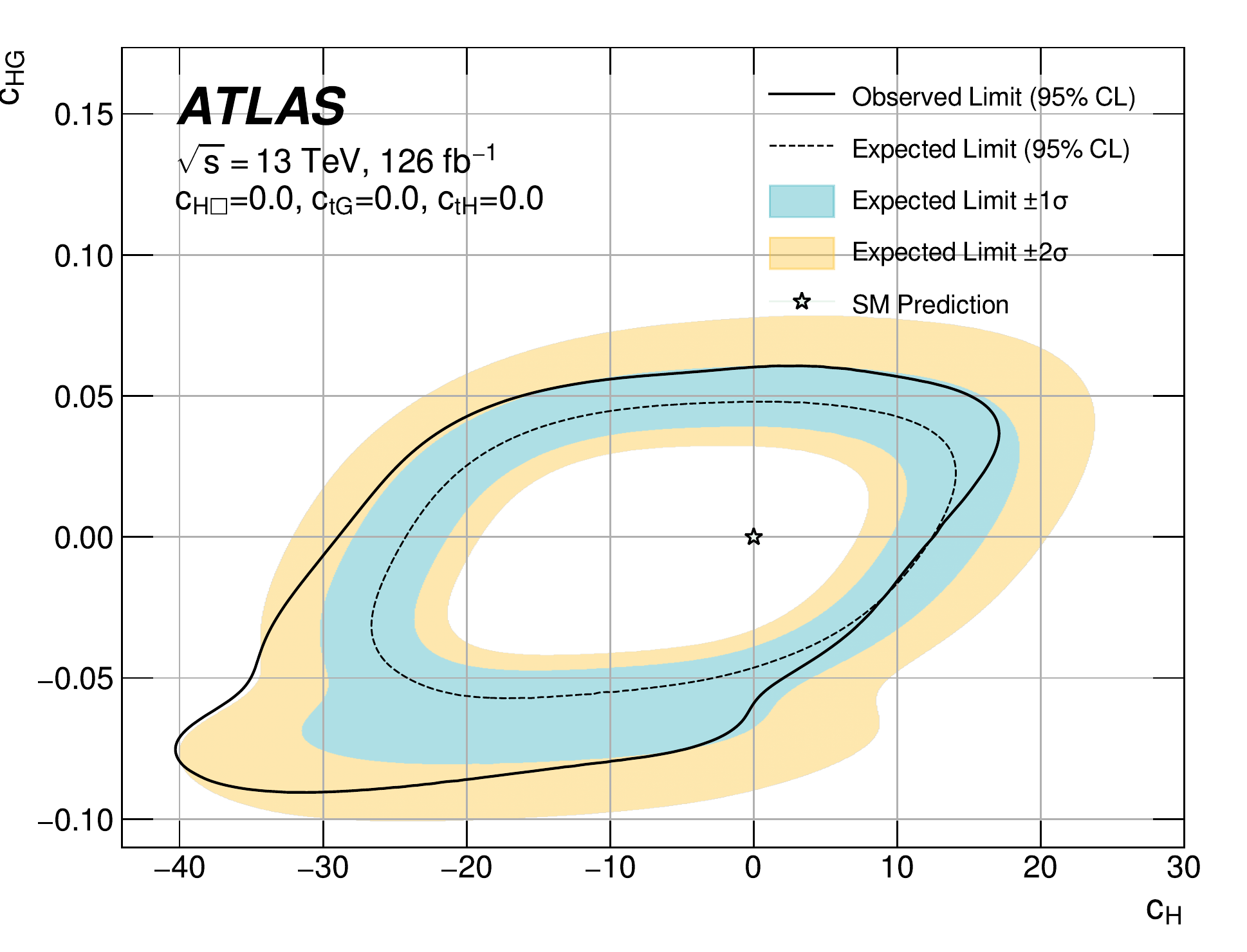}
}
\subfloat[]{
\includegraphics[width=0.45\textwidth]{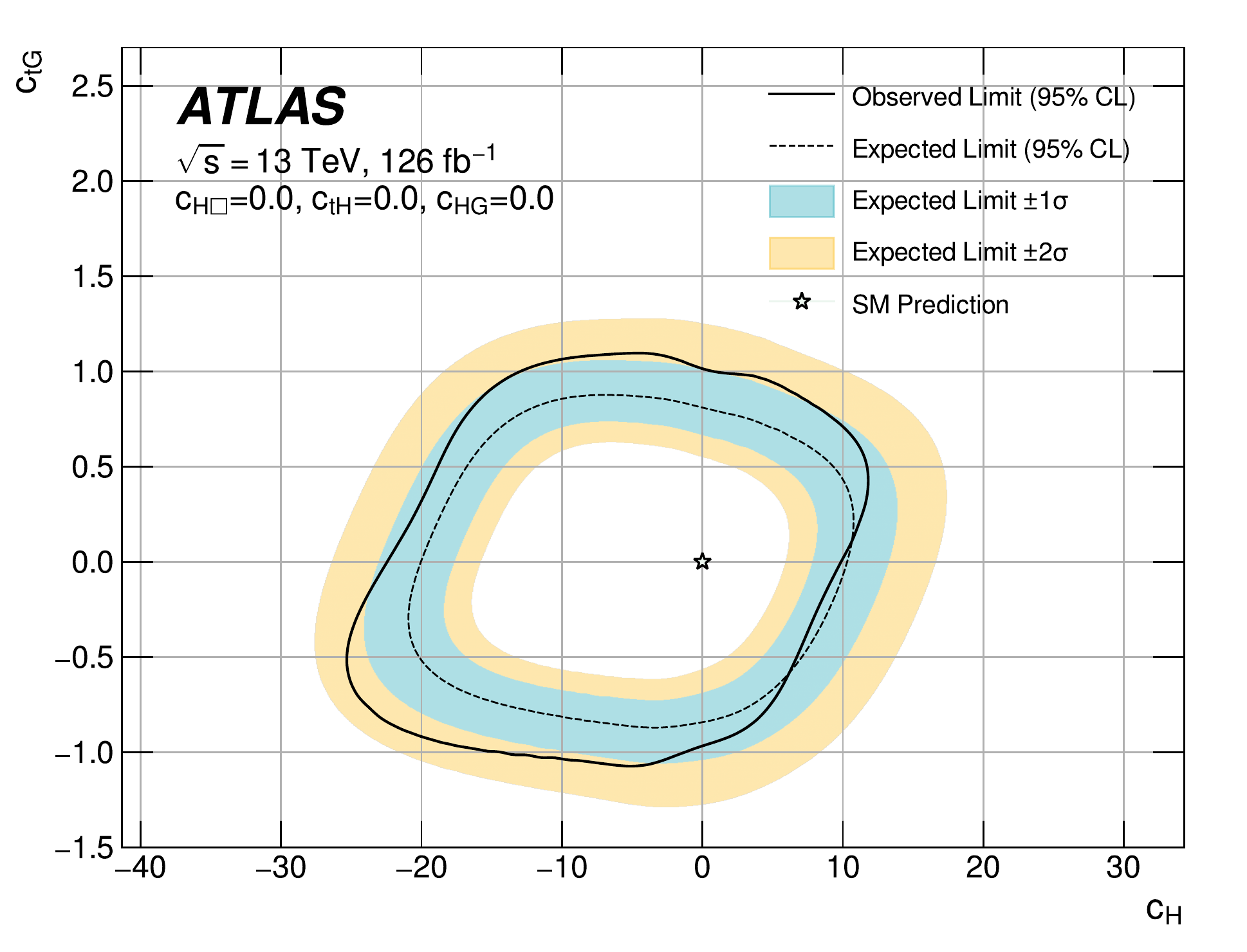}
}
\\
\subfloat[]{
\includegraphics[width=0.45\textwidth]{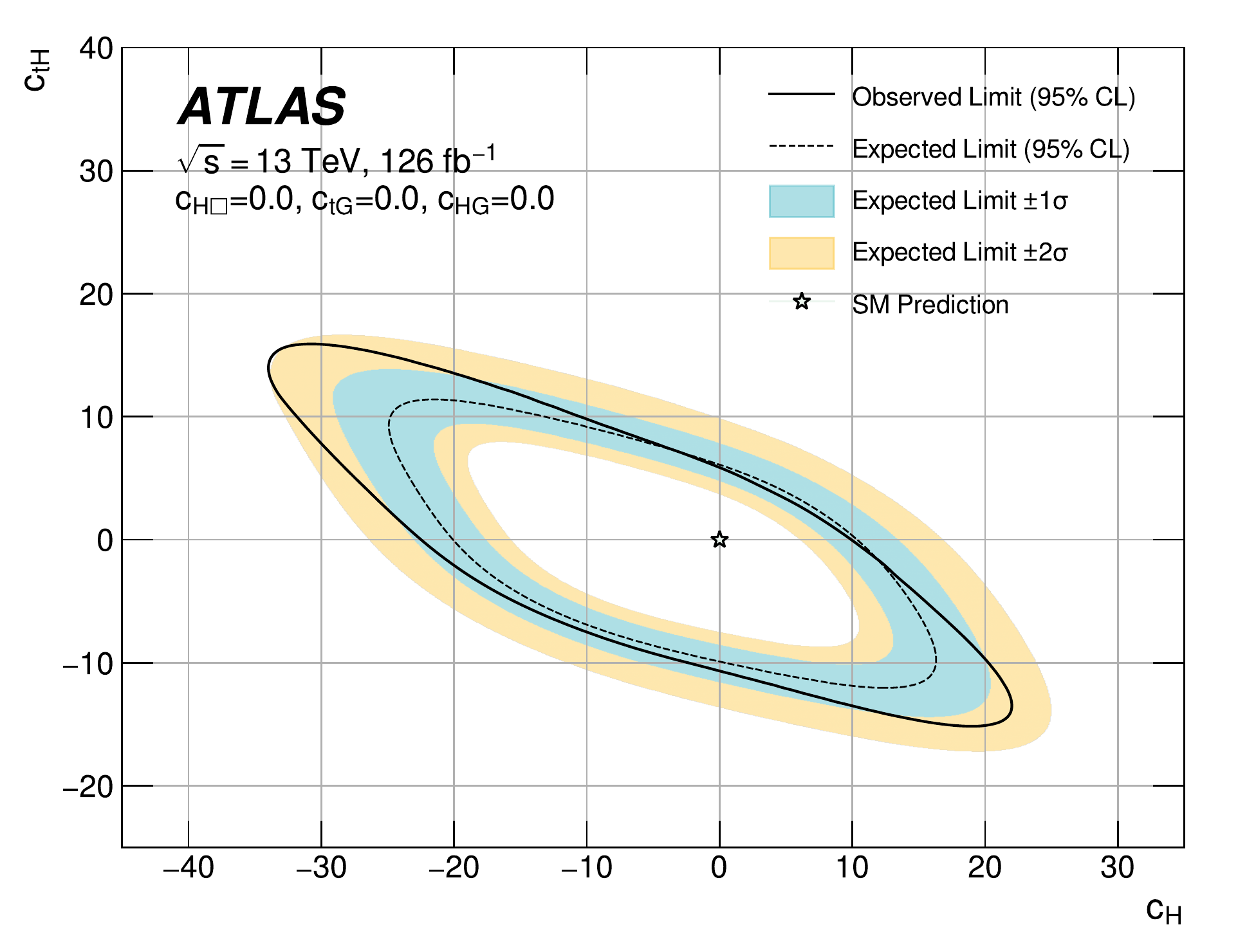}
}
\subfloat[]{
\includegraphics[width=0.45\textwidth]{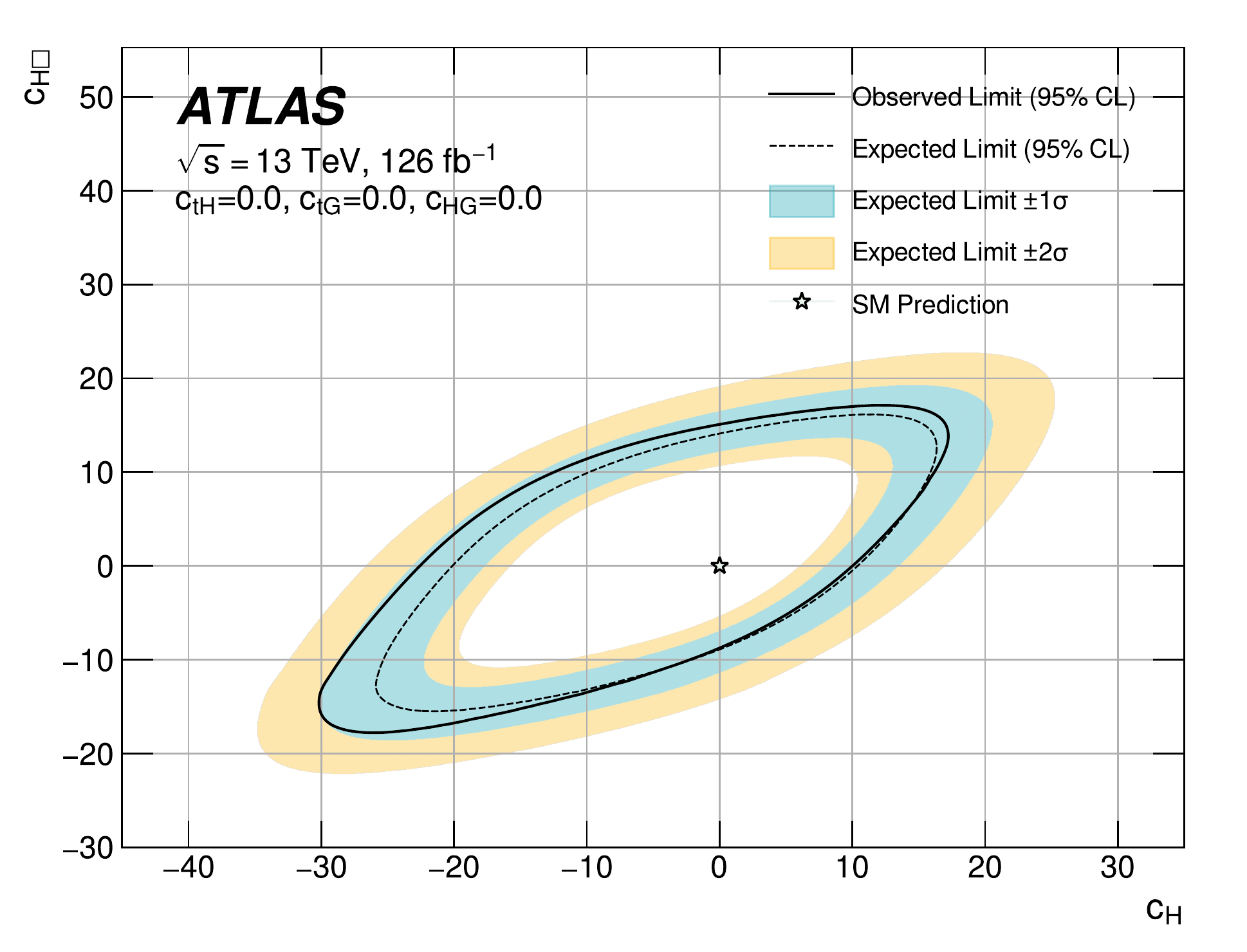}
}
\caption{
The observed $95\%$ CL exclusion limits on the SMEFT coefficients in the two-dimensional spaces (a) $c_{\higgs G}$ vs.\ $c_{\higgs}$, (b) $c_{tG}$ vs.\ $c_{\higgs}$, (c) $c_{t \higgs}$ vs.\ $c_{\higgs}$, and (d) $c_{\higgs\square}$ vs.\ $c_{\higgs}$, shown by the solid black lines.
The dashed black line indicates the expected $95\%$ CL exclusion limits.
The shaded blue band indicates the $\pm 1\sigma$ uncertainty of the exclusion limits, while the yellow band indicates the $\pm 2\sigma$ uncertainty.
The values of the other three coefficients for each plot are fixed to $0$. 
The VBF \HH process is ignored for this result.
}
\label{fig:smeft_2d_limits}
\end{figure}

\begin{figure}[!tbh]
\centering
\includegraphics[width=0.65\textwidth]{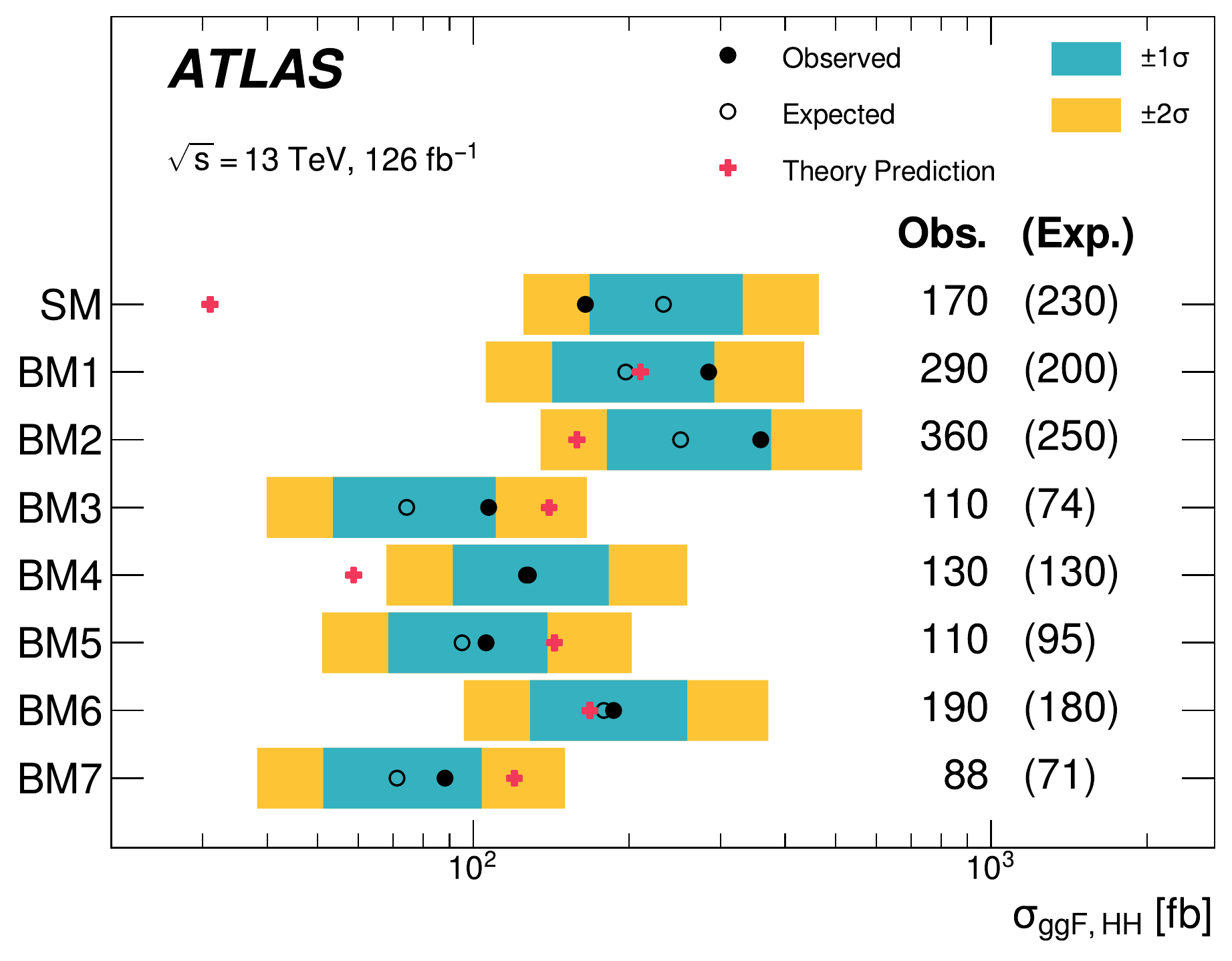}
\caption{The observed $95\%$ CL exclusion limits on the ggF \HH production cross-section in the SM and each of the seven HEFT benchmark models, given by the solid black points.
The blue and yellow bands show respectively the $1\sigma$ and $2\sigma$ bands around the expected upper limits, which are shown by the open circles.
The predicted ggF \HH production cross-section from each benchmark is indicated by a pink cross.
Benchmarks where the theory cross-section is higher than the exclusion limit (i.e. to the right) are excluded.
The VBF \HH process is ignored for this result.  }
\label{fig:heft_limits}
\end{figure}

In addition to constraints on the ggF and VBF \HH cross-sections and the \kl and \kvv coupling modifiers, constraints for relevant coefficients can be derived from the ggF selection of the analysis in the SMEFT and HEFT frameworks, as outlined in Section~\ref{sec:intro}.
The VBF \HH process was ignored for both the SMEFT and HEFT results; including the VBF \HH process as a background was found to have a negligible effect on the extracted parameter limits.
The slight dependence of the \hbb branching fraction on the SMEFT and HEFT coefficients is also ignored, as the impact on the analysis sensitivity is small.
Constraints on the SMEFT coefficients are extracted by considering the $95\%$ CL exclusion of the cross-section as a function of SMEFT parameter, as was done for the \kl and \kvv constraints discussed previously.
The extracted constraints on individual parameters in the scenario where the other parameters are fixed to $0$ are summarized in Table~\ref{tab:res_smeft}.
Limits approaching or exceeding $\pm 4\pi$ should be interpreted with caution because of the potential impact from effects such as missing higher-order model contributions.
The exclusion limits are also presented in two-dimensional SMEFT coefficient subspaces.
The exclusion limits for each coefficient versus the $c_{\higgs}$ coefficient (with the remaining three coefficients fixed to $0$) are shown in Figure~\ref{fig:smeft_2d_limits}.
The upper limits on the HEFT ggF \HH production cross-section in the seven benchmark models are presented in Figure~\ref{fig:heft_limits}.
The spread of sensitivity between the seven benchmark models reflects the different signal kinematics and, hence, shapes of the signal \mhh distributions.
The different variation between observed and expected limits is linked to the slight excess observed in the low \mhh region, as discussed earlier.
The red crosses in Figure~\ref{fig:heft_limits} indicate the predicted \HH cross-sections from the respective benchmark models.
As can be seen, BM3, BM5 and BM7 are observed to be excluded with more than 95\% confidence.
Constraints are placed on the values of $c_{gg\higgs\higgs}$ and $c_{t\overline{t}\higgs\higgs}$, with all other HEFT coefficients fixed to SM values.
The observed (expected) constraints on $c_{gg\higgs\higgs}$ are found to be $[-0.36, 0.78]$ ($[-0.42, 0.75]$), while the observed (expected) constraints on $c_{t\overline{t}\higgs\higgs}$ are found to be $[-0.55, 0.51]$ ($[-0.46, 0.40]$).

% The next lines are included from the .//tables/smeft_results.tex input file
\begin{table}[!tbh]
\centering
\caption{
The extracted upper and lower limits on the SMEFT parameters to which the analysis is sensitive.
For each parameter, the constraints are provided assuming the other parameters are fixed to $0$.
The VBF \HH process is ignored for this result.
}
\begin{tabular}{l SSSS[table-format=2.2,table-number-alignment=center]}
\toprule
\textbf{Parameter} & \multicolumn{2}{c}{\textbf{Expected Constraint}} & \multicolumn{2}{c}{\textbf{Observed Constraint}} \\
& {Lower} & {Upper}  & {Lower}  & {Upper}  \\
\hline
$c_{\higgs}$            & -20 & 11 &  -22 & 11 \\
$c_{\higgs G}$          & -0.056 & 0.049  & -0.067 & 0.060 \\
$c_{\higgs\square}$  		& -9.3 & 13.9 & -8.9 & 14.5 \\
$c_{t\higgs}$           & -10.0 & 6.4  & -10.7 & 6.2 \\
$c_{tG}$                & -0.97 & 0.94  & -1.12 & 1.15 \\
 
\bottomrule
\end{tabular}
\label{tab:res_smeft}
\end{table}

% End of text imported from the .//tables/smeft_results.tex input file

% End of text imported from the .//sections/result.tex input file

\FloatBarrier
 
\section{Conclusion}
\label{sec:conclusion}

% The next lines are included from the .//sections/conclusion.tex input file
A search for nonresonant pair production of Higgs bosons in the \bbbb final state was carried out, with dedicated analyses for the ggF and VBF production modes, using \SI{126}{\ifb} of \(\sqrt{s} = \SI{13}{\TeV}\) \(pp\) collision data collected by the ATLAS detector at the LHC.
The sensitivity of the analyses is improved relative to previous iterations by using more sophisticated background modeling techniques, event categorization and improved jet reconstruction and flavor identification algorithms, in addition to the increased integrated luminosity of the analyzed data.
 
No evidence of signal is found and the observed (expected) upper limit on the cross-section for nonresonant Higgs boson pair production is determined to be 5.4 (8.1) times the Standard Model predicted cross-section at 95\% confidence level.
Constraints are placed upon modifiers to the \HHH\ and \HHVV\ couplings.
The observed (expected) $2\sigma$ constraints on the \HHH\ coupling modifier, \kl, are determined to be $[-3.5, 11.3]$ ($[-5.4, 11.4]$), while the corresponding constraints for the \HHVV\ coupling modifier, \kvv, are $[-0.0, 2.1]$ ($[-0.1, 2.1]$). The results are also used to derive constraints on relevant coefficients in the SM effective field theory and the Higgs effective field theory frameworks.

% End of text imported from the .//sections/conclusion.tex input file

\section*{Acknowledgments}

% The next lines are included from the .//acknowledgements/Acknowledgements.tex input file

We thank CERN for the very successful operation of the LHC, as well as the
support staff from our institutions without whom ATLAS could not be
operated efficiently.
 
We acknowledge the support of
ANPCyT, Argentina;
YerPhI, Armenia;
ARC, Australia;
BMWFW and FWF, Austria;
ANAS, Azerbaijan;
CNPq and FAPESP, Brazil;
NSERC, NRC and CFI, Canada;
CERN;
ANID, Chile;
CAS, MOST and NSFC, China;
Minciencias, Colombia;
MEYS CR, Czech Republic;
DNRF and DNSRC, Denmark;
IN2P3-CNRS and CEA-DRF/IRFU, France;
SRNSFG, Georgia;
BMBF, HGF and MPG, Germany;
GSRI, Greece;
RGC and Hong Kong SAR, China;
ISF and Benoziyo Center, Israel;
INFN, Italy;
MEXT and JSPS, Japan;
CNRST, Morocco;
NWO, Netherlands;
RCN, Norway;
MEiN, Poland;
FCT, Portugal;
MNE/IFA, Romania;
MESTD, Serbia;
MSSR, Slovakia;
ARRS and MIZ\v{S}, Slovenia;
DSI/NRF, South Africa;
MICINN, Spain;
SRC and Wallenberg Foundation, Sweden;
SERI, SNSF and Cantons of Bern and Geneva, Switzerland;
MOST, Taiwan;
TENMAK, T\"urkiye;
STFC, United Kingdom;
DOE and NSF, United States of America.
In addition, individual groups and members have received support from
BCKDF, CANARIE, Compute Canada and CRC, Canada;
PRIMUS 21/SCI/017 and UNCE SCI/013, Czech Republic;
COST, ERC, ERDF, Horizon 2020 and Marie Sk{\l}odowska-Curie Actions, European Union;
Investissements d'Avenir Labex, Investissements d'Avenir Idex and ANR, France;
DFG and AvH Foundation, Germany;
Herakleitos, Thales and Aristeia programmes co-financed by EU-ESF and the Greek NSRF, Greece;
BSF-NSF and MINERVA, Israel;
Norwegian Financial Mechanism 2014-2021, Norway;
NCN and NAWA, Poland;
La Caixa Banking Foundation, CERCA Programme Generalitat de Catalunya and PROMETEO and GenT Programmes Generalitat Valenciana, Spain;
G\"{o}ran Gustafssons Stiftelse, Sweden;
The Royal Society and Leverhulme Trust, United Kingdom.
 
The crucial computing support from all WLCG partners is acknowledged gratefully, in particular from CERN, the ATLAS Tier-1 facilities at TRIUMF (Canada), NDGF (Denmark, Norway, Sweden), CC-IN2P3 (France), KIT/GridKA (Germany), INFN-CNAF (Italy), NL-T1 (Netherlands), PIC (Spain), ASGC (Taiwan), RAL (UK) and BNL (USA), the Tier-2 facilities worldwide and large non-WLCG resource providers. Major contributors of computing resources are listed in Ref.~\cite{ATL-SOFT-PUB-2021-003}.

% End of text imported from the .//acknowledgements/Acknowledgements.tex input file

\printbibliography
 
\clearpage
% ATLAS Collaboration author list
% Reference date of HDBS-2019-29 is 2022-09-07
% Author list last updated on date 29-JUN-23
% Data extracted on 29-Jun-2023 for paper reference HDBS-2019-29
% at 11:33am
 
\begin{flushleft}
\hypersetup{urlcolor=black}
{\Large The ATLAS Collaboration}

\bigskip

\AtlasOrcid[0000-0002-6665-4934]{G.~Aad}$^\textrm{\scriptsize 102}$,
\AtlasOrcid[0000-0002-5888-2734]{B.~Abbott}$^\textrm{\scriptsize 120}$,
\AtlasOrcid[0000-0002-1002-1652]{K.~Abeling}$^\textrm{\scriptsize 55}$,
\AtlasOrcid[0000-0002-8496-9294]{S.H.~Abidi}$^\textrm{\scriptsize 29}$,
\AtlasOrcid[0000-0002-9987-2292]{A.~Aboulhorma}$^\textrm{\scriptsize 35e}$,
\AtlasOrcid[0000-0001-5329-6640]{H.~Abramowicz}$^\textrm{\scriptsize 151}$,
\AtlasOrcid[0000-0002-1599-2896]{H.~Abreu}$^\textrm{\scriptsize 150}$,
\AtlasOrcid[0000-0003-0403-3697]{Y.~Abulaiti}$^\textrm{\scriptsize 117}$,
\AtlasOrcid[0000-0003-0762-7204]{A.C.~Abusleme~Hoffman}$^\textrm{\scriptsize 137a}$,
\AtlasOrcid[0000-0002-8588-9157]{B.S.~Acharya}$^\textrm{\scriptsize 69a,69b,p}$
% ...
\bigskip
\\

% ...
$^{29}$Physics Department, Brookhaven National Laboratory, Upton NY; United States of America.\\
% ...
$^{35}$$^{(a)}$Facult\'e des Sciences Ain Chock, R\'eseau Universitaire de Physique des Hautes Energies - Universit\'e Hassan II, Casablanca;$^{(b)}$Facult\'{e} des Sciences, Universit\'{e} Ibn-Tofail, K\'{e}nitra;$^{(c)}$Facult\'e des Sciences Semlalia, Universit\'e Cadi Ayyad, LPHEA-Marrakech;$^{(d)}$LPMR, Facult\'e des Sciences, Universit\'e Mohamed Premier, Oujda;$^{(e)}$Facult\'e des sciences, Universit\'e Mohammed V, Rabat;$^{(f)}$Institute of Applied Physics, Mohammed VI Polytechnic University, Ben Guerir; Morocco.\\
% ...
$^{55}$II. Physikalisches Institut, Georg-August-Universit\"{a}t G\"ottingen, G\"ottingen; Germany.\\
% ...
$^{69}$$^{(a)}$INFN Gruppo Collegato di Udine, Sezione di Trieste, Udine;$^{(b)}$ICTP, Trieste;$^{(c)}$Dipartimento Politecnico di Ingegneria e Architettura, Universit\`a di Udine, Udine; Italy.\\
% ...
$^{102}$CPPM, Aix-Marseille Universit\'e, CNRS/IN2P3, Marseille; France.\\
% ...
$^{117}$Department of Physics, New York University, New York NY; United States of America.\\
% ...
$^{120}$Homer L. Dodge Department of Physics and Astronomy, University of Oklahoma, Norman OK; United States of America.\\
% ...
$^{137}$$^{(a)}$Departamento de F\'isica, Pontificia Universidad Cat\'olica de Chile, Santiago;$^{(b)}$Millennium Institute for Subatomic physics at high energy frontier (SAPHIR), Santiago;$^{(c)}$Instituto de Investigaci\'on Multidisciplinario en Ciencia y Tecnolog\'ia, y Departamento de F\'isica, Universidad de La Serena;$^{(d)}$Universidad Andres Bello, Department of Physics, Santiago;$^{(e)}$Instituto de Alta Investigaci\'on, Universidad de Tarapac\'a, Arica;$^{(f)}$Departamento de F\'isica, Universidad T\'ecnica Federico Santa Mar\'ia, Valpara\'iso; Chile.\\
% ...
$^{150}$Department of Physics, Technion, Israel Institute of Technology, Haifa; Israel.\\
$^{151}$Raymond and Beverly Sackler School of Physics and Astronomy, Tel Aviv University, Tel Aviv; Israel.\\
% ...
$^{a}$ Also Affiliated with an institute covered by a cooperation agreement with CERN.\\
$^{b}$ Also at An-Najah National University, Nablus; Palestine.\\
$^{c}$ Also at Borough of Manhattan Community College, City University of New York, New York NY; United States of America.\\
$^{d}$ Also at Bruno Kessler Foundation, Trento; Italy.\\
$^{e}$ Also at Center for High Energy Physics, Peking University; China.\\
$^{f}$ Also at Center for Interdisciplinary Research and Innovation (CIRI-AUTH), Thessaloniki ; Greece.\\
% ...
$^{p}$ Also at Department of Physics, King's College London, London; United Kingdom.\\
% ...

\end{flushleft}

% Created with Glance <Atlas.Glance@cern.ch>

\end{document}